\documentclass[aps,pre]
{revtex4}
\usepackage[dvips]{graphics}
\usepackage{graphicx}
\usepackage{amsfonts}
\usepackage{amssymb}
\usepackage{amsmath}
\usepackage{subfigure}

\begin{document}

\title{A two-dimensional paradigm for symmetry breaking: \\
the nonlinear Schr{\"{o}}dinger equation with a four-well potential}
\author{C. Wang$^{1}$, G. Theocharis$^{1}$, P. G.\ Kevrekidis$^{1}$, 
N. Whitaker$^{1}$, K. J. H.\ Law$^{1}$, D. J.\ Frantzeskakis$^{2}$, 
and B.A. Malomed$^{3}$
}
\affiliation{
$^{1}$ Department of Mathematics and Statistics, University of Massachusetts,
Amherst MA 01003-4515, USA \\ 
$^{2}$ Department of Physics, University of Athens, Panepistimiopolis, Zografos,
Athens 15784, Greece \\ 
$^{3}$ Department of Physical Electronics, Faculty of Engineering, Tel Aviv
University, Tel Aviv 69978, Israel}

\begin{abstract}
We study the existence and stability of localized states in 
%a model based on
the two-dimensional (2D) nonlinear Schr{\"{o}}dinger (NLS)/Gross-Pitaevskii
equation with a symmetric four-well potential. Using a four-mode 
approximation, we are able to trace the parametric evolution of the trapped
stationary modes, starting from the corresponding linear limits, and thus
derive the complete bifurcation diagram for the families of these stationary
modes. The predictions based on the four-mode decomposition are found to be
in good agreement with the numerical results obtained from the 
%full 
NLS equation. Actually, the stability properties coincide with those suggested
by the corresponding discrete model in the large-amplitude
limit. The dynamics of the unstable modes is explored by means of direct
simulations. Finally, while we present the full analysis for the case of the
focusing nonlinearity, the bifurcation diagram for the defocusing case is
briefly considered  too.
\end{abstract}

\maketitle

\section{Introduction}

In the recent years, there has been a considerable effort 
%much effort has been focused 
on experimental and theoretical studies of Bose-Einstein condensates (BECs) 
\cite{book1,book2}. Many of these studies were focused on macroscopic nonlinear 
structures that arise in BECs, which often have counterparts in nonlinear optics 
\cite{agra}. One of the appealing features of this setting is the possibility 
to tailor the desirable geometry of magnetic, optical, or combined traps that 
confine the ultracold bosonic atoms. For this reason, the analysis of the existence,
stability and dynamical properties of nonlinear localized 
%modes 
states trapped in these geometries have become a focal point of research. 
The theoretical analysis is enabled by the fact that a very accurate description 
of 
%rarefied 
dilute atomic BECs is furnished, in the mean-field approximation, by the 
Gross-Pitaevskii (GP) equation, which is a variant of the nonlinear 
Schr{\"{o}}dinger (NLS) equation. The cubic nonlinearity in the GP equation 
originates from the 
%atomics collisions, 
interatomic interactions, accounted for through an effective mean-field.
%should be taken into account despite the rarefaction of the condensate. 
The NLS equation in this, as well as in somewhat different forms, is relevant
to a 
%number 
variety of alternative physical applications in nonlinear optics and 
other areas \cite{agra,ourbook,sulem}.

Among the trapping configurations available in current BEC experiments, one
that has drawn considerable attention is the double-well potential (DWP).
Its prototypical realization is provided by the a strong parabolic 
(harmonic) trap combined with a periodic potential, which can be created as
an ``optical lattice'', by a set of coherent laser beams illuminating the
condensate \cite{book1,book2,ourbook}. The DWP was realized experimentally 
in \cite{markus1}, using the magnetic field to induce the parabolic trap. 
The experiments reported in Ref. \cite{markus1} 
%had 
revealed a variety of fundamental effects, including tunneling and Josephson 
oscillations for a small number of atoms, or 
%the 
macroscopic quantum self-trapping leading to a stable asymmetric partition 
of atoms between the wells for a sufficiently large number of atoms. 
DWPs have also inspired theoretical studies of various topics, such 
as finite-mode reductions, finding exact analytical results for specially 
designed shapes of the potential, quantum effects 
\cite{smerzi,kiv2,mahmud,bam,Bergeman_2mode,infeld,todd,theo,carr}, 
and a nonlinear DWP (alias double-well pseudopotential), which is induced 
by the respective spatial modulation of the nonlinearity coefficient \cite{pseudo}.
It is relevant to mention that DWPs have also been studied in the context of
nonlinear optics, including twin-core self-guided laser beams in Kerr media 
\cite{HaeltermannPRL02} and optically-induced 
%dual-core 
waveguiding structures in photorefractive crystals \cite{zhigang}.

The aim of the present work is to extend the analysis of the DWP to a 
two-dimensional (2D) setting. Unlike previous studies of the trapping of 
quasi-2D BECs under the combined action of harmonic traps and 
optical-lattice potentials \cite{kody}, we implement a Galerkin-type 
few-mode reduction to deduce a discrete model, based on a symmetric set of 
four wells, which is the most natural configuration in the 2D case. The same 
model can also be realized in optics, using a bulk nonlinear medium with a 
set of four embedded waveguiding rods. In that setting, we use numerical 
methods to generate a bifurcation diagram for possible stationary states of 
the system. It is 
%worthy to note 
worth noting that all the states that are expected on 
the basis of a four-site discrete nonlinear Schr{\"{o}}dinger (DNLS)
reduction \cite{wannier} are also obtained in the continuum model
with the combined parabolic and periodic potential considered
herein. Furthermore, their stability, in the large
nonlinearity limit, coincides with what is expected from the DNLS model.

The paper is structured as follows. In section \ref{analytical}, we present
the model and the derivation of the four-mode approximation. Numerical
results are reported in section \ref{numerical}. We present complete
bifurcation diagrams of the possible stationary states for both the
underlying GP equation and for its four-mode reduction. Comparison between
them demonstrates very good agreement. In addition to the study of the
existence and stability, evolution of unstable modes is explored too 
by means of direct numerical simulations. 
Finally, we summarize 
%the 
our findings in section \ref{conclusion}, where we 
also discuss possible directions for further work.

\section{The model and the Galerkin approximation}

\label{analytical}

We start by presenting the basic model in the quasi-2D setting, namely the 
NLS/GP equation in $(2+1)$-dimensions, which is expressed in the following 
dimensionless form 
%is described by the following normalized NLS/GP equation 
\cite{book1,book2,ourbook}:
\begin{equation}
i\partial_{t}u=\hat{L}u+s|u|^{2}u-\mu u,  \label{eq1}
\end{equation}
where $u(x,y,t)$ is the normalized wave function, $\mu$ the chemical potential 
($-\mu $ is the propagation constant in the optical realization of the model), 
$s=\pm 1$ corresponds, respectively, to repulsive or attractive interatomic interactions 
in BEC (alias self-defocusing or self-focusing Kerr nonlinearity, in terms of nonlinear
optics), and $\hat{L}$ is the single-particle operator given by:
\begin{equation}
\hat{L}=-\frac{1}{2}\Delta + V(x,y).  \label{eq2}
\end{equation}
In Eq. (\ref{eq2}), $\Delta \equiv \partial_{x}^2 + \partial_{y}^2$ is the $2D$ Laplacian,
%is the operator in the 
while $V(x,y)$ is the four-well potential, assumed to take the following form,
\begin{equation}
V(x,y)=\frac{1}{2}\Omega ^{2}r^{2}+V_{0}\left[ \mathrm{cos}(2kx)+\mathrm{cos}
(2ky)\right],  \label{eq3}
\end{equation}
with $r^{2} \equiv x^{2}+y^{2}$. It is clear that $V(x,y)$ is composed of a harmonic 
trap of strength $\Omega$ and a periodic (OL) potential with strength $V_0$ and period 
$d=\pi/k$.
In the following analysis 
%presented 
below, we adopt the following representative
values for parameters of the potential: $\Omega =0.21$, $V_{0}=0.5$ and 
$k=0.3$, in which case the four smallest eigenvalues of operator $\hat{L}$
are found to be
\begin{equation}
\omega _{0}=0.3585, 
\,\,\,\, 
\omega _{1}=\omega _{2}=0.3658, 
\,\,\,\, 
\omega _{3}=0.3731.
\label{omega}
\end{equation}

\begin{figure}[tbph]
\centering
\includegraphics[width=.24\textwidth]{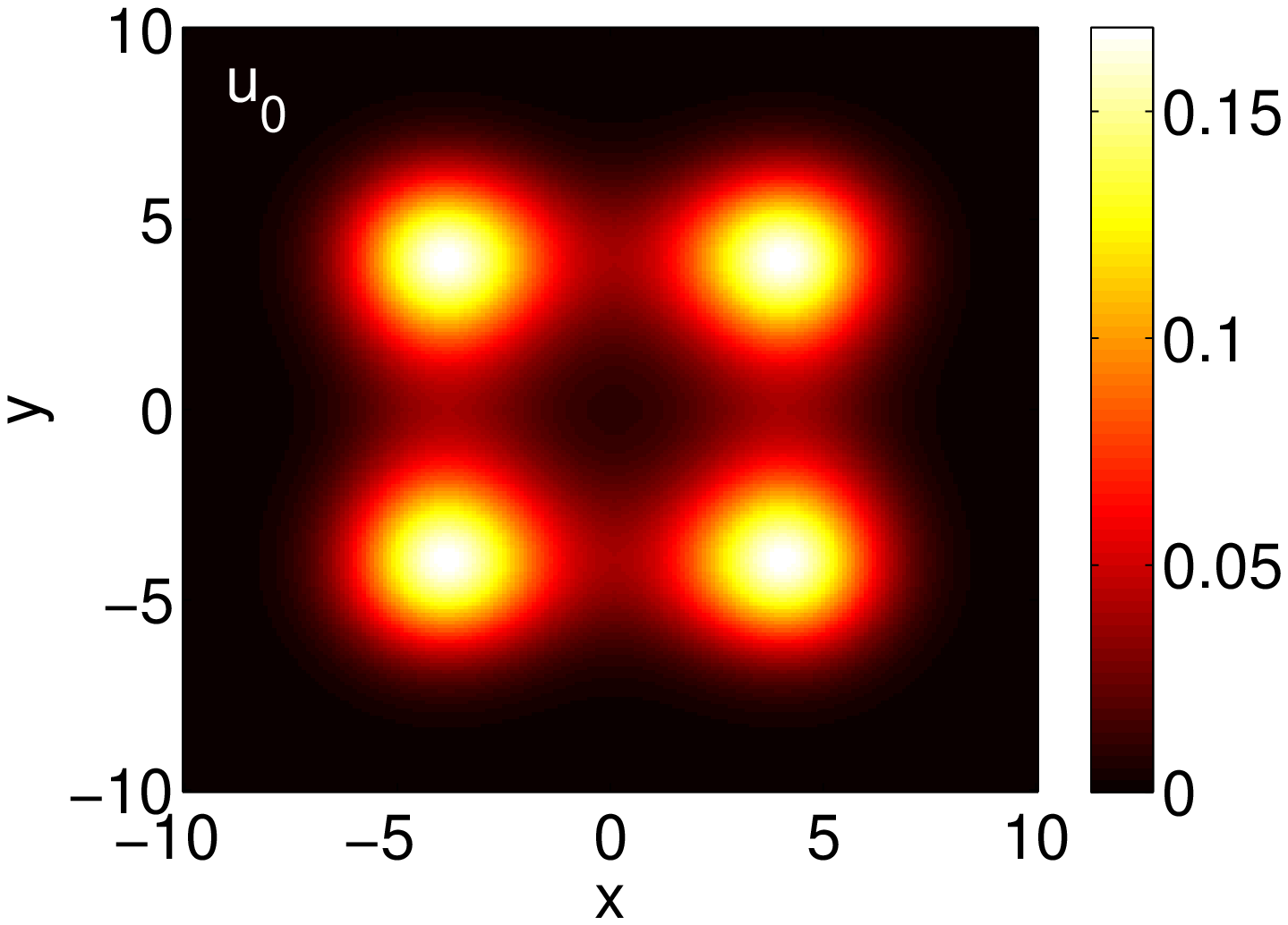} %
\includegraphics[width=.24\textwidth]{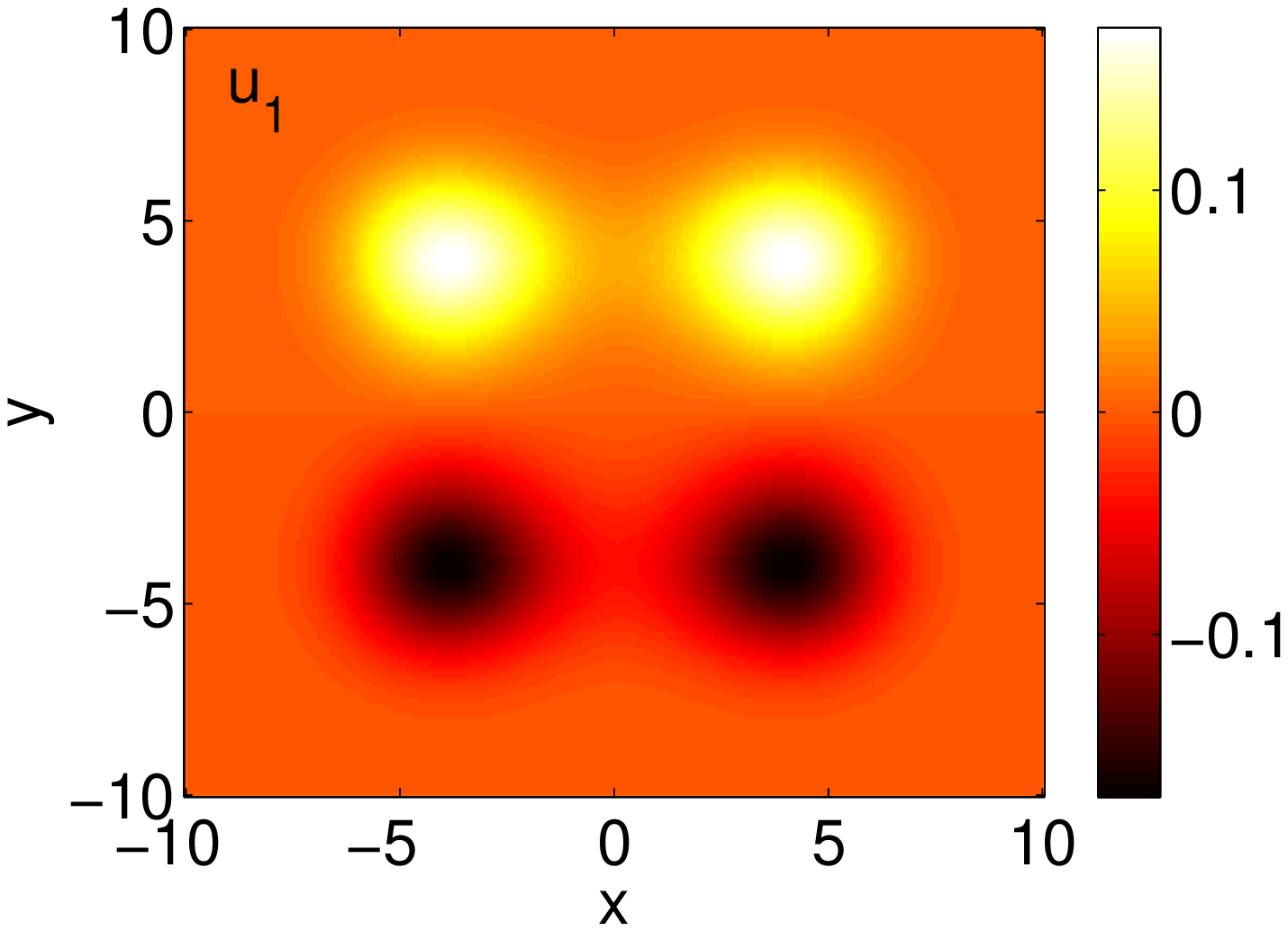}\newline
\includegraphics[width=.24\textwidth]{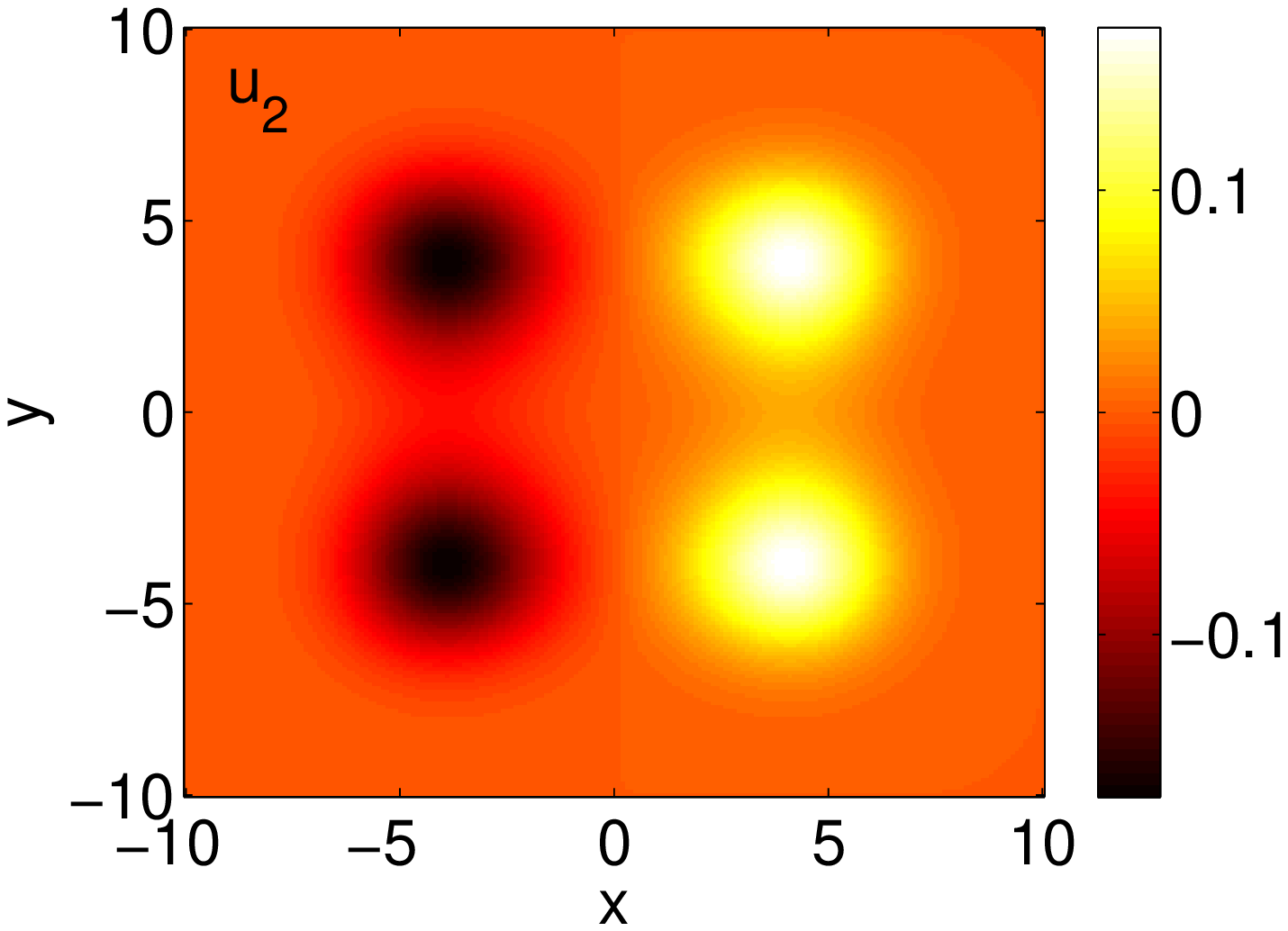} %
\includegraphics[width=.24\textwidth]{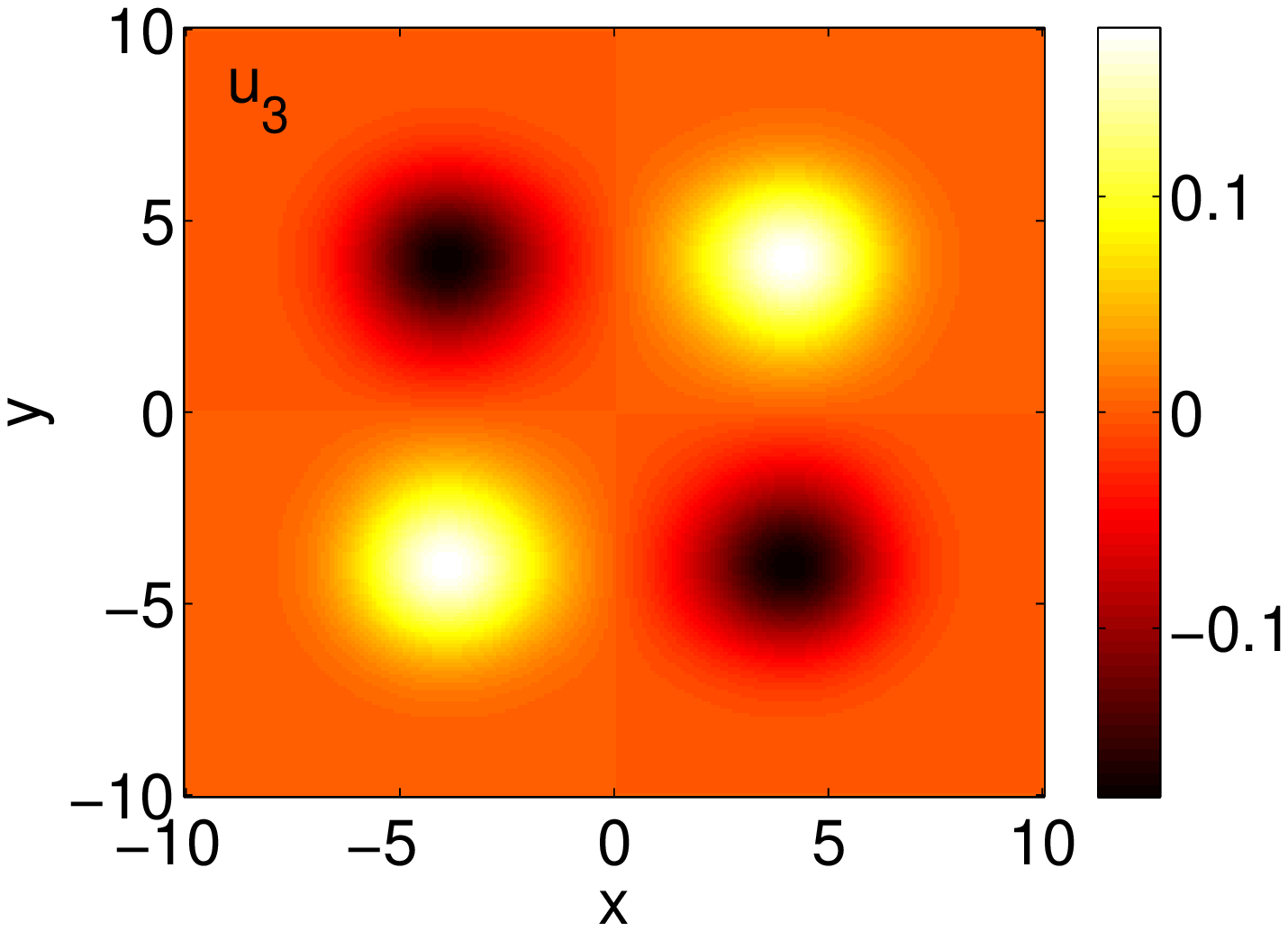}\newline
\caption{(Color online) The wave functions of the ground state, $u_{0}$, and
the first three excited states, $u_{1}$, $u_{2}$ and $u_{3}$, for the
four-well potential of Eq. (\protect\ref{eq3}) with $\Omega =0.21$, $%
V_{0}=0.5$ and $k=0.3$. Note the difference in the grayscale (color, in the
online version) bars in the first and three others panels, related to the
fact that the wave function of the ground state is positive, while the
excited states feature sign-changing patterns.}
\label{figEst}
\end{figure}

In a weakly nonlinear setting, we implement the natural possibility of a
four-mode approximation, based on a Galerkin-type expansion of $u(x,y,t)$
and truncation of the higher-order modes. We denote the ground and first
three excited eigenstates of the linear operator $\hat{L}$, shown in Fig. %
\ref{figEst}, as $u_{0}$ and $u_{1,2,3}$. This set constitutes a natural
minimum basis for the Galerkin approximation in the system of four potential
wells coupled by tunneling. Eigenstates $u_{j}$ ($j=0,1,2,3$) can be chosen
to be real, given the Hermitian nature of the operator $\hat{L}$. Then,
solutions of Eq. (\ref{eq1}) for values of the chemical potential in a
vicinity of linear eigenvalues (\ref{omega}), may be approximated by linear
combinations of the four eigenfuctions.

\begin{figure}[tbph]
\centering
\includegraphics[width=.24\textwidth]{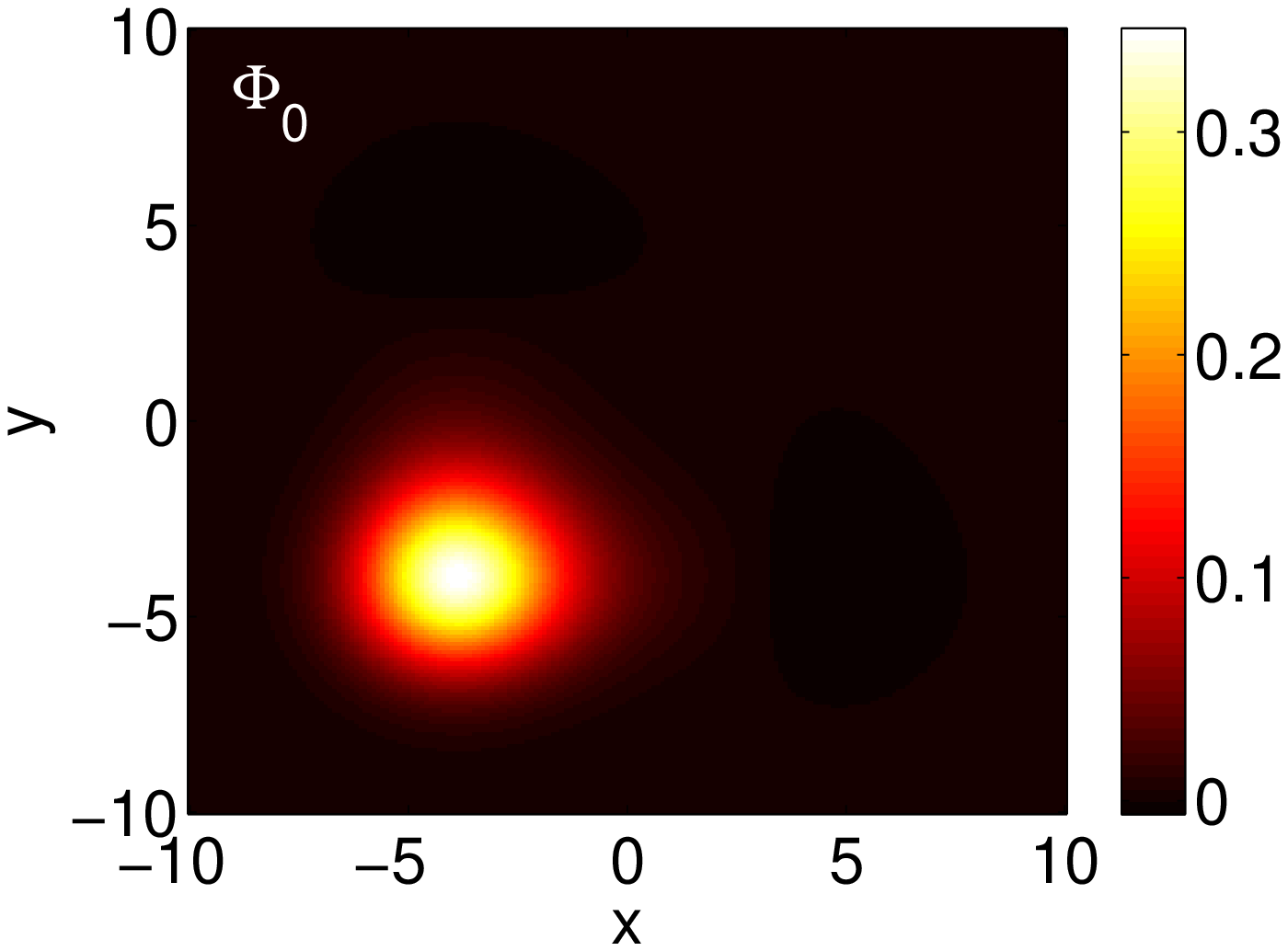} \includegraphics[width=.24%
\textwidth]{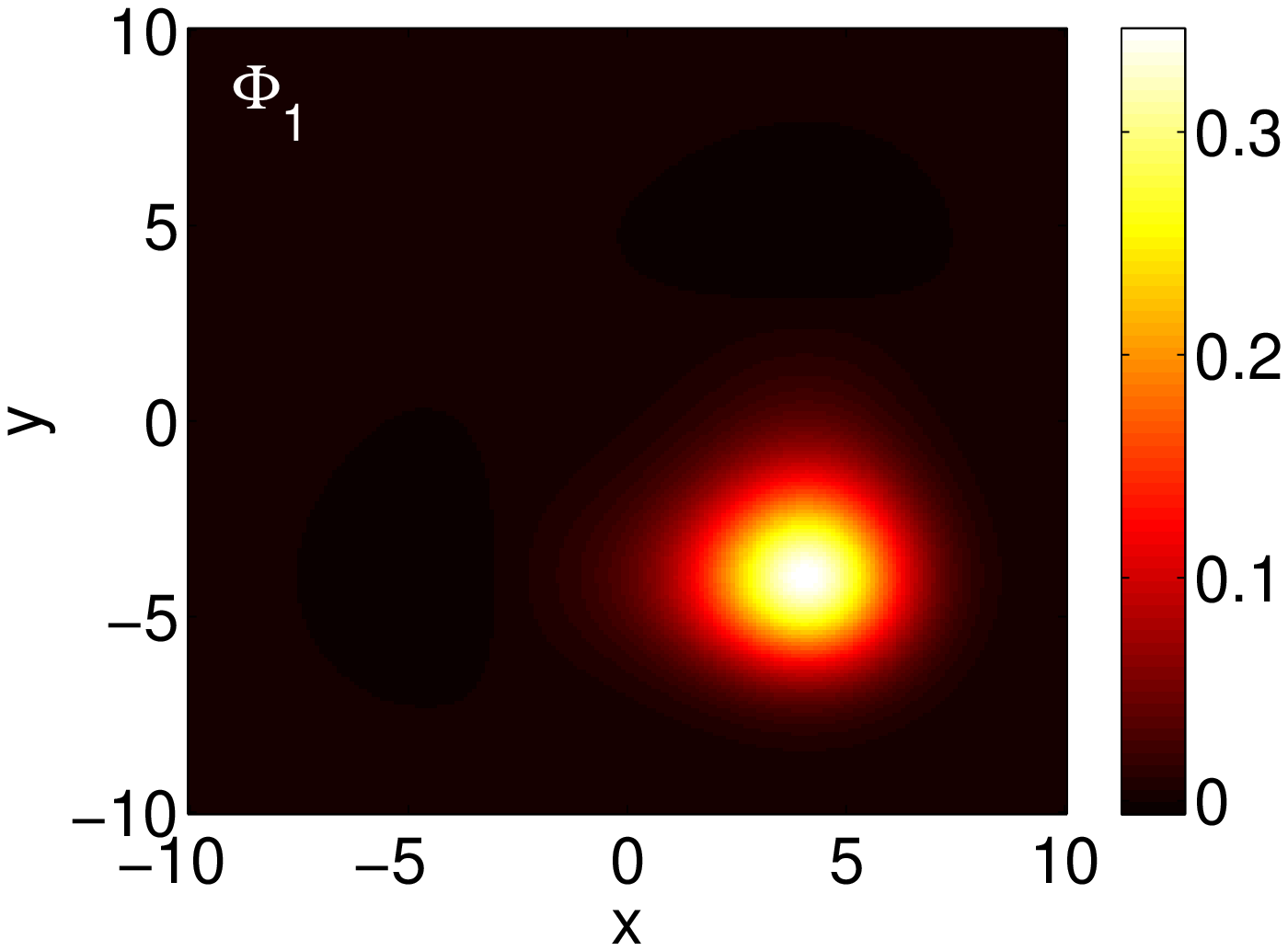}\newline
\includegraphics[width=.24\textwidth]{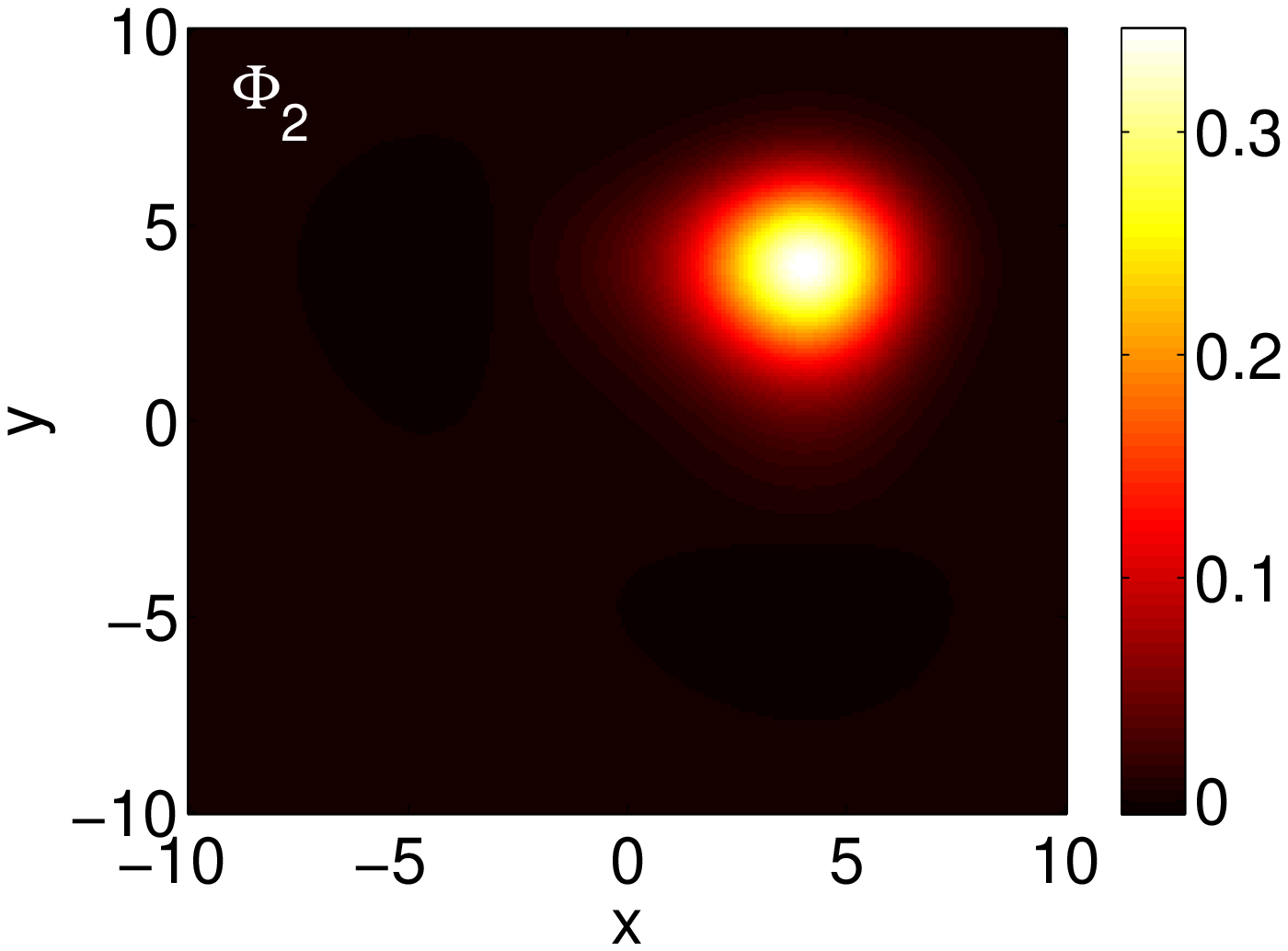} \includegraphics[width=.24%
\textwidth]{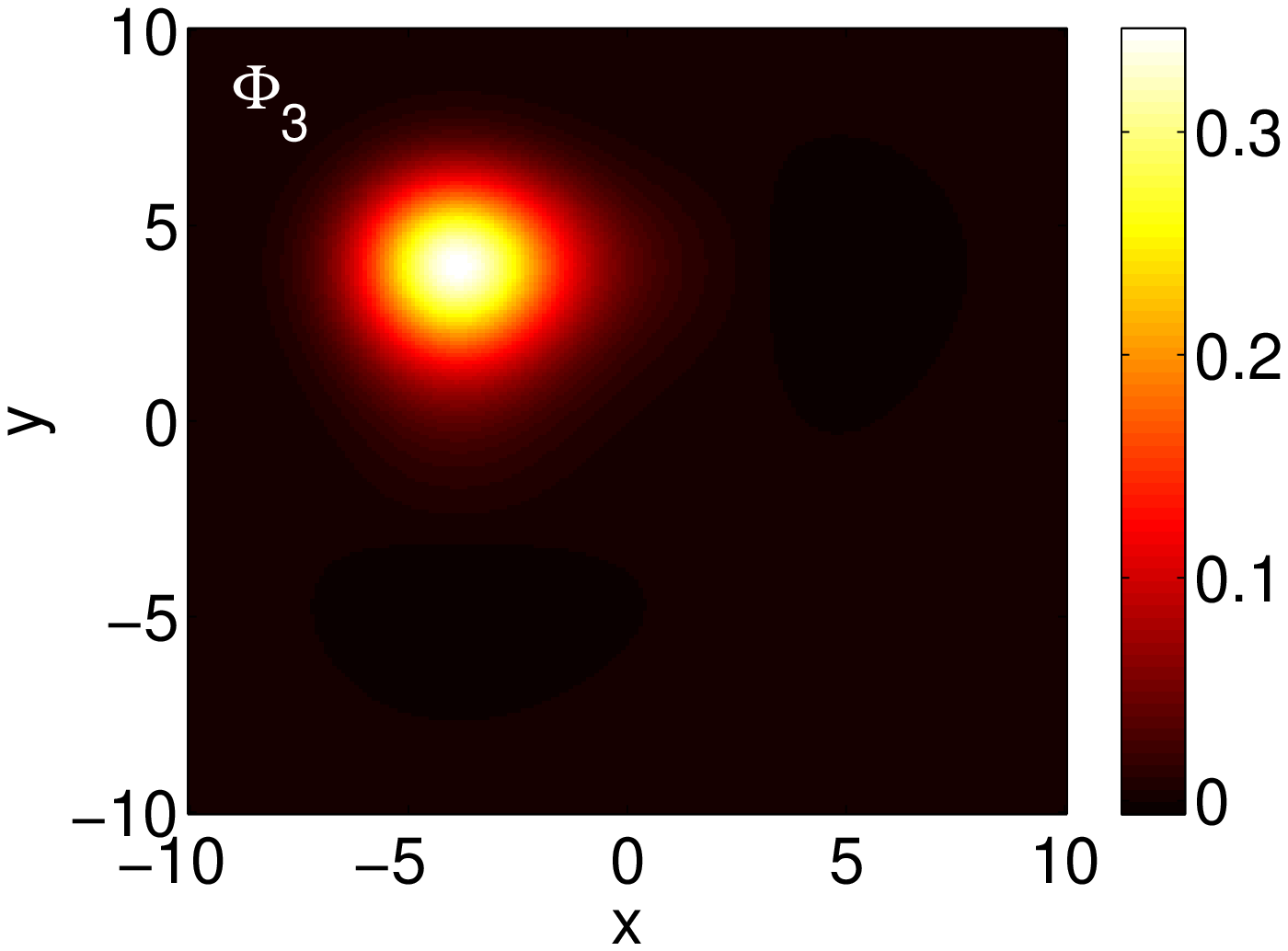}\newline
\caption{(Color online) Basis modes \{$\Phi _{0}$, $\Phi _{1}$, $\Phi _{2}$,
$\Phi _{3}$\} that are localized in each of the wells.}
\label{figPhi}
\end{figure}

Actually, it is more convenient to use a transformed basis, \{$\Phi _{0}$, $%
\Phi _{1}$, $\Phi _{2}$, $\Phi _{3}$\}, as shown in Fig. \ref{figPhi}, which
is based on populations of the four wells. This basis is generated from the original
set by a linear transformation:
\begin{equation}
\begin{bmatrix}
\Phi _{0} & \Phi _{1} & \Phi _{2} & \Phi _{3}%
\end{bmatrix}%
=%
\begin{bmatrix}
u_{0} & u_{1} & u_{2} & u_{3}%
\end{bmatrix}%
\mathrm{T},  \label{eq4}
\end{equation}%
where the appropriate transformation matrix is
\begin{equation}
\mathrm{T}=\dfrac{1}{2}%
\begin{pmatrix}
1 & 1 & 1 & 1 \\
-1 & -1 & 1 & 1 \\
-1 & 1 & 1 & -1 \\
1 & -1 & 1 & -1%
\end{pmatrix}%
.  \label{eqT}
\end{equation}
Each mode $\Phi _{j}$ ($j=0,1,2,3$) is localized in one of the four wells,
with the four of them constituting an orthonormal set.

Using the new basis, we can readily reformulate the four-mode decomposition as
\begin{equation}
u(x,y,t)=\sum_{j=0}^{3}c_{j}(t)\Phi _{j}(x,y),  \label{eq5}
\end{equation}
with time-dependent complex amplitude $c_{j}(t)$, $j=0,1,2,3$. Substituting
Eq. (\ref{eq5}) into Eq. (\ref{eq1}) and projecting onto the orthonormal
basis \{$\Phi _{0}$, $\Phi _{1}$, $\Phi _{2}$, $\Phi _{3}$\}, we derive, by
means of straightforward algebra, the following system of four ordinary
differential equations (ODEs),
\begin{equation}
\begin{split}
& i\dot{c}_{j}=\tilde{\omega}_{j}+sA_{j}|c_{j}|^{2}c_{j}+s\sum_{k\neq
j}B_{jk}\,(2|c_{k}|^{2}c_{j}+c_{k}^{2}c_{j}^{\ast }) \\
& +s\sum_{k\neq
j}[D_{kj}|c_{k}|^{2}c_{k}+D_{jk}\,(2|c_{j}|^{2}c_{k}+c_{j}^{2}c_{k}^{\ast
})]+s\!\sum_{k\neq l\neq j\neq
k}E_{kjl}\,(2|c_{k}|^{2}c_{l}+c_{k}^{2}c_{l}^{\ast }) \\
& +s\!\sum_{k\neq l\neq j\neq k}\!E_{jkl}\,(c_{j}^{\ast
}c_{k}c_{l}+c_{j}c_{k}^{\ast }c_{l}+c_{j}c_{k}c_{l}^{\ast })+s\,G\!\sum_{%
\begin{array}{c}
_{k\neq l\neq m\neq k} \\
_{k,l,m\neq j}%
\end{array}%
}c_{k}^{\ast }c_{l}c_{m},
\end{split}
\label{eq6}
\end{equation}
with the summation performed over $k,l,m=0,1,2,3$. To cast these equations
in a more compact form, we have defined
\begin{equation}
\begin{bmatrix}
\tilde{\omega}_{0} \\
\tilde{\omega}_{1} \\
\tilde{\omega}_{2} \\
\tilde{\omega}_{3}%
\end{bmatrix}%
=\frac{1}{4}%
\begin{bmatrix}
\gamma _{0}-4\mu & \gamma _{1} & \gamma _{3} & \gamma _{2} \\
\gamma _{1} & \gamma _{0}-4\mu & \gamma _{2} & \gamma _{3} \\
\gamma _{3} & \gamma _{2} & \gamma _{0}-4\mu & \gamma _{1} \\
\gamma _{2} & \gamma _{3} & \gamma _{1} & \gamma _{0}-4\mu%
\end{bmatrix}%
\begin{bmatrix}
c_{0} \\
c_{1} \\
c_{2} \\
c_{3}%
\end{bmatrix}%
,  \label{eq7}
\end{equation}
where $\gamma _{0}\equiv \omega _{0}+\omega _{1}+\omega _{2}+\omega _{3}$, $%
\gamma _{1}=\omega _{0}+\omega _{1}-\omega _{2}-\omega _{3}$, $\gamma
_{2}\equiv \omega _{0}-\omega _{1}+\omega _{2}-\omega _{3}$, and $\gamma
_{3}\equiv \omega _{0}-\omega _{1}-\omega _{2}+\omega _{3}$. Notice that,
for the underlying eigenvalues (\ref{omega}), one has $\gamma _{1}=\gamma
_{2}=\omega _{0}-\omega _{3}$ and $\gamma _{3}\approx 0$. Furthermore, 
Eqs. (\ref{eq6}) 
involve nonlinear coefficients given by overlap integrals, \textit{%
viz.}, $A_{n}\equiv \int \!\!\!\int \!\Phi _{n}^{4}\,dxdy$, $B_{mn}\equiv
\int \!\!\!\int \!\Phi _{m}^{2}\Phi _{n}^{2}\,dxdy$, $D_{mn}\equiv \int
\!\!\!\int \!\Phi _{m}^{3}\Phi _{n}\,dxdy$, $E_{lmn}\equiv \int \!\!\!\int
\!\Phi _{l}^{2}\Phi _{m}\Phi _{n}\,dxdy$, $G\equiv \int \!\!\!\int \!\Phi
_{0}\Phi _{1}\Phi _{2}\Phi _{3}\,dxdy$, with $l,m,n=0,1,2,3$; these indices
must be mutually different wherever they appear in the coefficients before
the nonlinear terms.

For our choice of the parameters of the potential, the overlapping between
modes $\Phi_{i}$ is weak (see Fig. \ref{figPhi}), therefore all the overlap
integrals are much smaller than the $A_{n}$'s. Neglecting these small
overlap terms leads to the following simplification of Eq. (\ref{eq6}):
\begin{equation}
i\dot{c}_{j}=\tilde{\omega}_{j}+A_{j}|c_{j}|^{2}c_{j},~j=0,1,2,3.
\label{eq8}
\end{equation}
It has been checked that the latter reduction of the four-mode equations
very slightly affects the accuracy of the solutions, while it renders the
identification of various bifurcation branches significantly easier. 
Furthermore, this reduction is more convenient in simulations as we may use 
%including the use of 
these solutions as inputs for generating numerical solutions of
the full GP system, as explained below.

In this 2D setting, we seek both real and complex stationary solutions to
the ODE system. Substituting $c_{j}(t)\equiv \rho _{j}(t)e^{i\varphi
_{j}(t)} $ into Eq. (\ref{eq8}), we split them into real equations for $\rho
_{j}$ and $\varphi _{j}$:
\begin{align}
\dot{\rho}_{0}& =\frac{1}{4}\gamma _{1}[\rho _{1}\sin (\varphi _{1}-\varphi
_{0})+\rho _{3}\sin (\varphi _{3}-\varphi _{0})],  \label{eq9} \\
\dot{\varphi}_{0}& =(\mu -\frac{1}{4}\gamma _{0})-sA_{0}\rho _{0}^{2}-\frac{1%
}{4}\gamma _{1}[\frac{\rho _{1}}{\rho _{0}}\cos (\varphi _{1}-\varphi _{0})+%
\frac{\rho _{3}}{\rho _{0}}\cos (\varphi _{3}-\varphi _{0})],  \label{eq10}
\\
\dot{\rho}_{1}& =\frac{1}{4}\gamma _{1}[\rho _{0}\sin (\varphi _{0}-\varphi
_{1})+\rho _{2}\sin (\varphi _{2}-\varphi _{1})],  \label{eq11} \\
\dot{\varphi}_{1}& =(\mu -\frac{1}{4}\gamma _{0})-sA_{1}\rho _{1}^{2}-\frac{1%
}{4}\gamma _{1}[\frac{\rho _{0}}{\rho _{1}}\cos (\varphi _{0}-\varphi _{1})+%
\frac{\rho _{2}}{\rho _{1}}\cos (\varphi _{2}-\varphi _{1})],  \label{eq12}
\end{align}%
with the equations for $\rho _{2,3}$ and $\varphi _{2,3}$ obtained by
interchanging the indices, $0\longleftrightarrow 2$ and $%
1\longleftrightarrow 3$, except for in $\gamma _{0}$ and $\gamma _{1}$.

Looking for solutions with constant $\rho _{j}$ and $\varphi _{j}$
which are integer multiples of $\pi $, we reduce Eqs. (\ref{eq9}) -
(\ref{eq12})
%four algebraic equations for $\rho_{j}$, $j=0,1,2,3$:
%\begin{align}
%\mu\rho_{0}
%&=\frac{1}{4}[\gamma_{0}\rho_{0}+\gamma_{1}(\rho_{1}+\rho_{3})]+sA_{0}\rho_{0}^3,\label{eq13}\\
%\mu\rho_{1}
%&=\frac{1}{4}[\gamma_{0}\rho_{1}+\gamma_{1}(\rho_{0}+\rho_{2})]+sA_{1}\rho_{1}^3,\label{eq14}\\
%\mu\rho_{2}
%&=\frac{1}{4}[\gamma_{0}\rho_{2}+\gamma_{1}(\rho_{1}+\rho_{3})]+sA_{2}\rho_{2}^3,\label{eq15}\\
%\mu\rho_{3}
%&=\frac{1}{4}[\gamma_{0}\rho_{3}+\gamma_{1}(\rho_{0}+\rho_{2})]+sA_{3}\rho_{3}^3 \label{eq16}\\
%\end{align}
to a set of four algebraic equations for $\rho _{j}$, which can be used to
derive a complete set of stationary solutions of the four-mode truncation.
These were further used as initial guesses to find numerical solutions of
the full system of the GP equations. Moreover, our analysis of the four-mode
system indicates that nontrivial complex solutions in this setting are only
possible in the form of \textit{discrete vortices}, i.e., solutions with
phase sets $\varphi _{j}=\pi j/2$, $j=0,1,2,3$ \cite{peli2d}, which have
been studied in detail in Refs. \cite{kody} and \cite{todd_chaos} (we also
briefly consider them here).

\section{Numerical results}

\label{numerical}

\subsection{Attractive interactions}

\label{attractive} 

\begin{figure}[tbph]
\centering
\includegraphics[width=.4\textwidth]{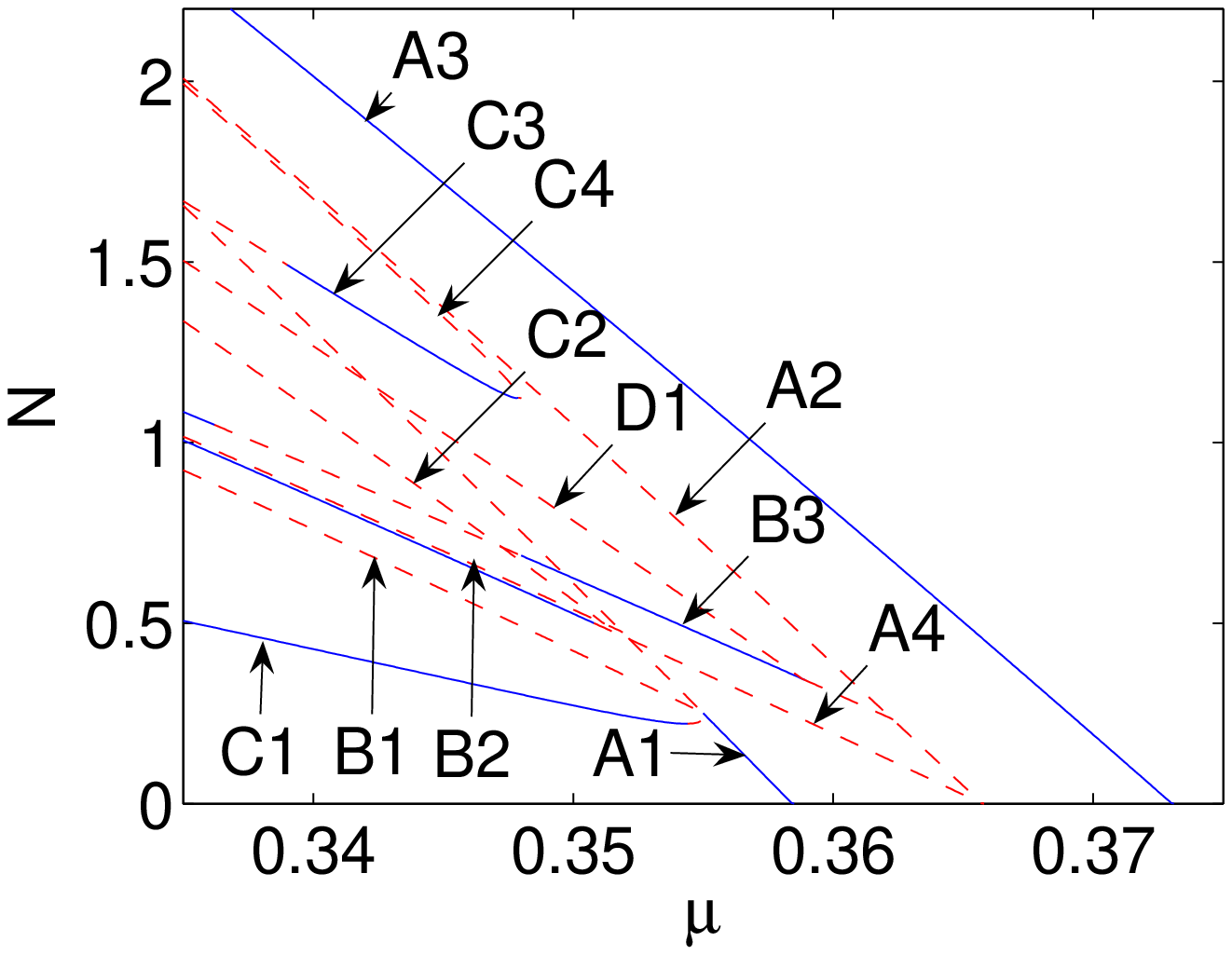} %
\includegraphics[width=.4\textwidth]{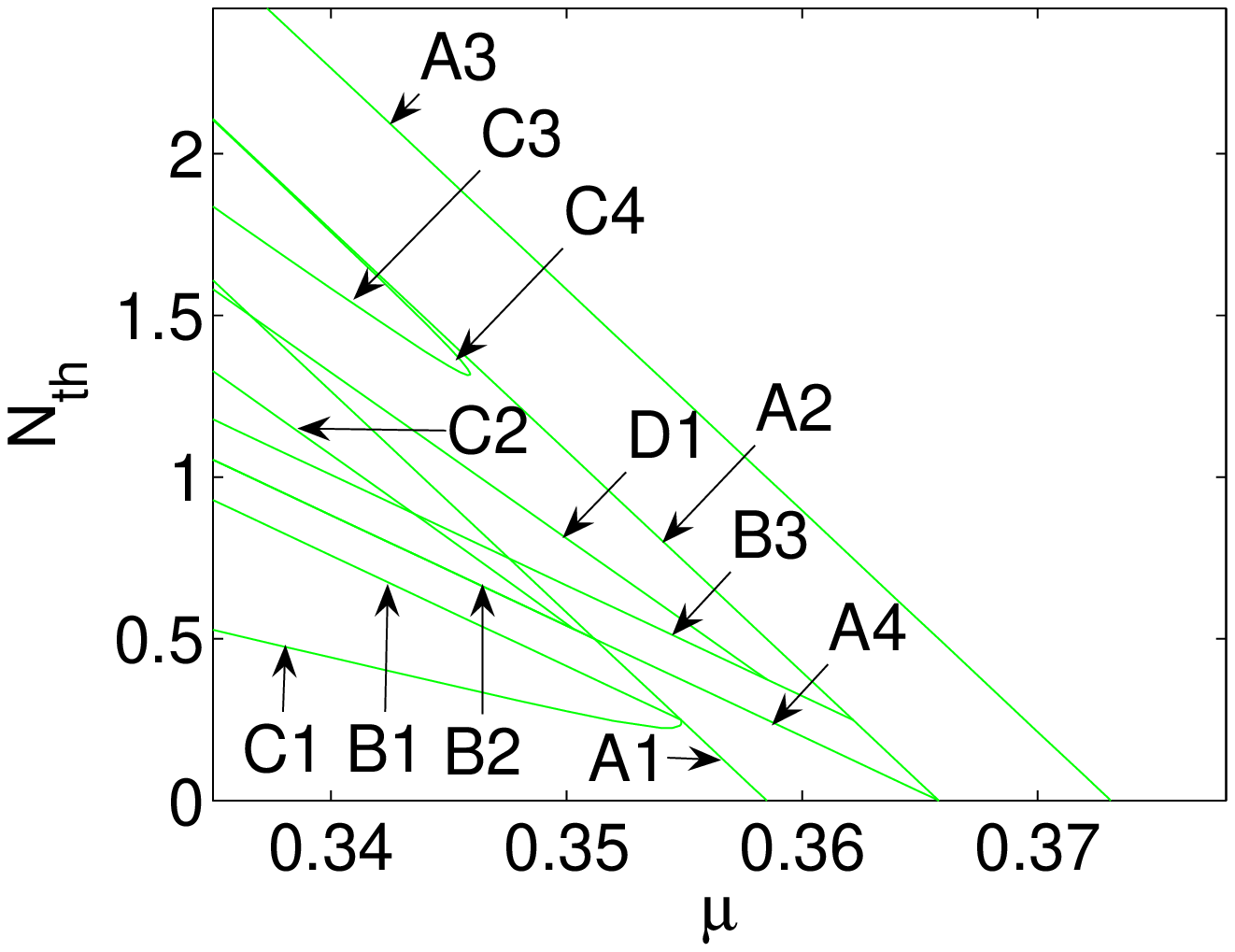}\newline
\includegraphics[width=.3\textwidth]{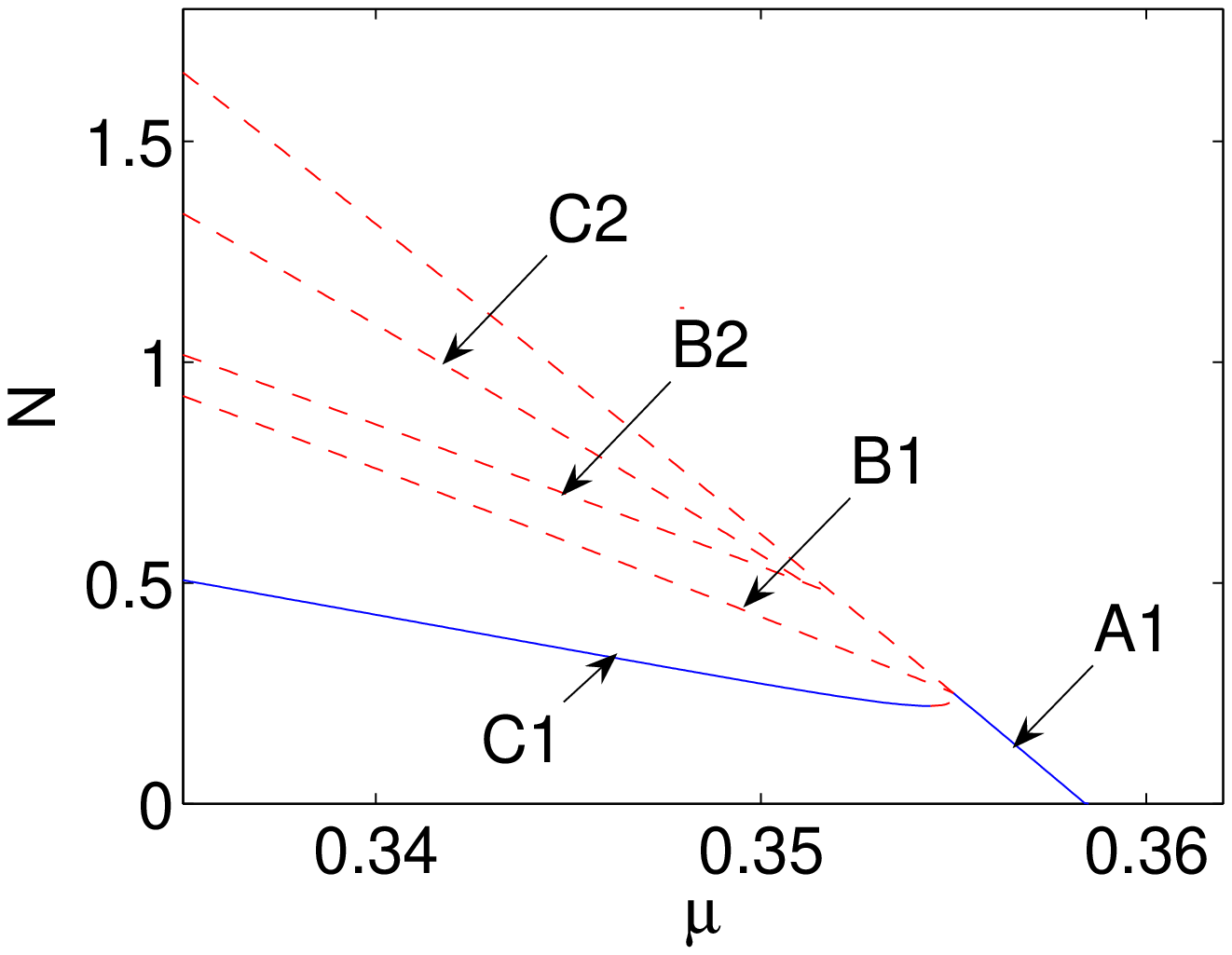} %
\includegraphics[width=.3\textwidth]{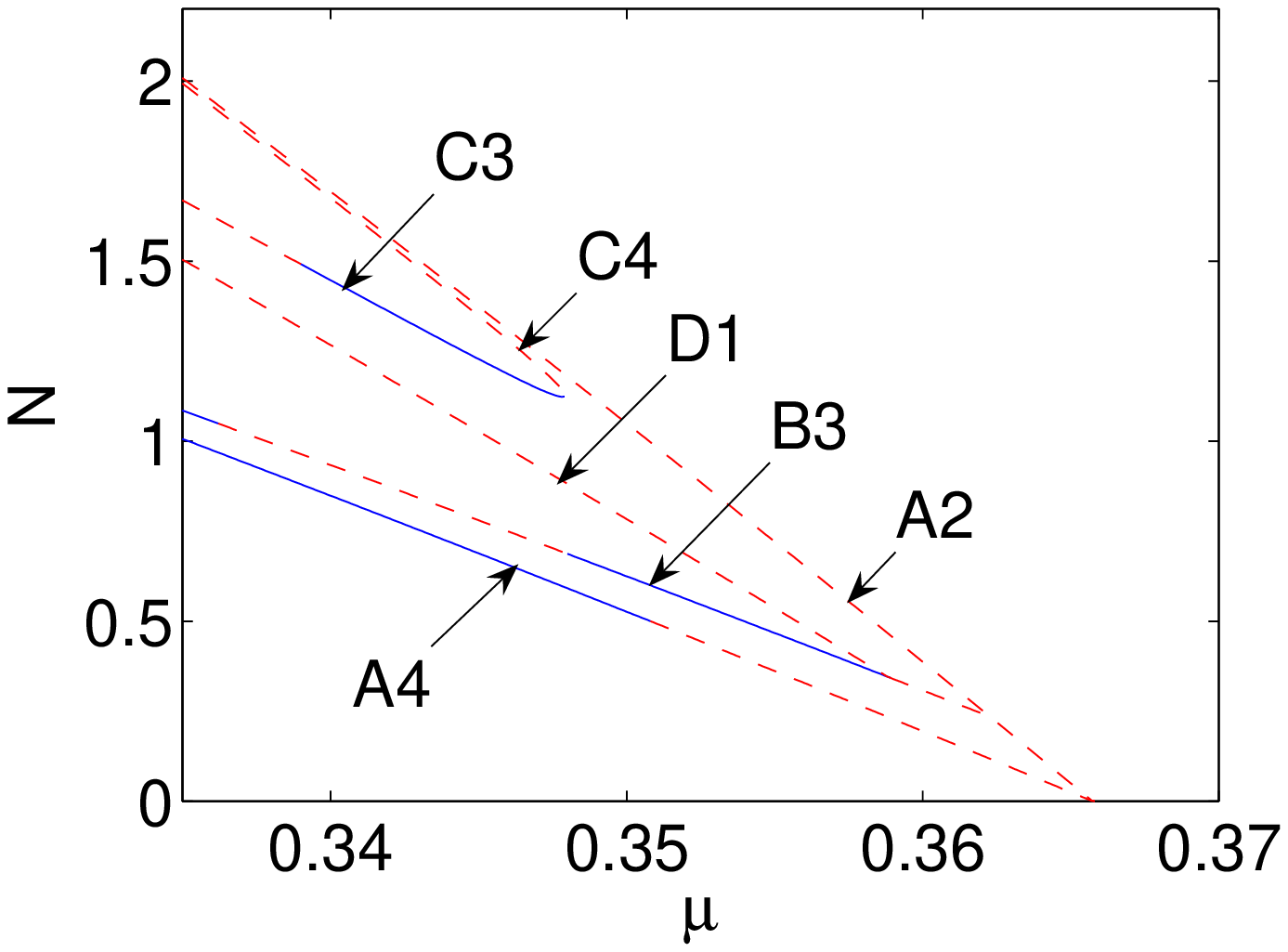}\newline
\caption{(Color online) Top panels: squared 
norm $N$ (normalized number of atoms 
in BECs or power in optics)
%and optics, respectively) 
of numerically found solutions of Eq. (\protect\ref{eq1}) (left), 
and their counterparts predicted by the four-mode approximation (right), for 
%the case of the
attractive interatomic interactions ($s=-1$), as a function of $\protect\mu $, 
i.e., respectively, the chemical potential or propagation constant. The
bottom panels are segments of the top left panel. The (blue) solid lines and 
(red) dashed lines denote stable and unstable solutions, respectively. The 
branches are explained in the text and their profiles and stability are 
detailed in Figs. \protect\ref{figA}-\protect\ref{figD}.}
\label{fig_foc}
\end{figure}

Let us first present results of numerical simulations pertaining to 
%We begin the numerical analysis with 
attractive interactions (alias self-focusing nonlinearity) 
case, i.e., $s=-1$ in Eq.(\ref{eq1}).\ Our basic bifurcation diagram, shown
in Fig. \ref{fig_foc}, displays the squared $L^{2}$ norm of the solution
(which physically describes the number of atoms in BECs or the power in
optics), $N=\int \!\!\!\int \!|u(x,y,t)|^{2}\,dxdy$, as a function of the 
chemical potential $\mu $. The left panel of Fig. \ref{fig_foc} presents the
full numerical bifurcation diagram, which involves twelve real and one
complex solutions (for the latter solution, $N$ is the same as one of the
real branches, hence this branch is not visible as a separate curve in the
diagram). The companion diagram in the right panel is obtained from the
above-mentioned algebraic system for stationary solutions produced by the
the four-mode reduction, and 
%[Eqs.(\ref{eq13})-(\ref{eq16})]
%It 
demonstrates good agreement with its numerical counterpart.

\begin{figure}[tbph]
\centering
\includegraphics[width=.24\textwidth]{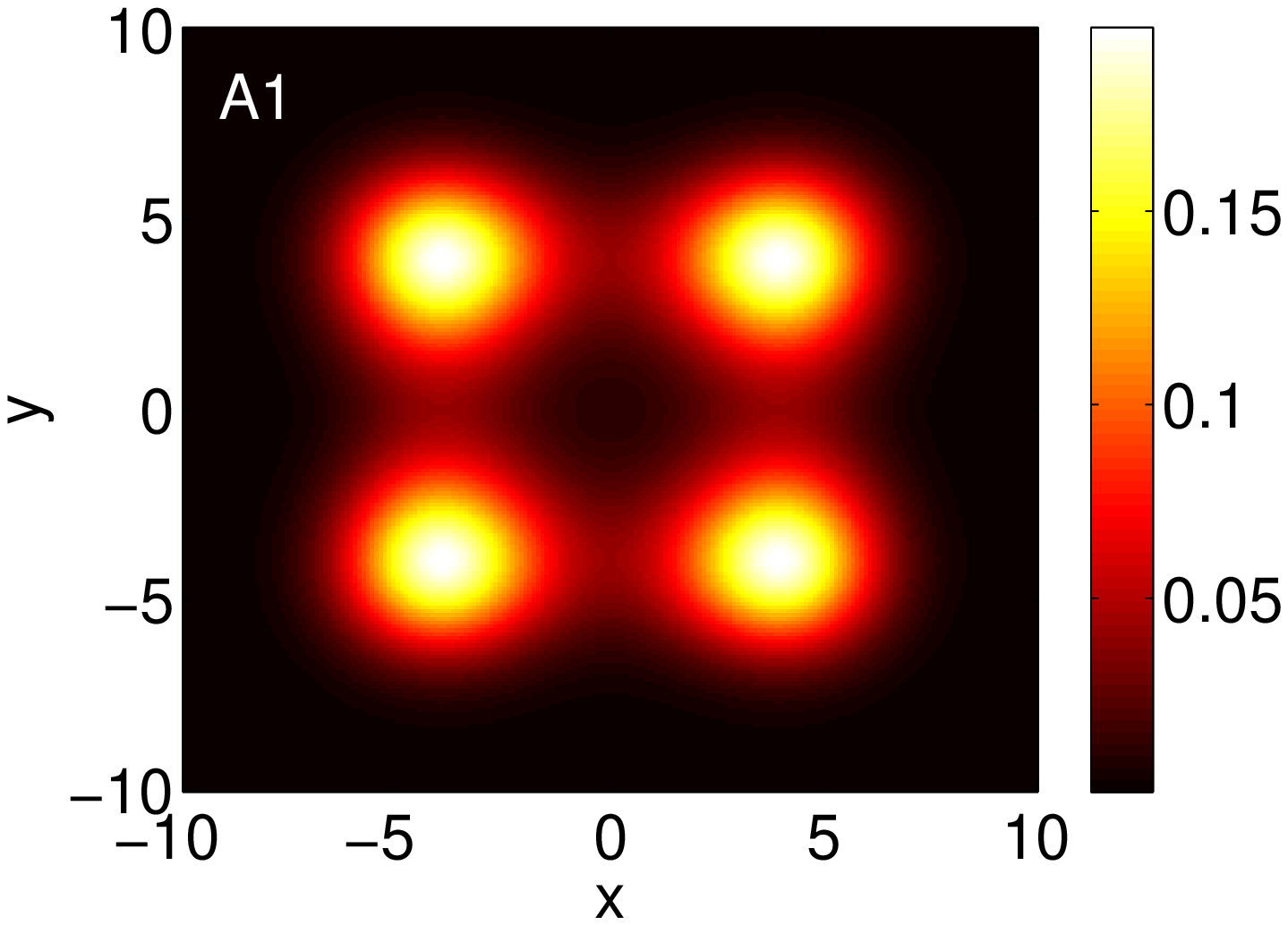} %
\includegraphics[width=.24\textwidth]{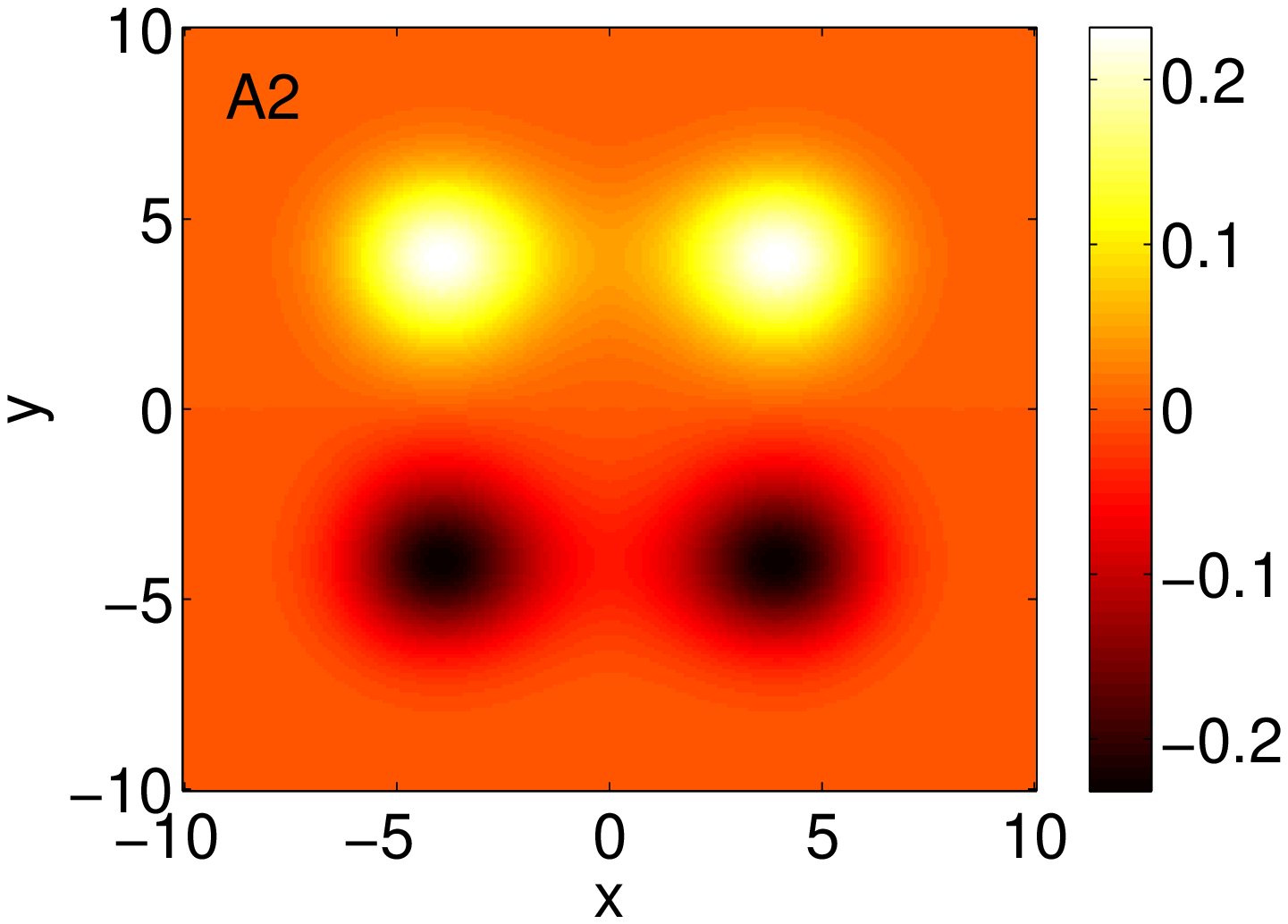} %
\includegraphics[width=.24\textwidth]{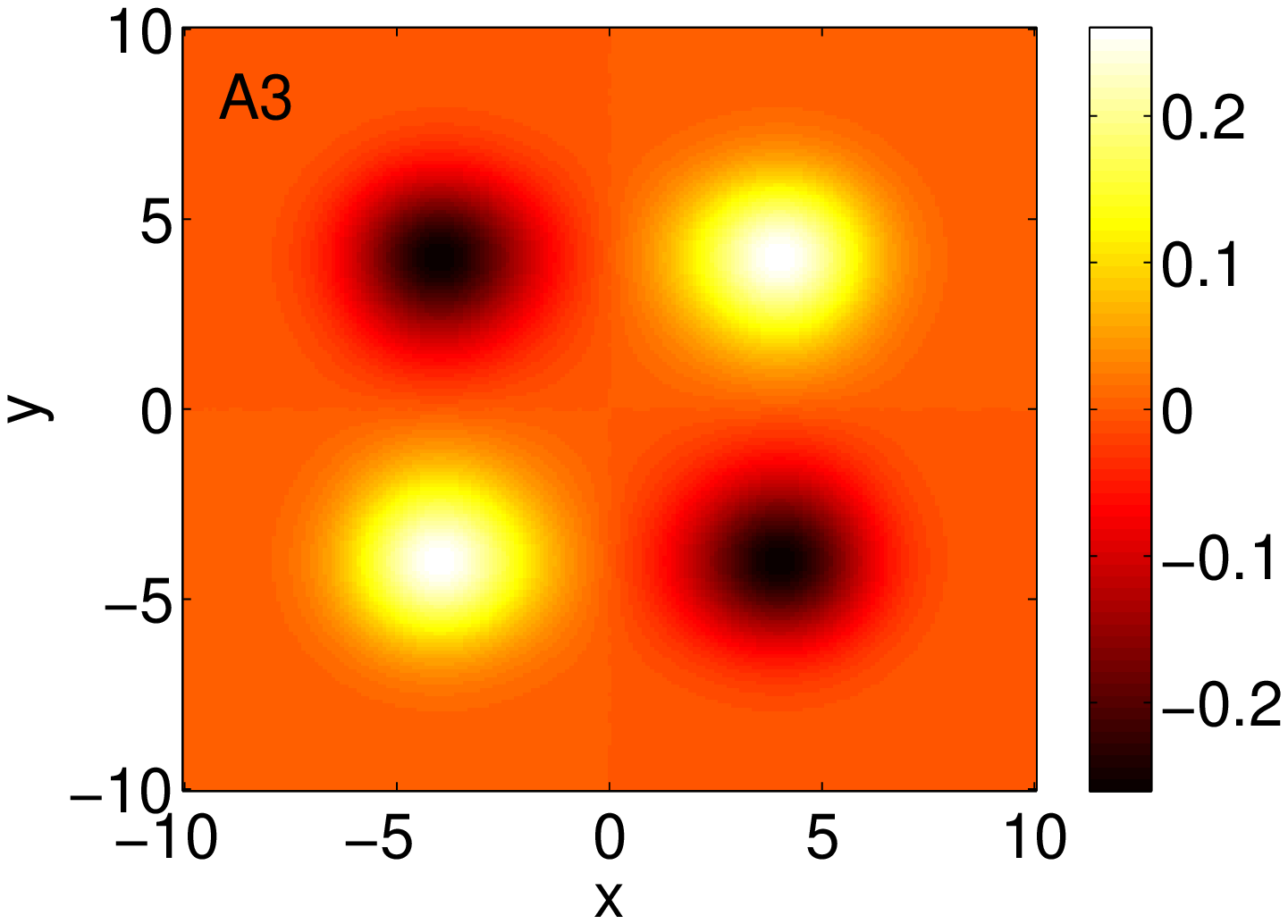} %
\includegraphics[width=.24\textwidth]{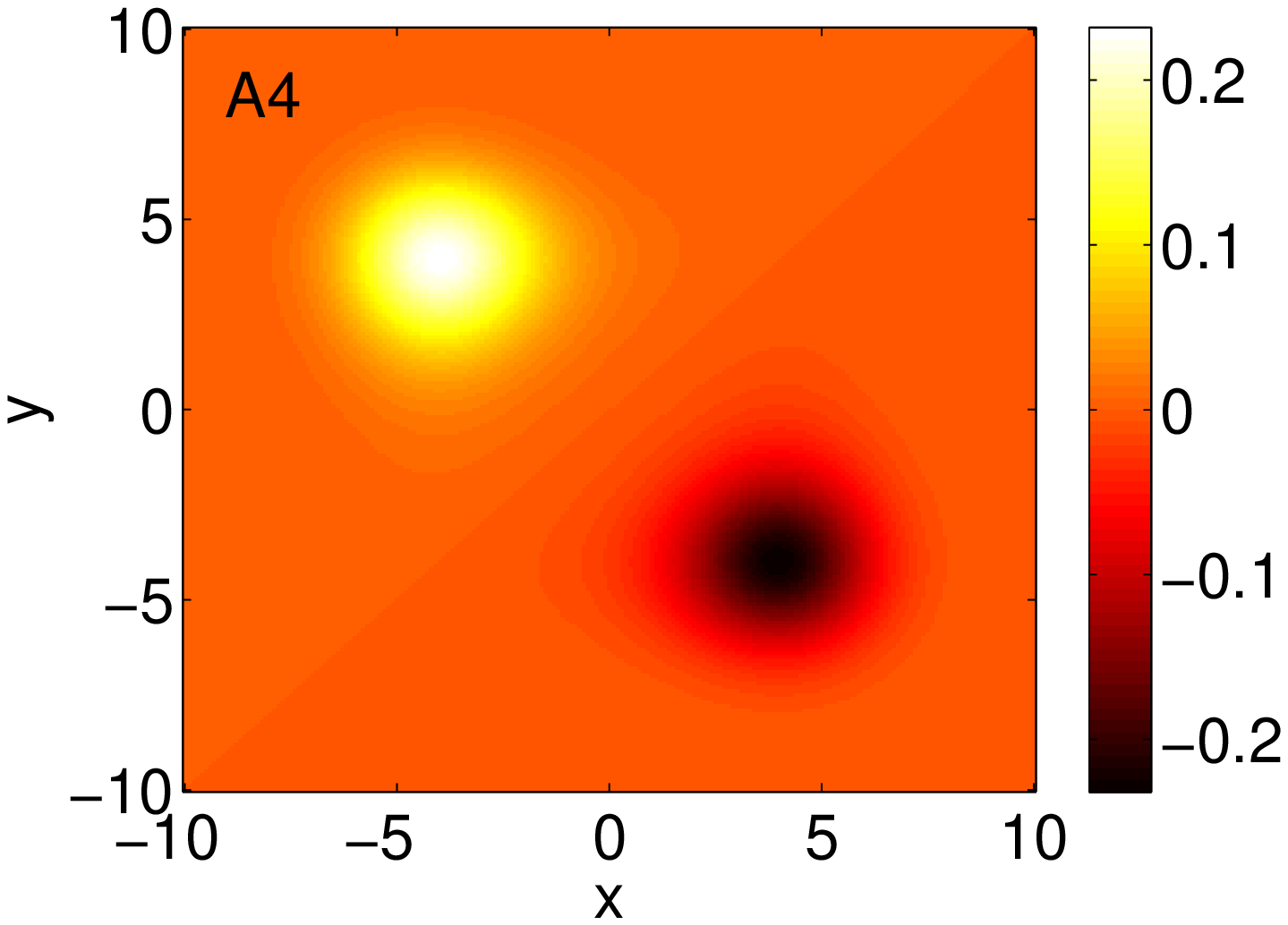}\newline
\includegraphics[width=.24\textwidth]{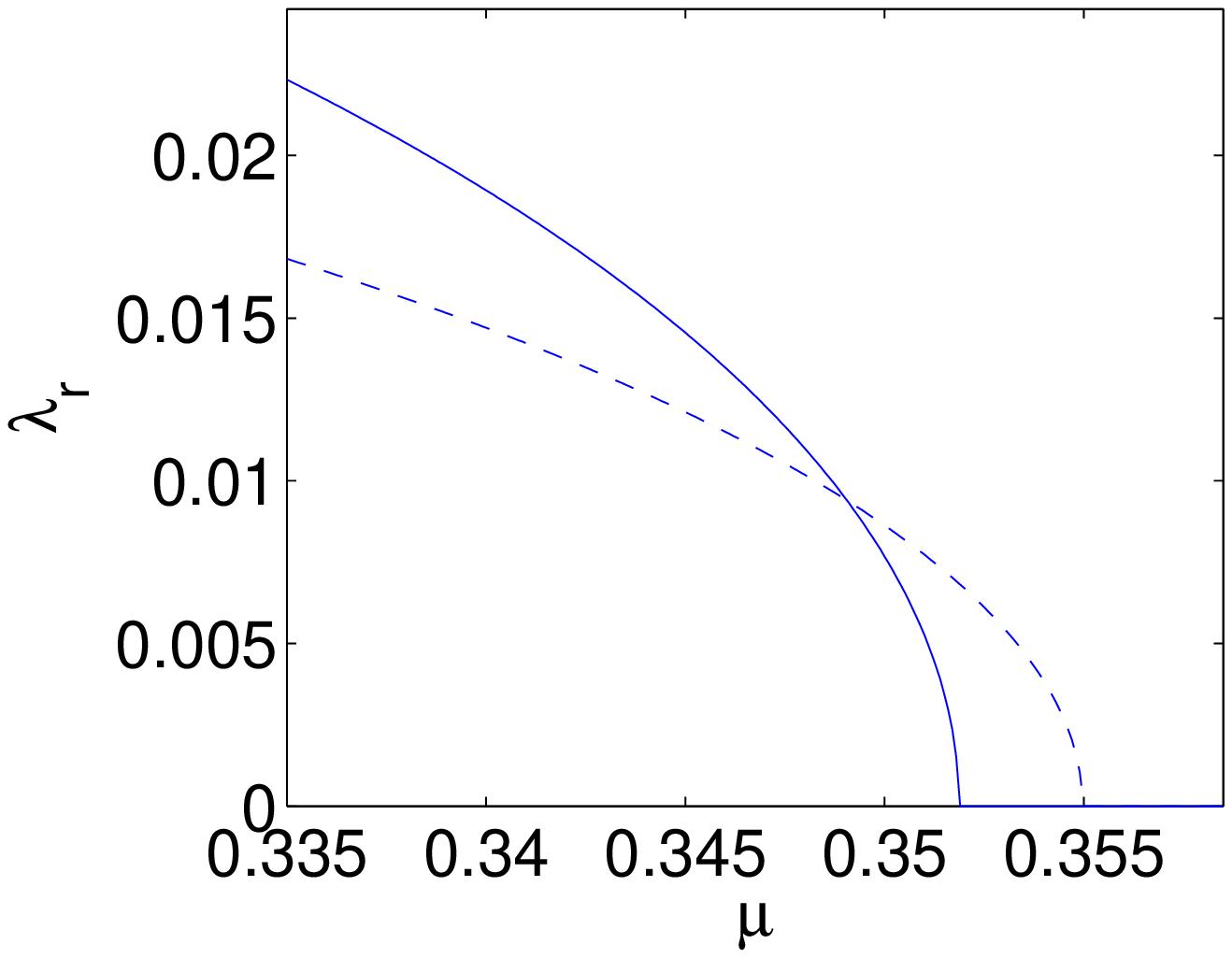} %
\includegraphics[width=.24\textwidth]{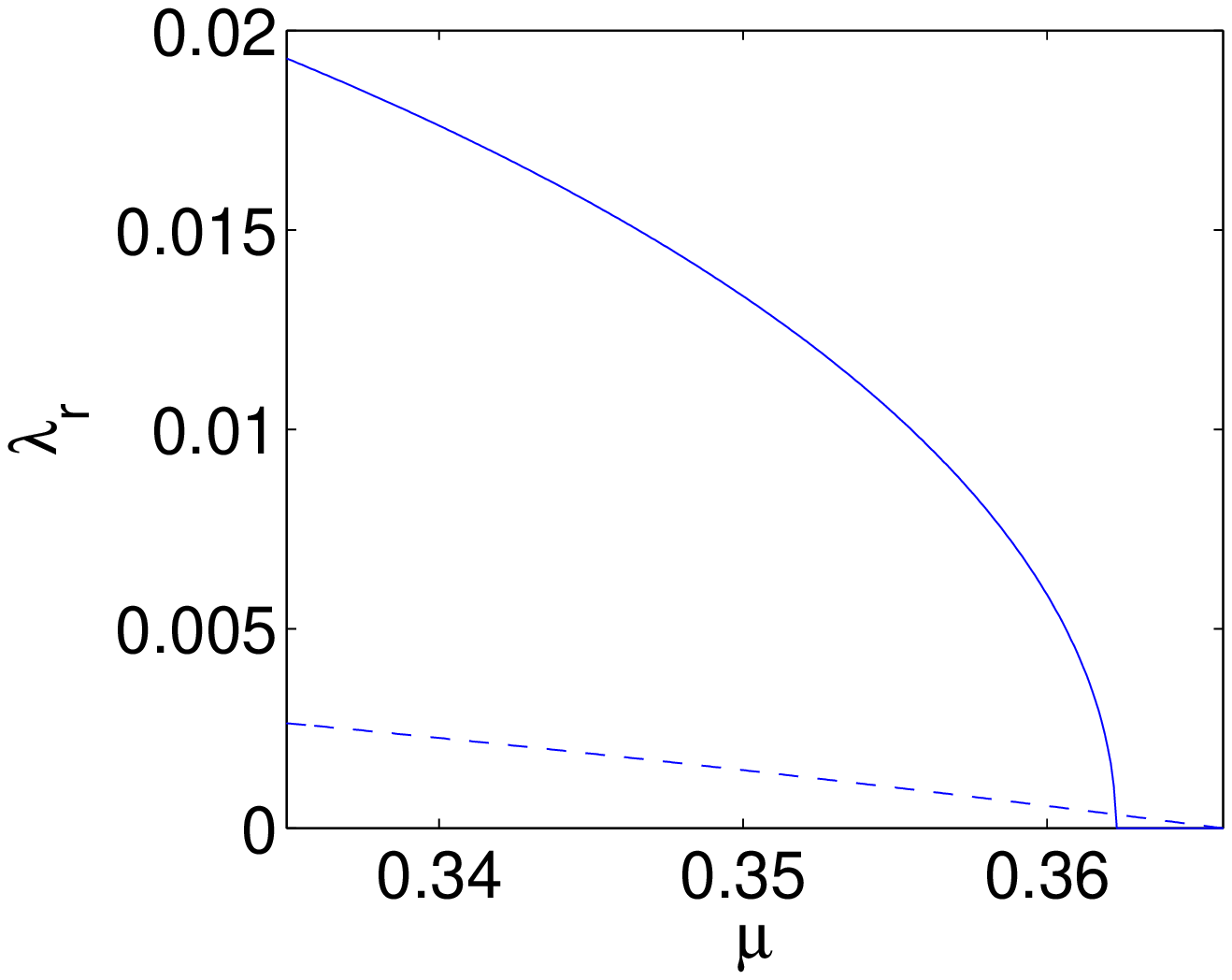} %
\includegraphics[width=.24\textwidth]{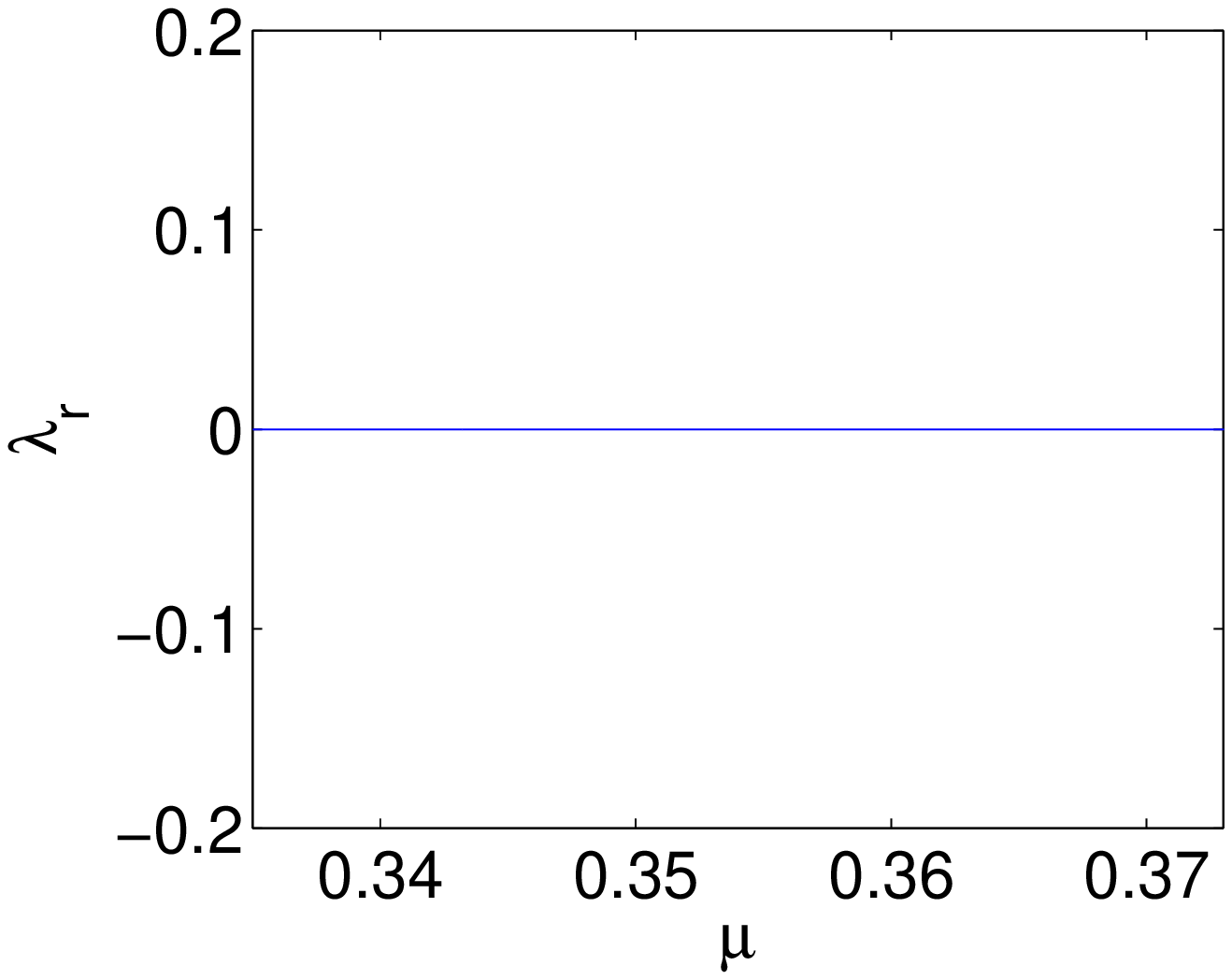} %
\includegraphics[width=.24\textwidth]{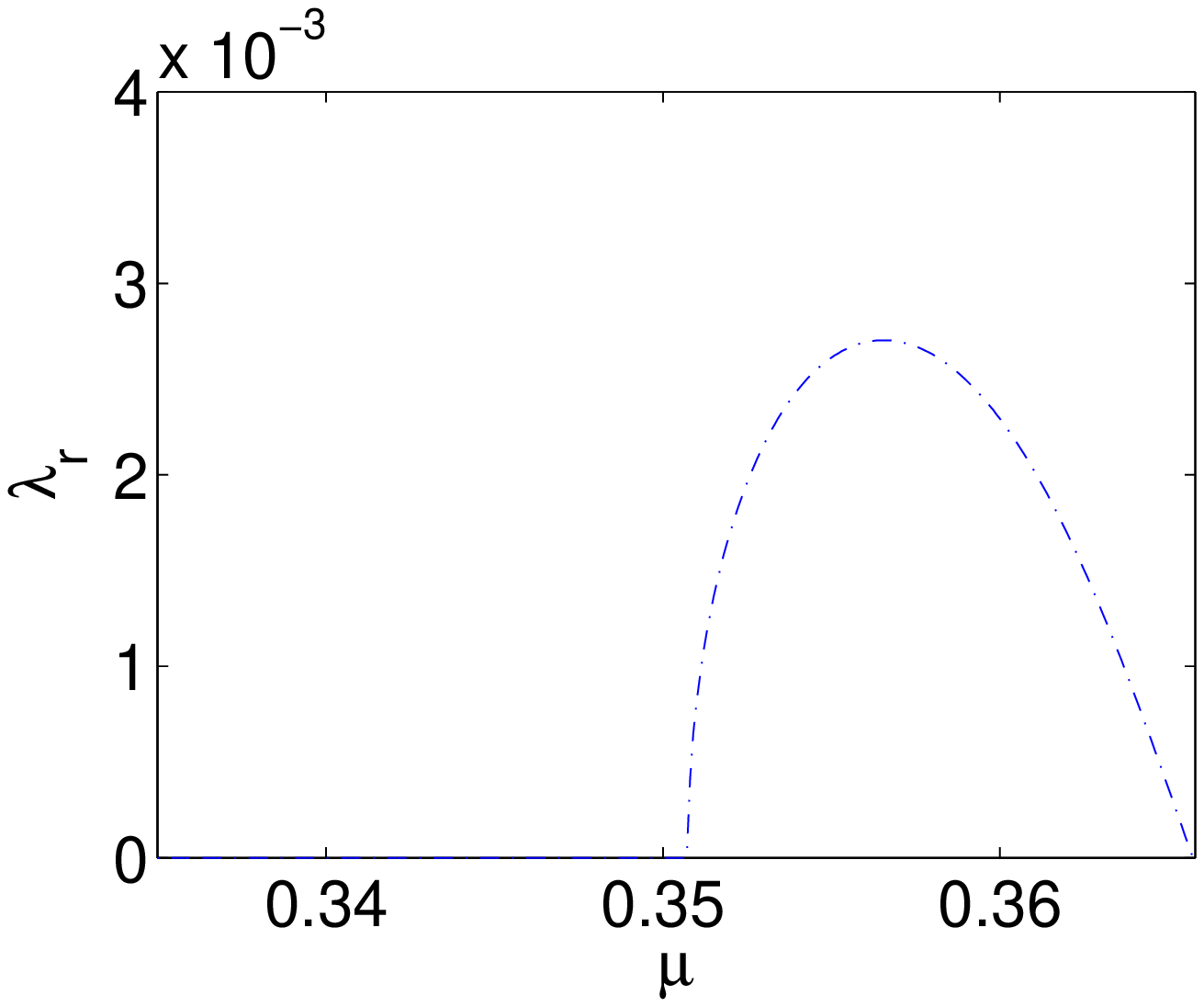}\newline
\caption{(Color online) Top panels show the 
%waveforms 
profiles of wave functions of branches A1, A2, A3
and A4 (from left to right) at $\protect\mu =0.34$. The bottom panels
display real parts of the unstable eigenvalues of the respective branches as
functions of the parameter $\protect\mu$.
}
\label{figA}
\end{figure}

\begin{figure}[tbph]
\centering
\includegraphics[width=.24\textwidth]{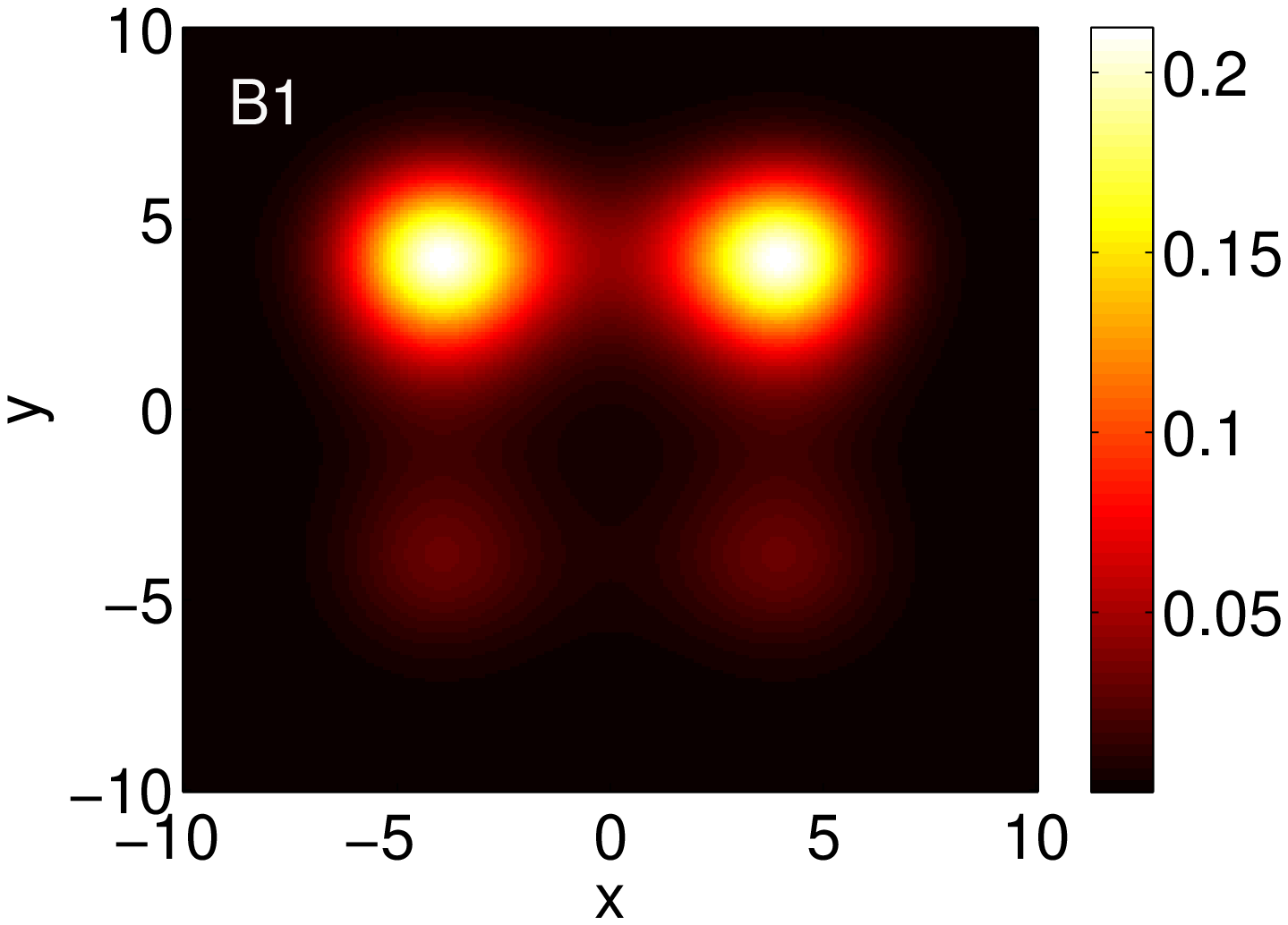} %
\includegraphics[width=.24\textwidth]{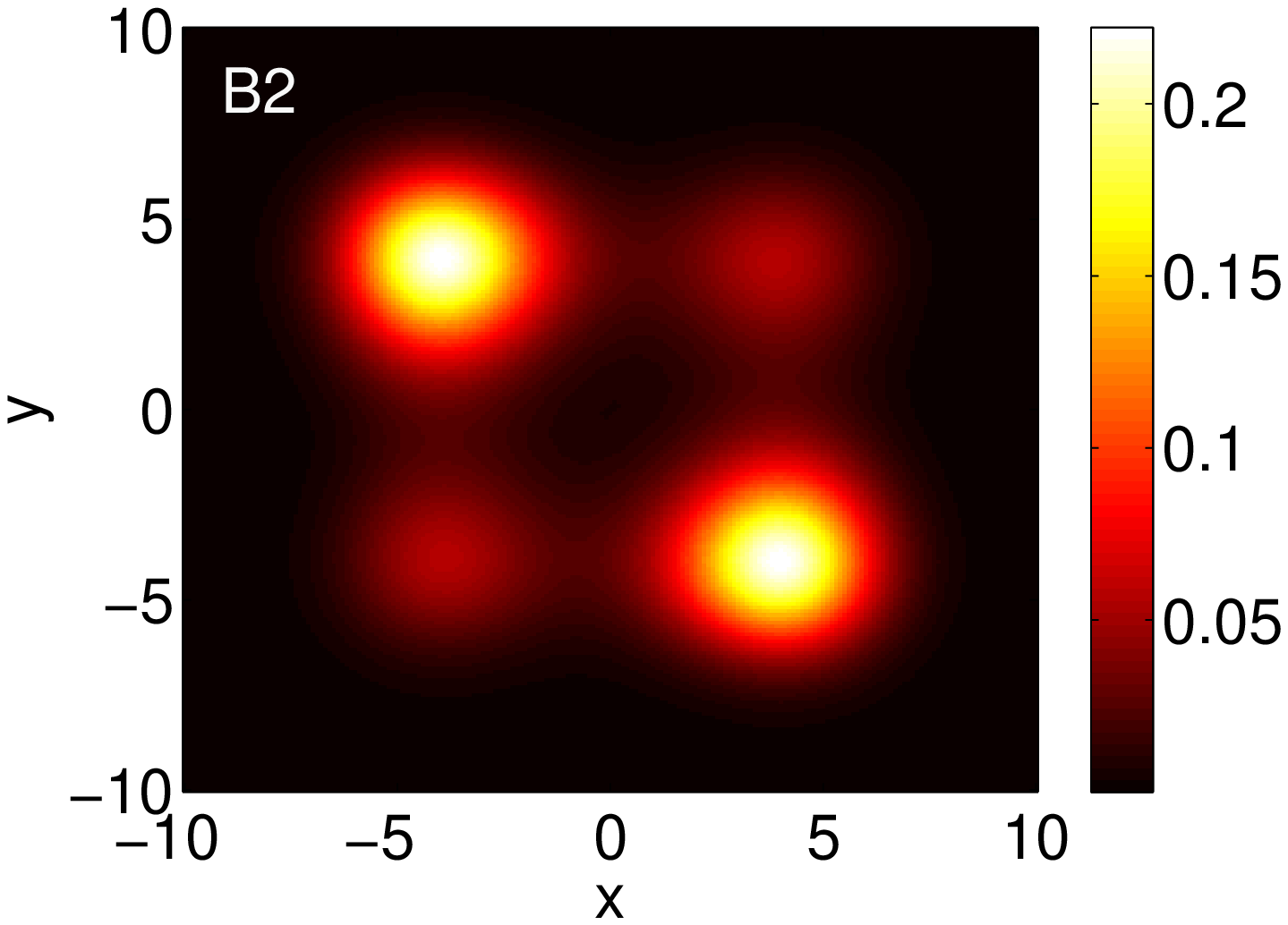} %
\includegraphics[width=.24\textwidth]{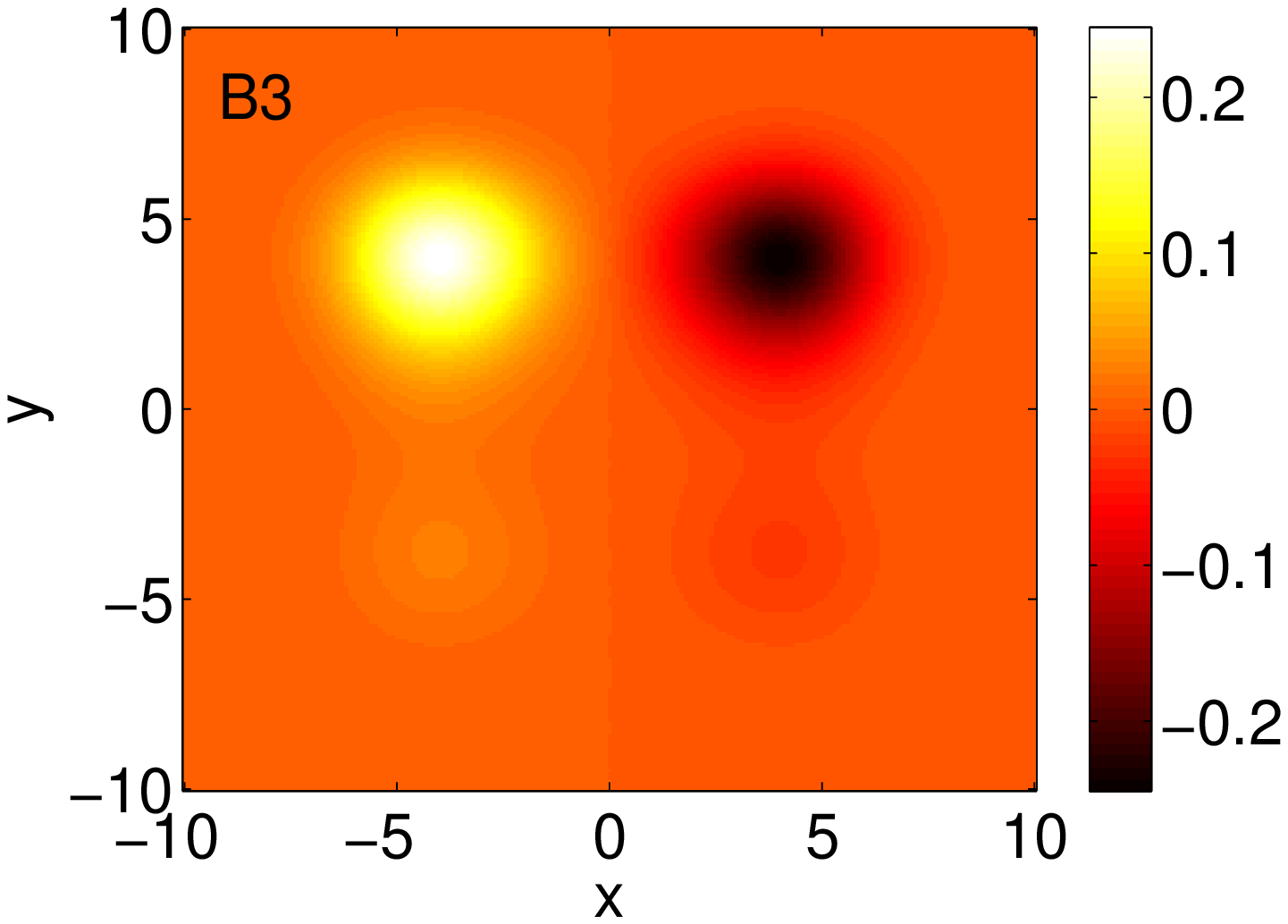}\newline
\includegraphics[width=.24\textwidth]{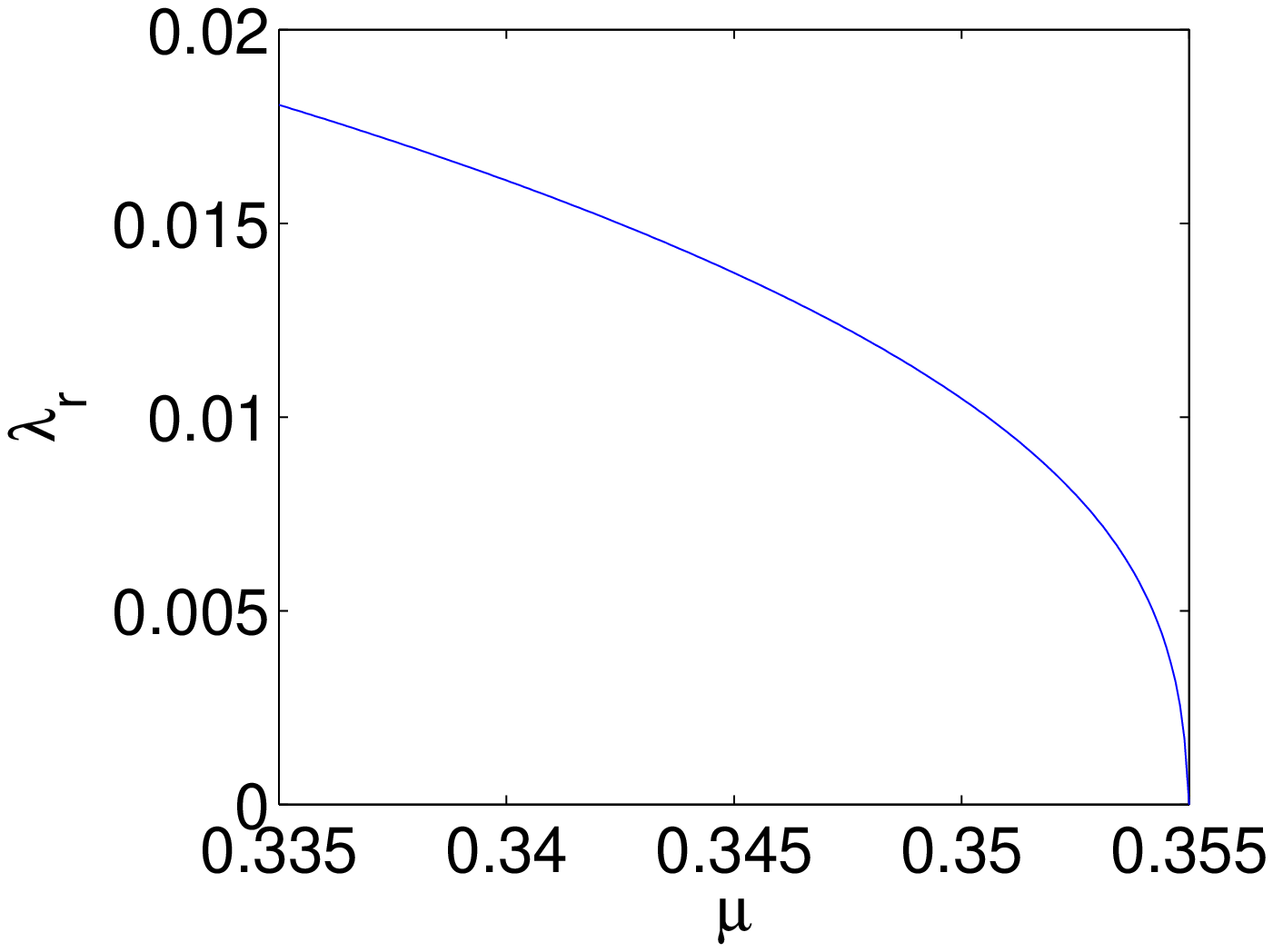} %
\includegraphics[width=.24\textwidth]{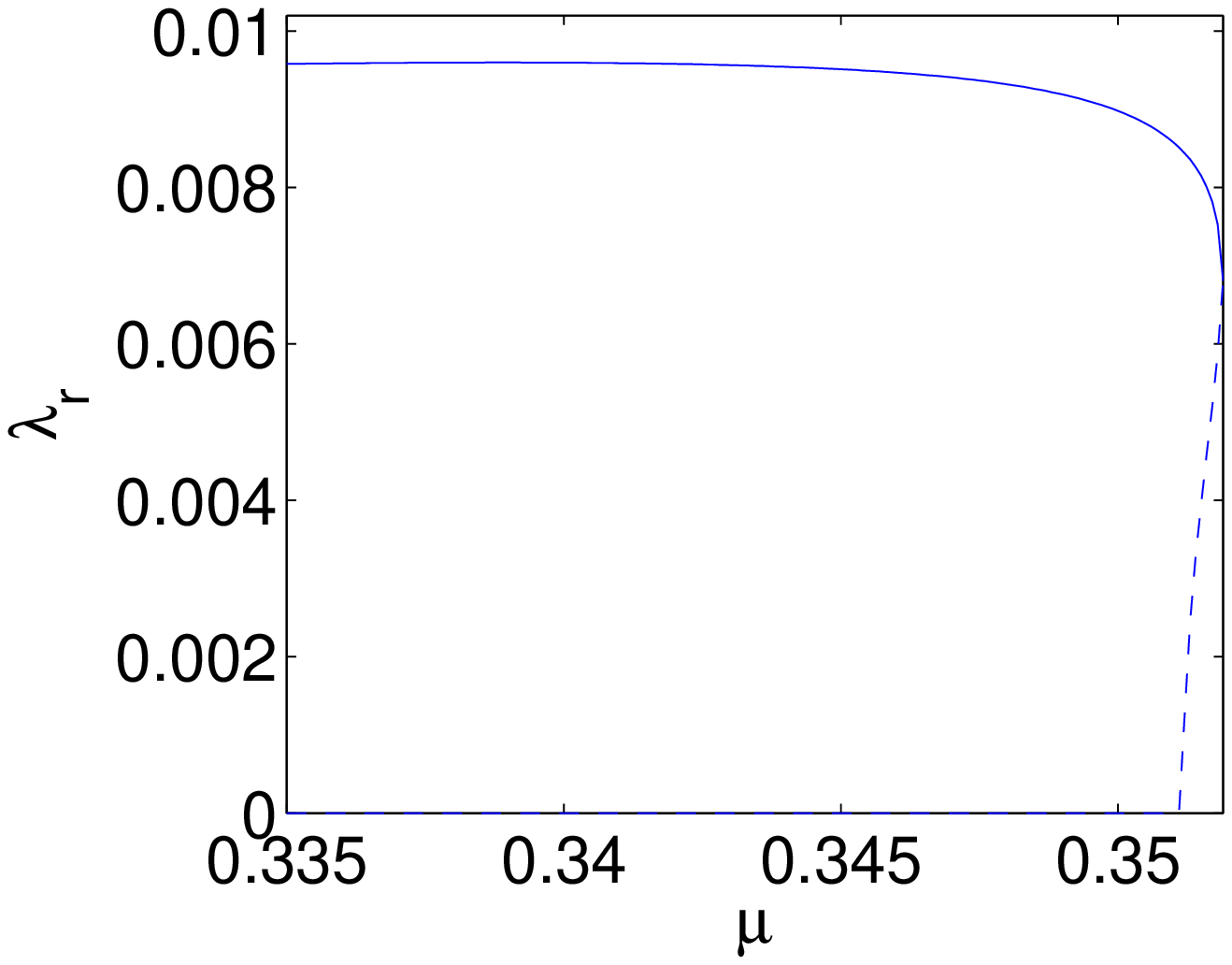} %
\includegraphics[width=.24\textwidth]{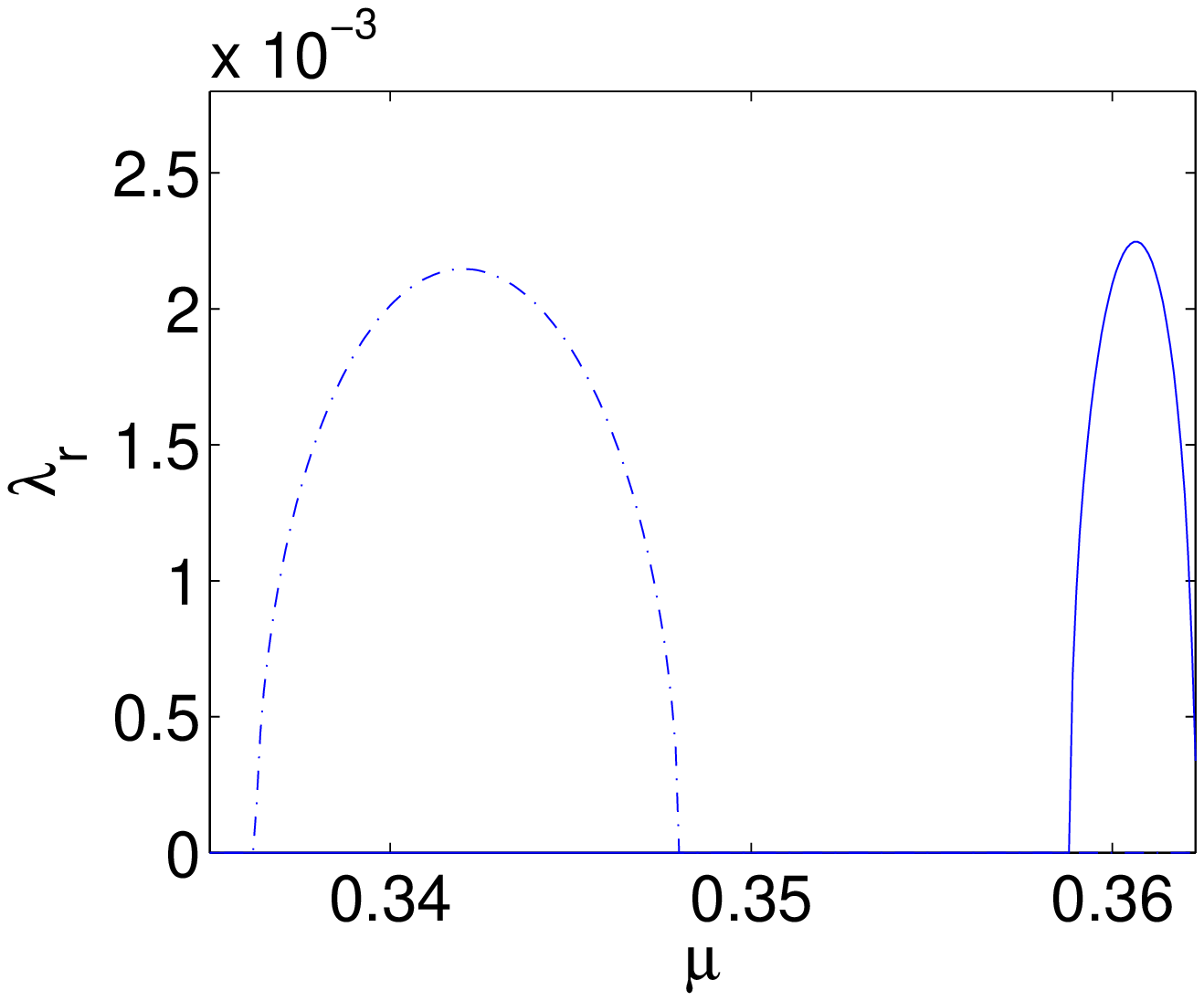}\newline
\caption{(Color online) Top: profiles of wave functions that represent
branches B1, B2 and B3 (from left to right) at $\protect\mu =0.34$. Bottom:
real parts of unstable eigenvalues of the corresponding branches as a
function of $\protect\mu $. The dashed-dotted line in the bottom right panel
indicates a complex quartet of eigenvalues.}
\label{figB}
\end{figure}

\begin{figure}[tbph]
\centering
\includegraphics[width=.24\textwidth]{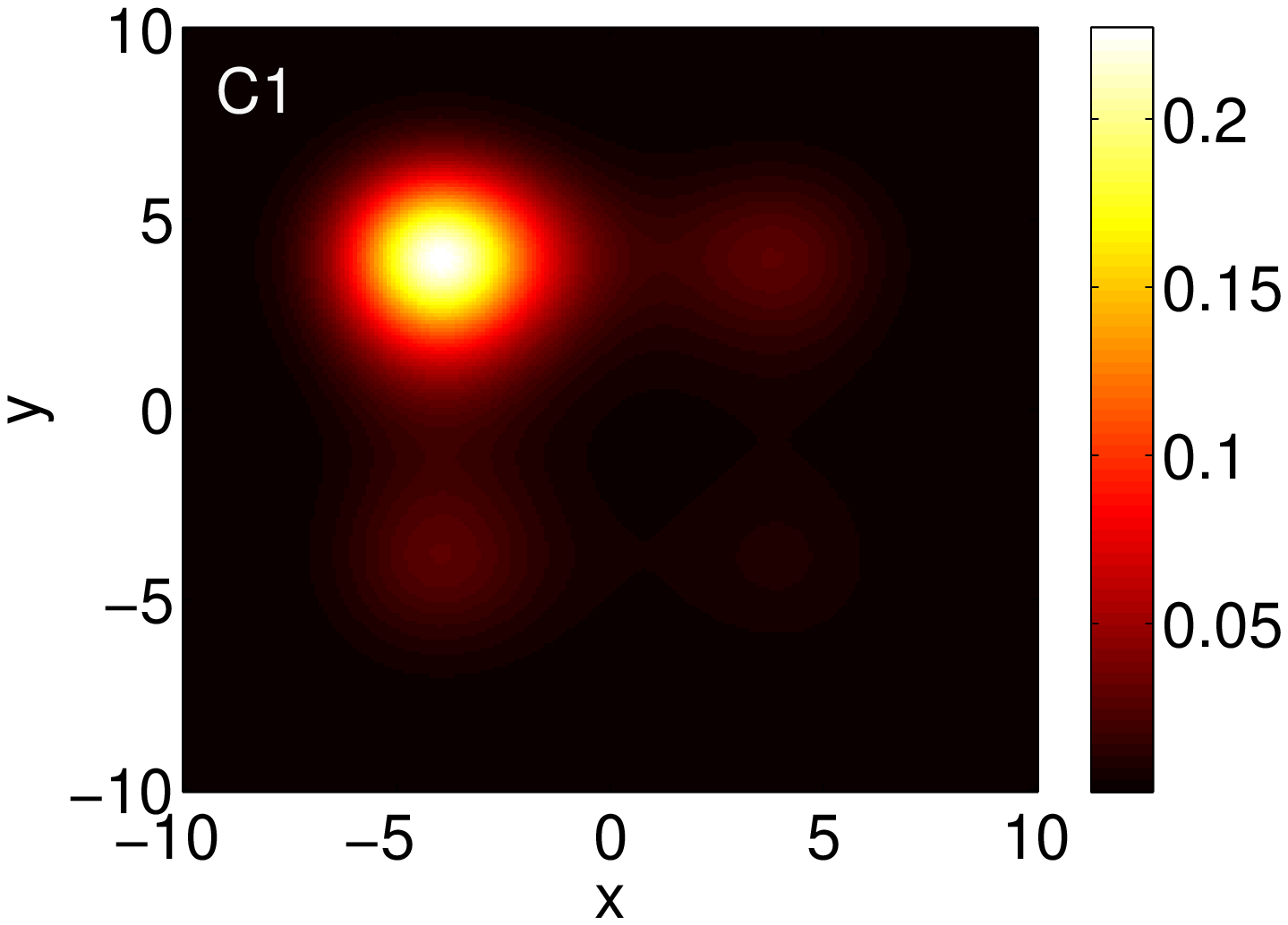} %
\includegraphics[width=.24\textwidth]{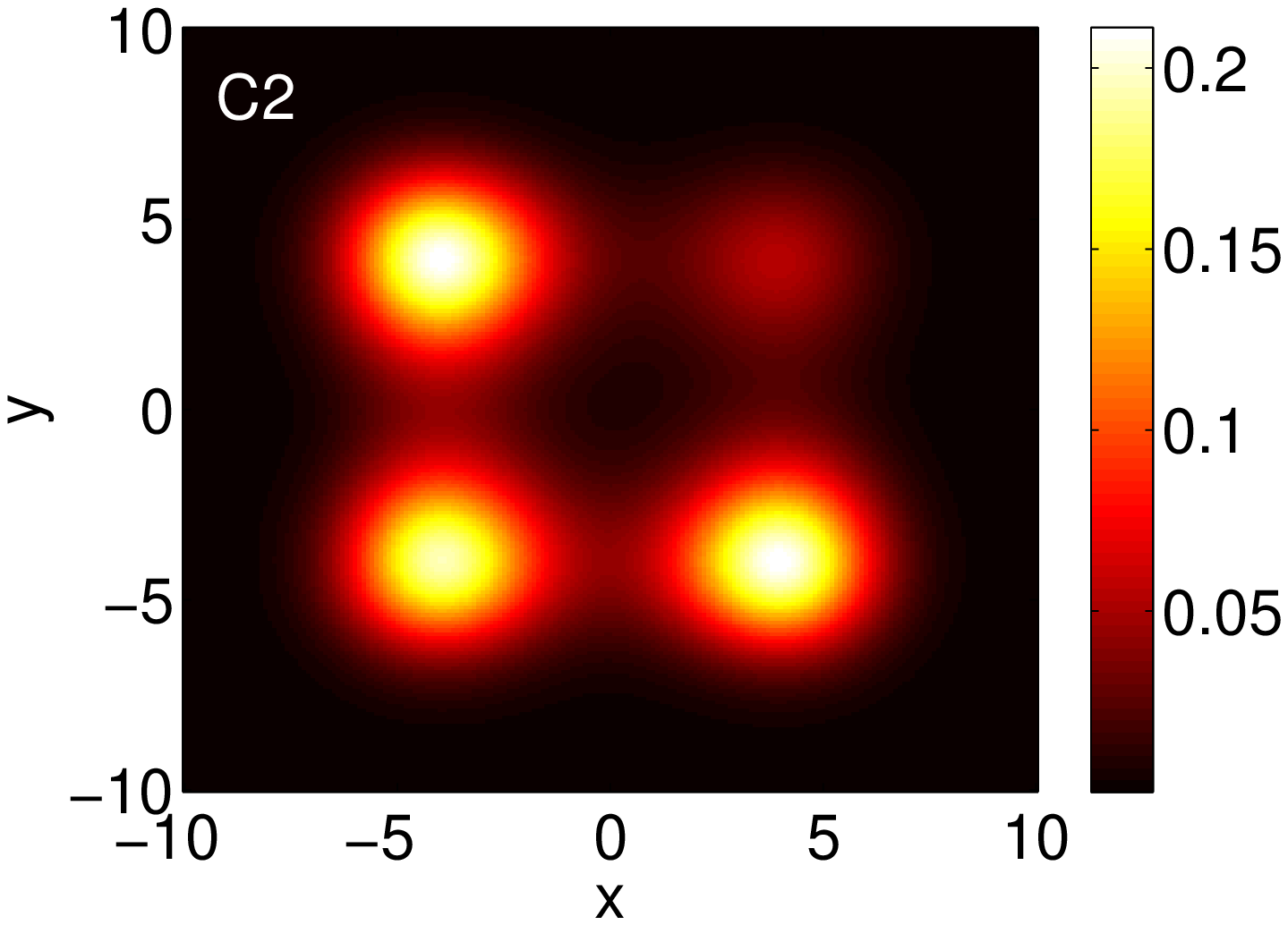} %
\includegraphics[width=.24\textwidth]{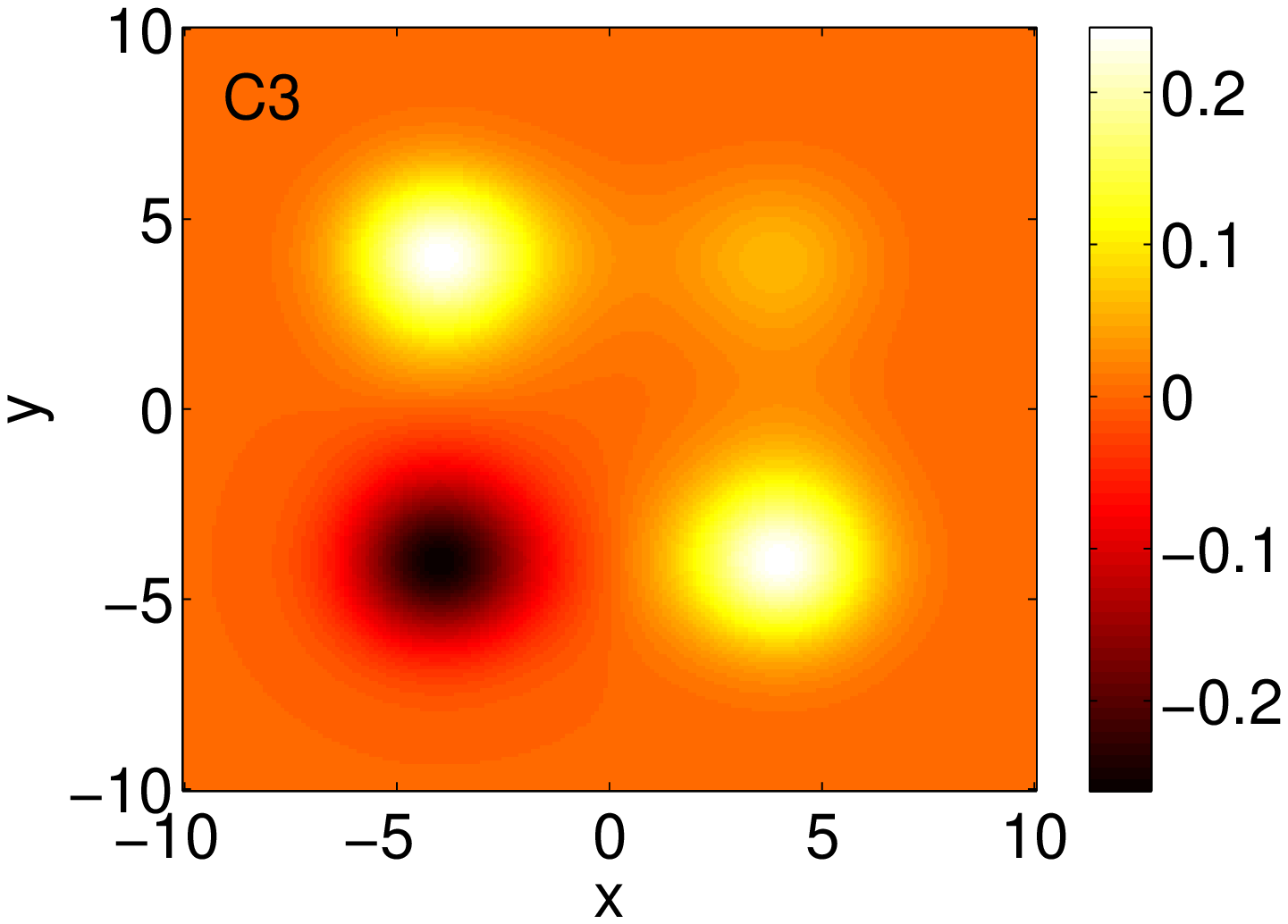} %
\includegraphics[width=.24\textwidth]{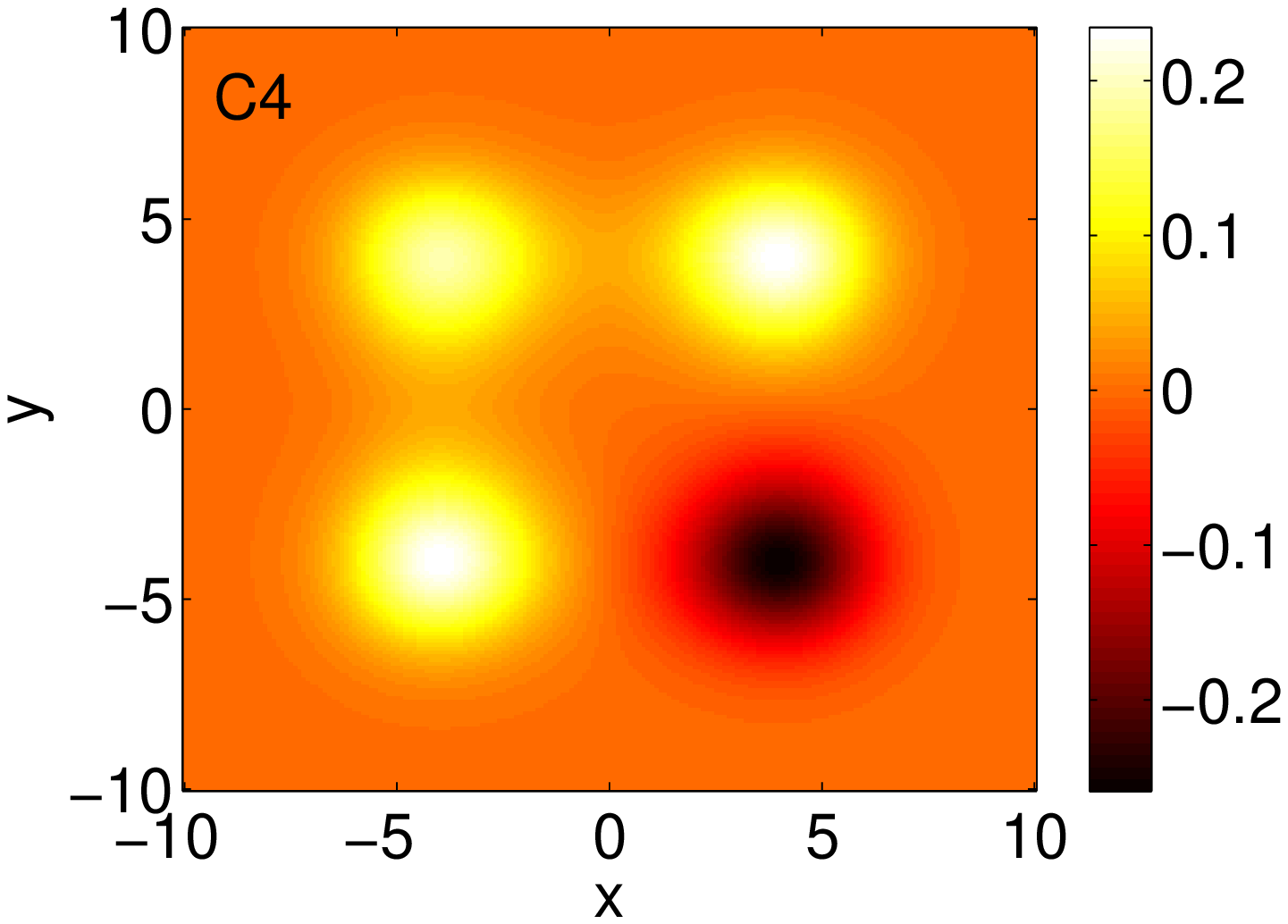}\newline
\includegraphics[width=.24\textwidth]{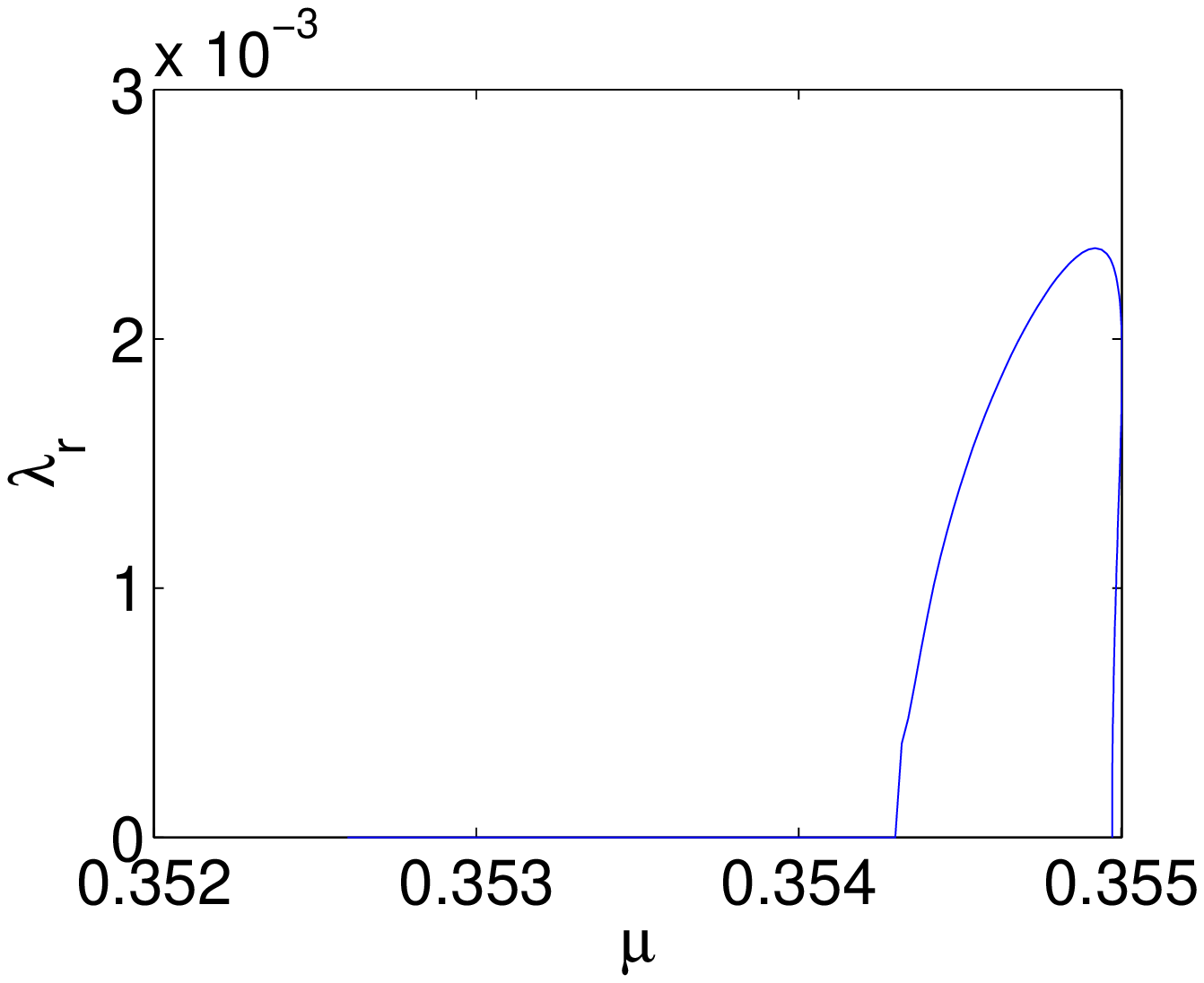} %
\includegraphics[width=.24\textwidth]{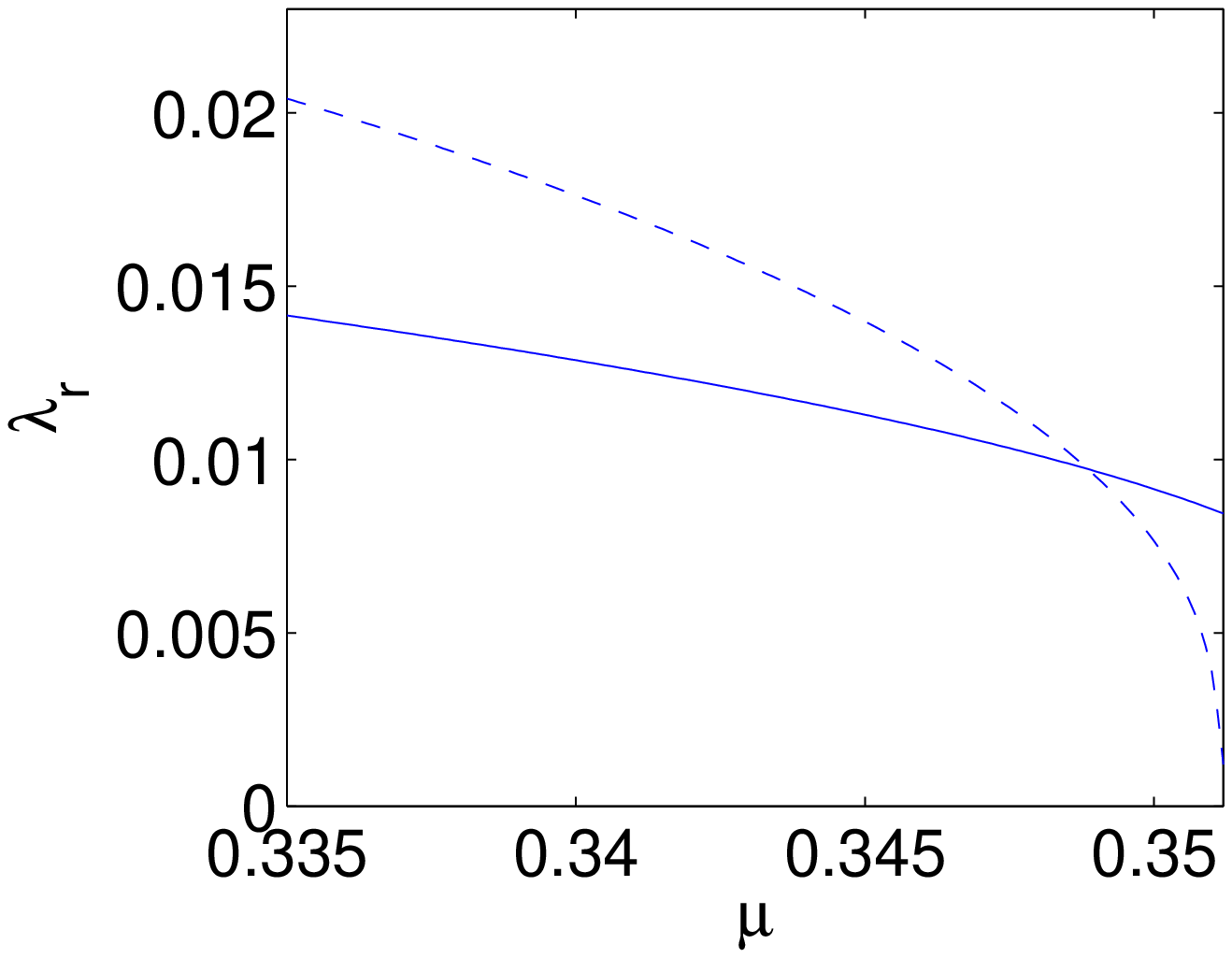} %
\includegraphics[width=.24\textwidth]{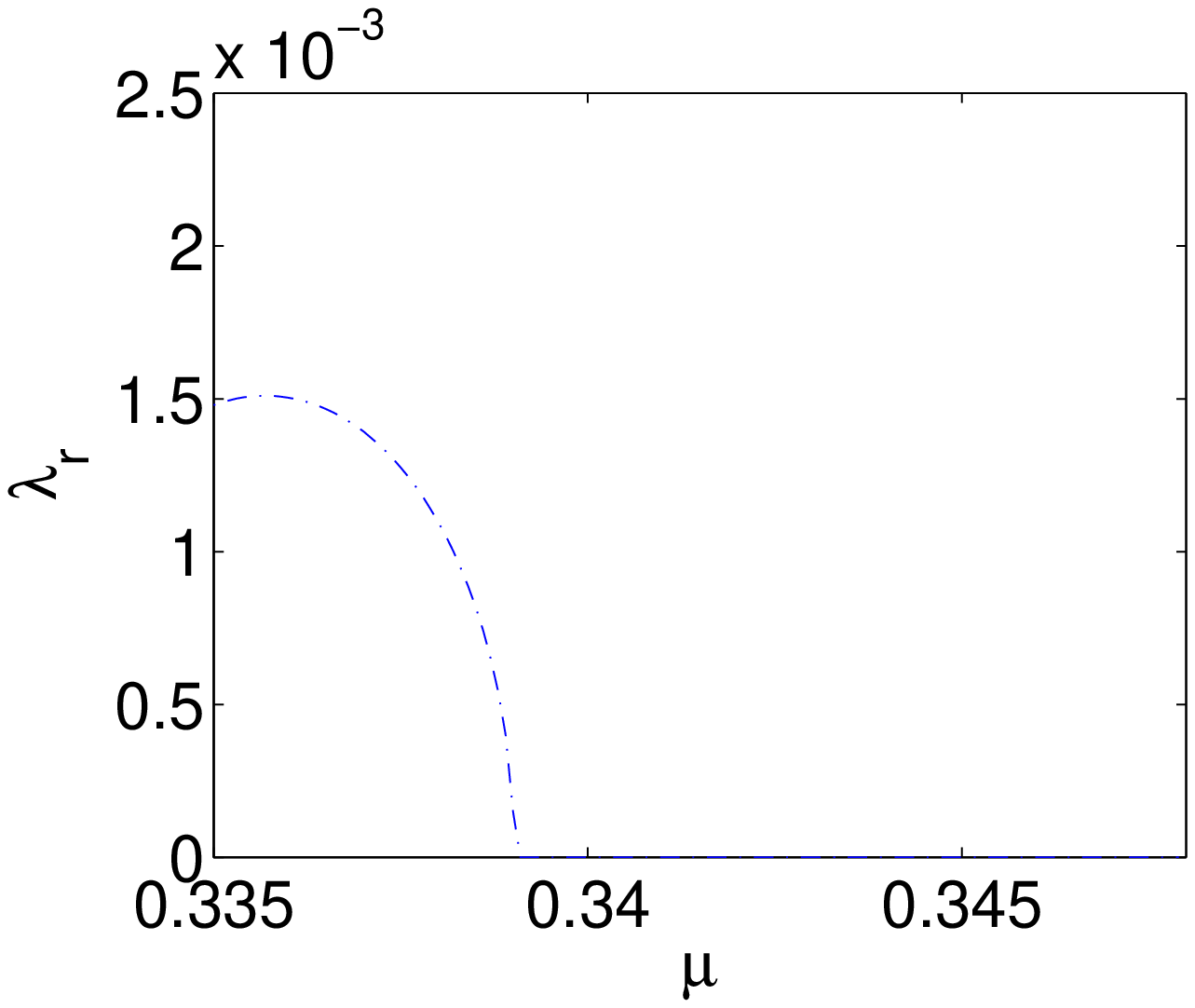} %
\includegraphics[width=.24\textwidth]{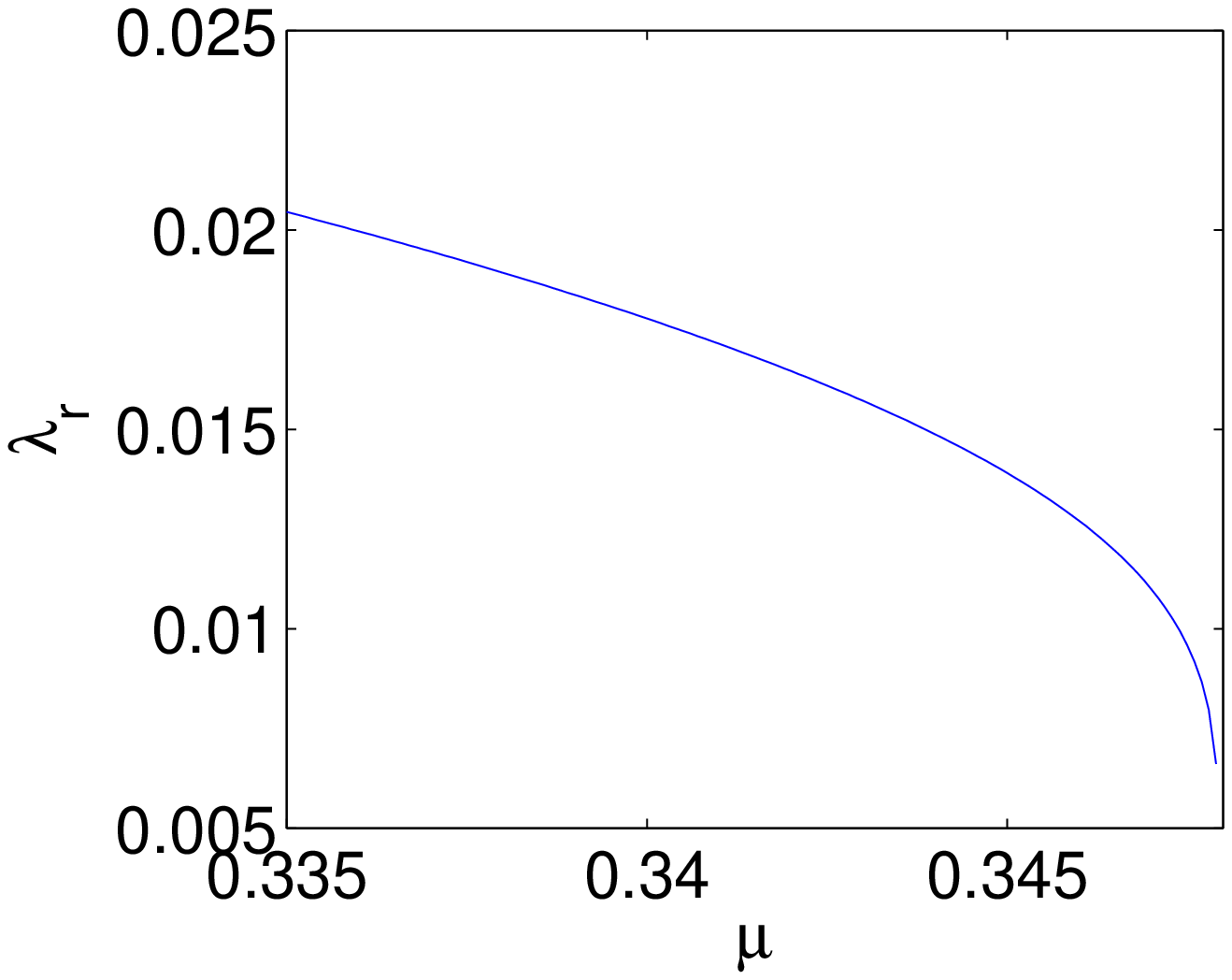}\newline
\caption{(Color online) Top: profiles of wave functions of branches C1, C2, C3 and C4 (from
left to right) at $\protect\mu =0.34$. Bottom: real parts of unstable
eigenvalues of the corresponding branches as a function of the eigenvalue
parameter $\protect\mu $.}
\label{figC}
\end{figure}

\begin{figure}[tbph]
\centering
\includegraphics[width=.24\textwidth]{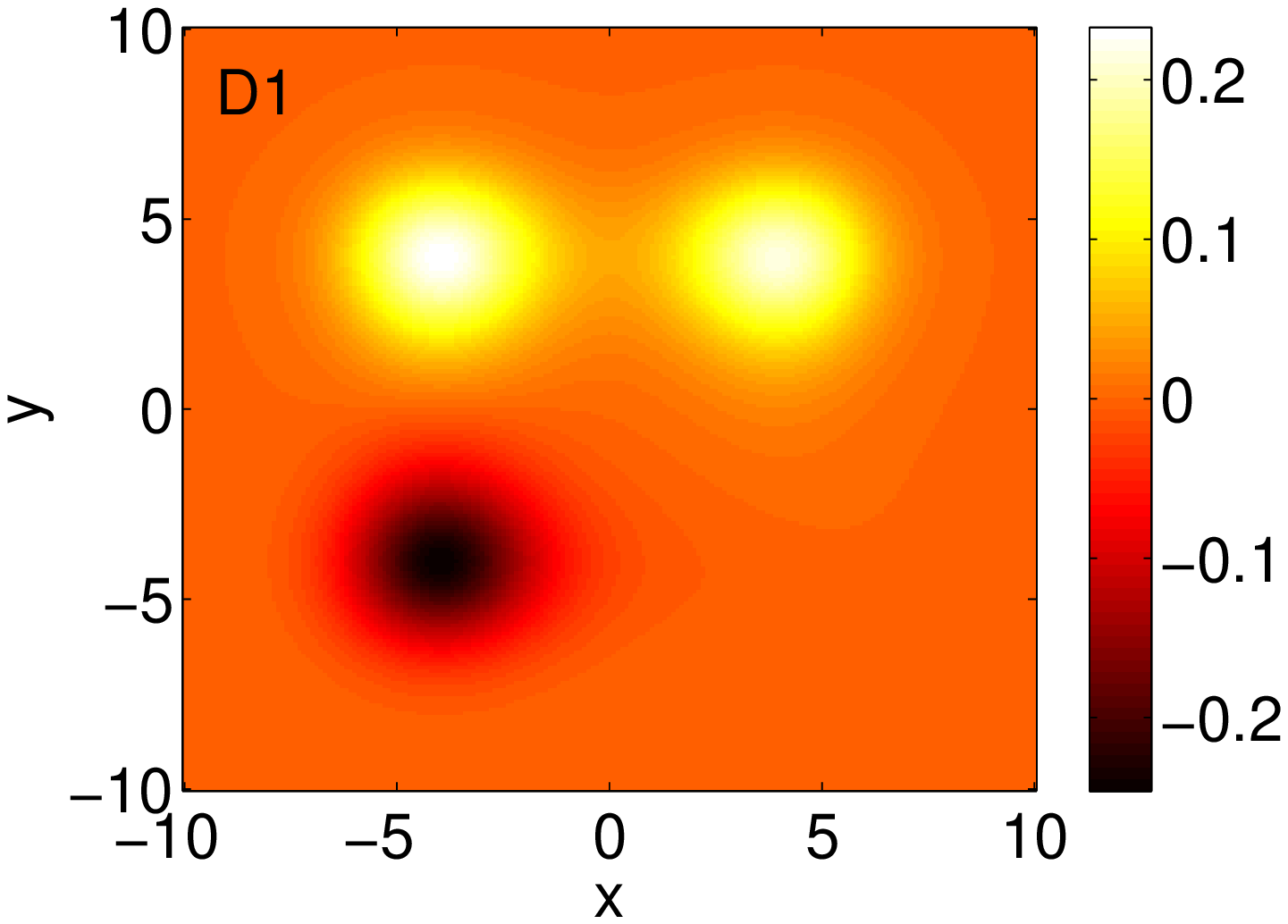} %
\includegraphics[width=.24\textwidth]{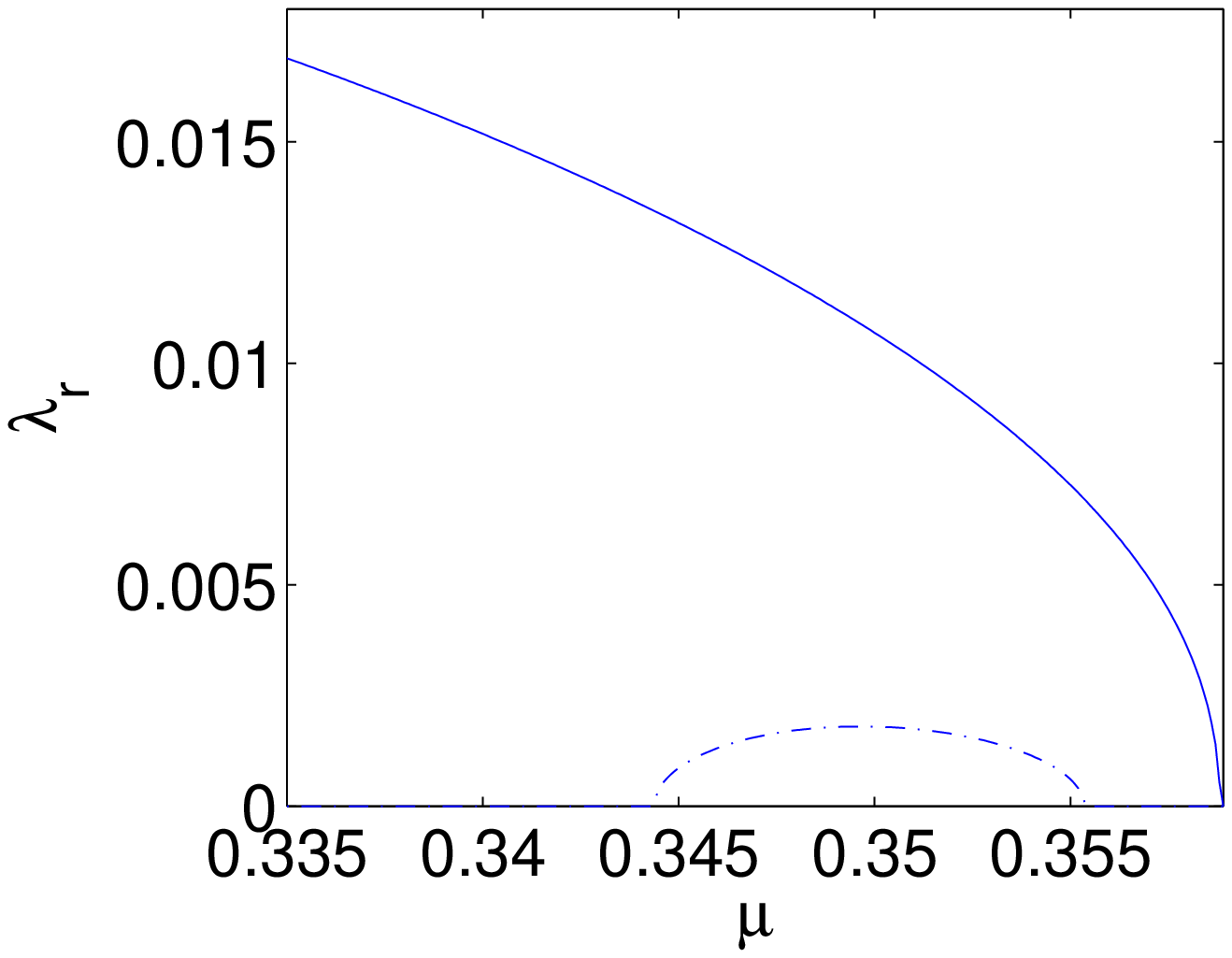}\newline
\caption{(Color online) Left: 
the profile 
%eigenstates belonging to 
of wave function of branch D1 at $\protect\mu =0.34$. 
Right: real parts of its unstable eigenvalues as a function of
the eigenvalue parameter $\protect\mu $. The dashed-dotted line in the right
panel indicates a complex quartet of eigenvalues.}
\label{figD}
\end{figure}

The twelve real branches are labeled mainly according to their relation to 
the populations of the four wells. To support our explanation, we introduce 
a symbolic representation that we developed in the form of $2\times 2$ 
matrices, labeling different waveforms that arise in the diagram, as 
follows: $A1\equiv 
\begin{pmatrix}
1 & 1 \\
1 & 1%
\end{pmatrix}%
$, $A2\equiv
\begin{pmatrix}
1 & 1 \\
-1 & -1%
\end{pmatrix}%
$, $A3\equiv
\begin{pmatrix}
-1 & 1 \\
1 & -1%
\end{pmatrix}%
$, $A4\equiv
\begin{pmatrix}
1 & 0 \\
0 & -1%
\end{pmatrix}%
$, $B1\equiv
\begin{pmatrix}
1 & 1 \\
\varepsilon & \varepsilon%
\end{pmatrix}%
$, $B2\equiv
\begin{pmatrix}
1 & \varepsilon \\
\varepsilon & 1%
\end{pmatrix}%
$, $B3\equiv
\begin{pmatrix}
1 & -1 \\
\varepsilon & -\varepsilon%
\end{pmatrix}%
$, $C1\equiv
\begin{pmatrix}
1 & \varepsilon \\
\varepsilon & \varepsilon%
\end{pmatrix}%
$, $C2\equiv
\begin{pmatrix}
1 & \varepsilon \\
1-\varepsilon & 1%
\end{pmatrix}%
$, $C3\equiv
\begin{pmatrix}
1 & \varepsilon \\
-1-\varepsilon & 1%
\end{pmatrix}%
$, $C4\equiv
\begin{pmatrix}
1-\varepsilon & 1 \\
1 & -1-\varepsilon%
\end{pmatrix}%
$, and $D1\equiv
\begin{pmatrix}
1 & 1-\varepsilon \\
-1-\varepsilon & -\varepsilon%
\end{pmatrix}$. 
In this representation, $1$, $-1$ and $0$ have the obvious meaning, by
indicating that a particular well is or is not populated, and the phase of the
wave function in it being $0$ or $\pi $ in the cases of $+1$ and $-1$,
respectively, when populated. Symbol $\varepsilon $ denotes either a small
(but nonzero) population in one of the wells, or a symmetry-breaking effect
(when some of the density peaks feature values $\pm 1\pm \varepsilon $, thus
being slightly different from $\pm 1$). The labeling is then defined as 
%the following: 
follows: branches A1-A4 have the same amplitude at the four wells as long
as they are populated, branches B1-B3 feature two pairs of peaks with
different amplitudes, branches C1-C4 have three different amplitudes, while
D1 has all of its four peaks different. The waveforms in the top rows of
Figs. \ref{figA}-\ref{figD} display prototypical realizations of the
relevant branches. Their stability properties are illustrated, as a function
of eigenvalue parameter $\mu $, in the bottom rows.
%will help understanding
%the above representation and labeling.

We will now explain in detail solutions appearing in the full bifurcation
diagram, starting from the linear limits ($N\rightarrow 0$).\ First, we look
at the group of solutions related to branch A1, as shown in the bottom left
panel of Fig. \ref{fig_foc}. This branch arises from the symmetric linear
mode at $\mu =\omega _{0}$, i.e., the ground state in the linear limit, $%
u_{0}$. Accordingly, A1 features four identically populated wells. The
analysis demonstrates that it is stable near the linear limit, but is soon
destabilized, due to the emergence of branches C1 and B1, through
subcritical and supercritical pitchfork bifurcations, respectively, 
around $\mu =0.355$. 
In other words, there are two consecutive \textit{steady-state bifurcations}, 
in the language of Ref. \cite{dellnitz}, in two different subspaces, in
which one unstable solution, C1, collides with A1 and, simultaneously, a
pair of eigenvalues emerges on the real axis for the A1 branch with the
decrease of $\mu $ (in a subcritical pitchfork); then, a super-critical
pitchfork takes place in another subspace, in which B1 retains only one real
pair, while another pair passes through the origin along the A1 branch. The
actual ``pitchfork'' cannot be visualized here in the usual manner, because
any of the four equivalent versions of B1 (obtained by $\pi /2$ rotation)
have the same value of $N$, and so are represented by the same branch in the
graph. Branch B1, which is unstable due to a pair of real eigenvalues in the
linearization around it throughout its domain of existence, features two of
the wells on one side being less populated than the other two. 
Configuration B1 
%gets 
becomes increasingly more asymmetric as it deviates from A1. A noteworthy
feature is shown by branch C1, which bifurcates from A1 at almost the same
place as B1: after having emerged, it tends to be located on the left 
of A1 as are all
other branches bifurcating from A1. 
However, within a narrow interval of $\mu $, its norm
decreases ($dN/d\mu >0$) slightly before starting to grow as usual 
($dN/d\mu <0$). Naturally, when the norm decreases the solution is 
destabilized and then it remains stable after the turning point, which is 
explained by the well-known Vakhitov-Kolokolov criterion \cite{vk}. 
Branch A1 is endowed with two identical pairs of real eigenvalues by B1 
and C1 upon their bifurcation 
(which is shown as the dashed line in the bottom left panel of 
Fig. \ref{figA}.) As $N$ grows further, a subsequent bifurcation, 
at $\mu =0.3519$, leading to the emergence of branch B2, adds yet another 
real eigenvalue pair to A1; 
%which 
this means that A1 possesses in total three real eigenvalue pairs 
%in total for $N$ large enough. 
for sufficiently large vaues of $N$. Branch B2 features two wells on the 
diagonal which are less populated than the other two, and it is unstable, 
with two pairs of real eigenvalues, near the point where it is generated 
by the bifurcation from A1; however, one of the pairs is eliminated by the 
emergence of a new branch, C2, from B2 through a pitchfork shortly afterwards. 
Branch B2 then remains unstable with one real pair, while C2 (with three 
principal sites, one of which is of lesser amplitude than the other two) 
is unstable due to two real eigenvalue pairs.

This description encompasses all branches of stationary solutions which can be
traced back to the ground state of the linear system. Detailed information
for the 
%waveforms 
wave function profiles and the development of the real eigenvalues associated to
them is presented in Figs. \ref{figA}-\ref{figC}.

Next we turn to 
%solutions 
the states originating from the second linear mode, as shown
in the bottom right panel of Fig. \ref{fig_foc}. Branch A2 starts from the
respective eigenvalue, $\mu =\omega _{1}=\omega _{2}$, which pertains to the
first and second (degenerate at the linear limit) 
excited states. This branch emerges as an unstable\ one,
carrying a real eigenvalue pair. 
%Its waveform 
The respective wave function profile features four wells populated
with the same amplitude but $\pi$ out-of-phase between the two sides, see 
Fig. \ref{figA}. Branch B3 emerges from A2 through a supercritical pitchfork 
at $\mu =0.3623$, lending another real eigenvalue pair to A2. 
Similar to the case of B1 (as it separates from A1), in the B3 state two 
wells on the one side tend to be less populated than the other two, as this 
branch moves further from A2. Branch B3 remains unstable with one real 
eigenvalue pair, until getting stabilized by another pitchfork bifurcation 
which takes place at $\mu =0.3589$;
% which 
this simultaneously gives rise to a new 
unstable branch, D1, that has different populations in all four wells. 
Notice that both B3 and D1 pass through a Hamiltonian-Hopf bifurcation 
(alias 1:1 resonance, in terms of Ref. \cite{dellnitz}), which means that, 
in the relevant parametric interval ($0.3362<\mu <0.348$ for B3; 
$0.3444<\mu <0.3553$ for D1), B3 is destabilized by a complex quartet 
of small eigenvalues in the linearization around the stationary solution, 
while D1 remains unstable, but with one real eigenvalue pair and a complex quartet 
%destabilizes B3 and
%causes D1 to remain unstable but with
%three pairs of eigenvalues in total (in the relevant
%parametric interval) whose real parts are positive.
(the dashed-dotted lines in the bottom right panels of Figs. \ref{figB} and %
\ref{figD} refer to this effect).

Furthermore, branch A4 bifurcates from the same linear mode as A2. It is
unstable near the linear limit due to a Hamiltonian-Hopf bifurcation, but
with the increase of $N$ it becomes stable. Branch A4 features a waveform in
which only two wells lying on the diagonal are populated, with the same 
amplitude but opposite signs.
%Among all the solutions, it is the only one that has some of the wells
%not populated.

Branch A3 arises from the third excited linear mode at $\mu =\omega_{3}$.
In this case, two wells on the diagonal are populated with equal amplitudes,
while in the other two the amplitudes are of opposite signs. It is the
unique stationary solution which remains stable across the entire
bifurcation diagram (despite the fact that it has three pairs of purely 
imaginary 
eigenvalues with negative Krein signature \cite{sandstede}, 
which in principle, can give rise to Hamiltonian-Hopf bifurcations).

Finally, branches C3 and C4, which are located slightly below A2 in Fig. \ref%
{fig_foc}, correspond to a pair of 
%solutions 
states which arise through a
saddle-node bifurcation at some critical value of chemical potential ($\mu
\approx 0.348$, for the model's parameters chosen in the present case.)
Branch C4 (the one with higher values of $N$) is unstable with a real
eigenvalue pair, while C3 remains stable, except inside a short instability
interval, which is accounted for by a Hamiltonian-Hopf bifurcation.

\begin{figure}[tbph]
\centering
\includegraphics[width=.24\textwidth]{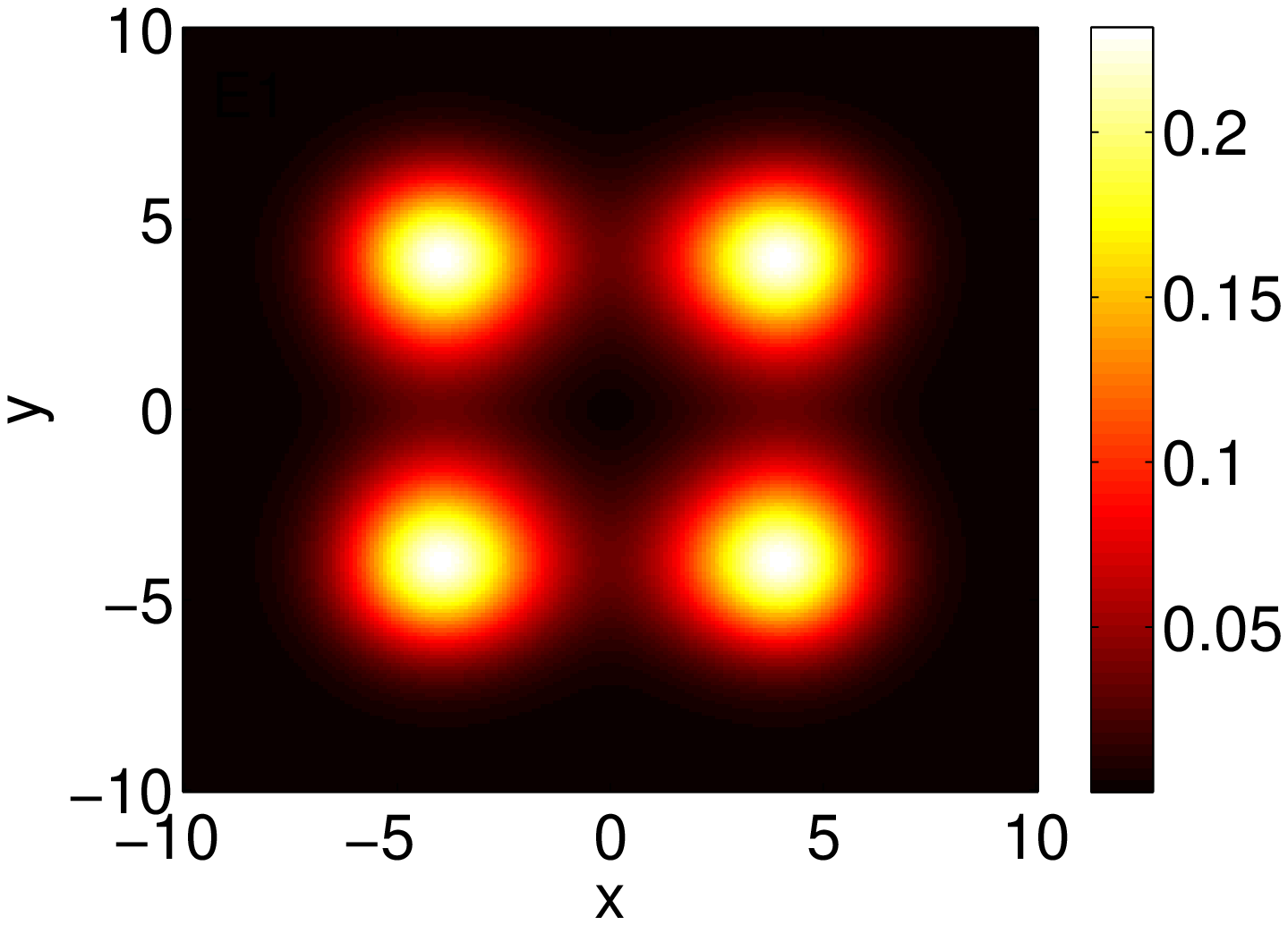} %
\includegraphics[width=.24\textwidth]{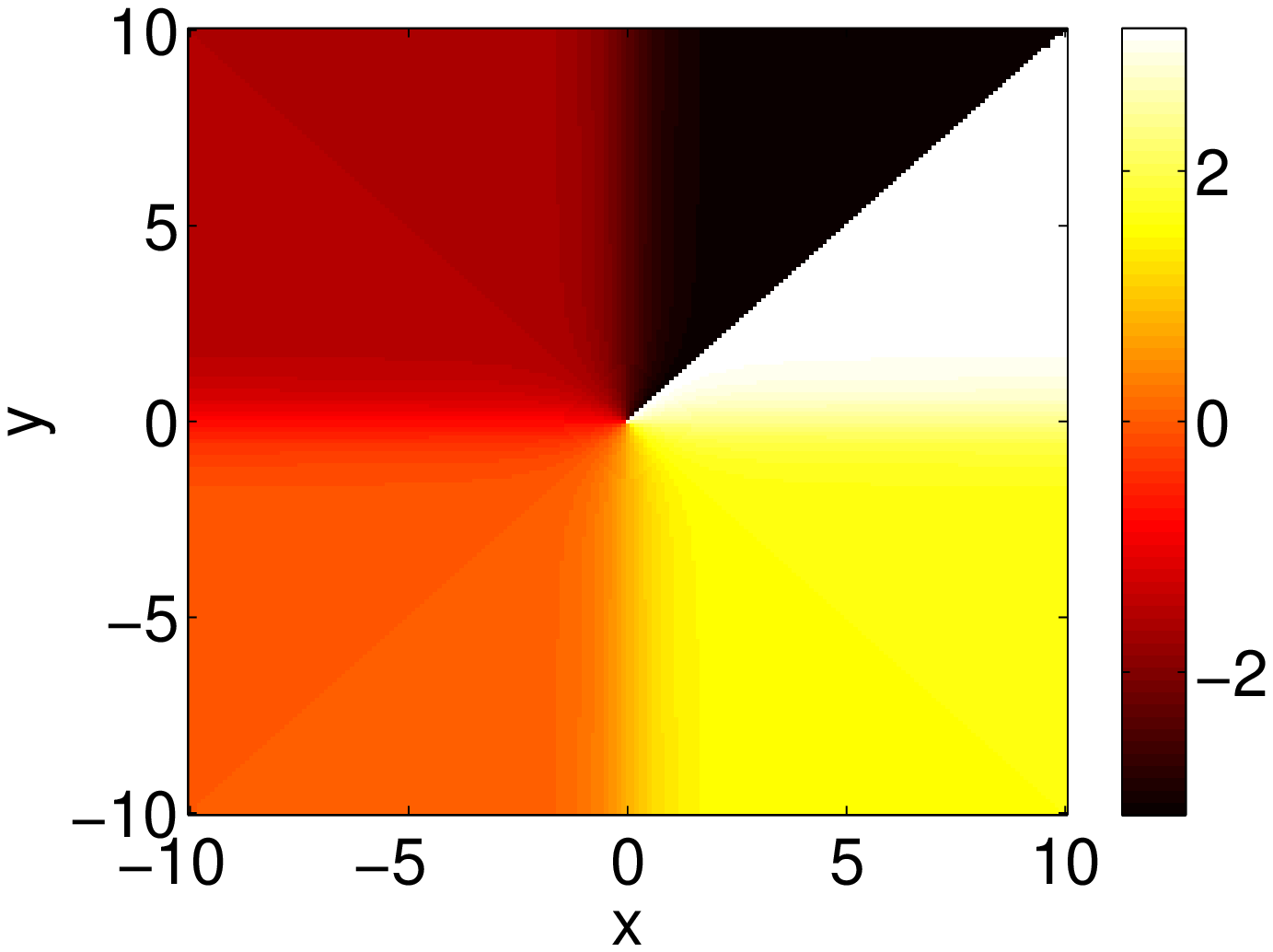}\newline
\caption{(Color online) The absolute value (left) and phase (right) of a vortex state
%vortex at 
for $\protect\mu =0.34$.}
\label{figE}
\end{figure}

In addition to the above real stationary states, we have also found complex
solutions in the form of vortices \cite{kody,todd_chaos}. A typical example
of such a solution is shown in Fig. \ref{figE}. Throughout the regime of
parameters considered herein, such solutions have been found to be \emph{%
linearly stable}.

It is interesting to note that, for all the solutions considered herein, in
the large-$N$ limit their stability characteristics coincide with what can
be suggested by the DNLS model considered in Ref. \cite{peli2d} (see also
Refs. \cite{peli1d,peli3d} for the corresponding 1D and 3D stability
results). Gross features of these findings are that, whenever two adjacent
sites are in-phase, a real eigenvalue pair is expected to emerge due to
their interaction, while whenever such sites are out-of-phase, the relevant
eigenvalue is expected to be imaginary \cite{multip}, but with 
%the 
negative
%sign of the so-called 
Krein signature \cite{peli2d}, which implies a 
potential for a Hamiltonian-Hopf bifurcation. It should also be noted that,
in the limit of the infinite lattice, it is naturally expected that the
asymmetries observed herein in many of the branches will disappear (i.e.,
the amplitudes in different wells will be equal) -- see also a relevant
discussion in Ref. \cite{3well}.
%%%%%%%%%%%%%%%%%%%%%%%%%%%%%%%%
%%%%%%%%%%%%%%%%%%%%%%%%%%%%%%%%

\subsection{Repulsive interactions}

\label{repulsive}

\begin{figure}[tbph]
\centering
\includegraphics[width=.4\textwidth]{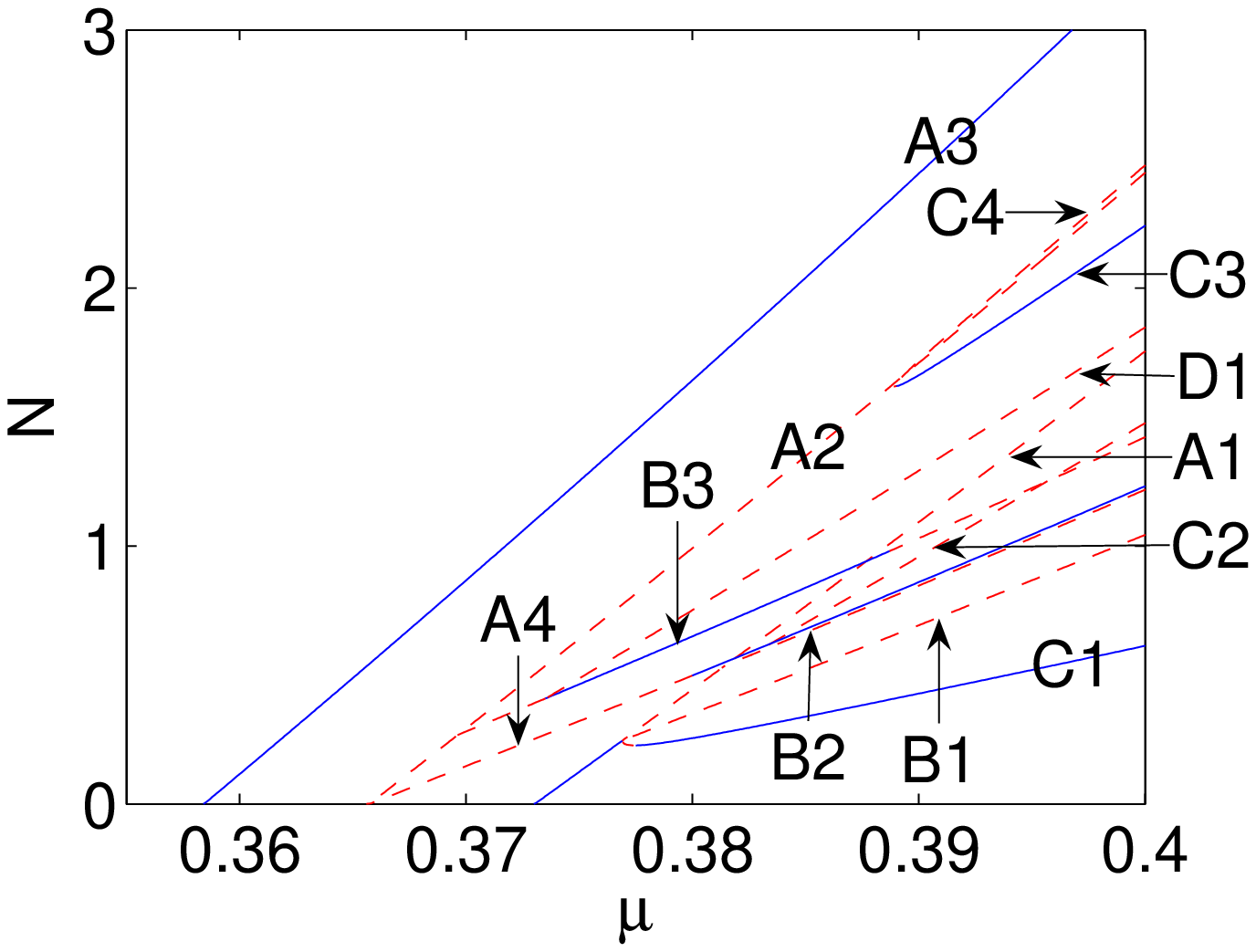} %
\includegraphics[width=.4\textwidth]{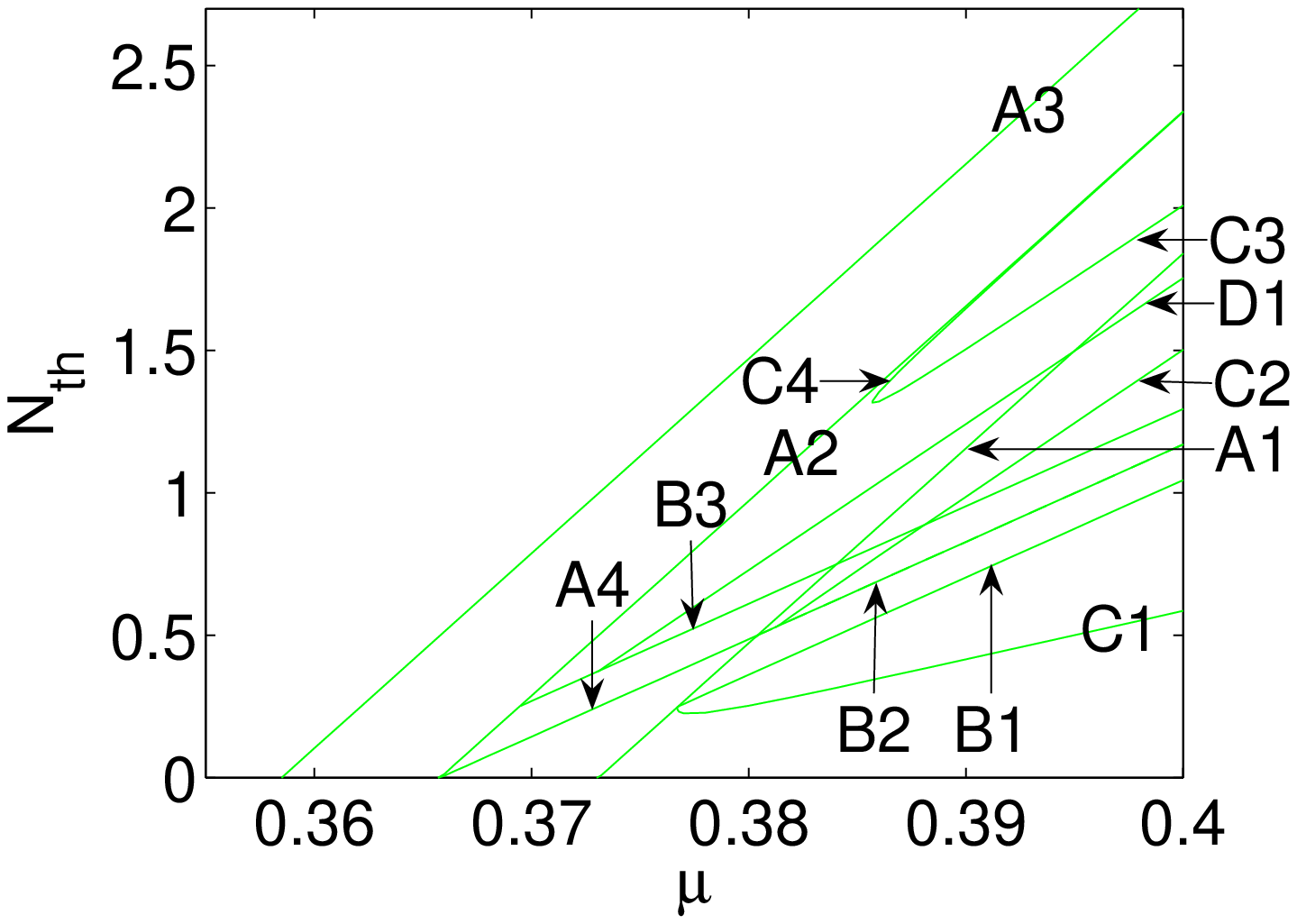}\newline
\caption{(Color online) The norm of the numerical (left) and approximate (right) stationary
solutions to Eq. (\protect\ref{eq1}) with the self-defocusing 
%repulsive 
nonlinearity ($s=+1$), 
as a function of the parameter $\protect\mu $ (the chemical potential in BECs or
propagation constant in optics). The labels of the branches are explained in 
detail in the text.}
%Labels attached to the branches are consistent with
%the ones in Fig. \protect\ref{fig_foc}, in the sense of the \textit{staggering transformation}.}
\label{fig_rep}
\end{figure}

We now briefly discuss the case of repulsive interactions 
(alias self-defocusing nonlinearity), corresponding to $s=+1$
in Eq. (\ref{eq1}), with an objective to highlight its similarities with 
and
differences from the 
%self-focusing 
case of attractive interactions. The bifurcation diagram
for the model is displayed in Fig. \ref{fig_rep}. 

The solutions are labeled so as to match the self-focusing case, 
by means of the appropriate \textit{staggering transformation} \cite{hadistuff}. 
The latter effectively converts the defocusing nonlinearity into a focusing one 
by changing the relative phase of nearest-neighbors from $0$ to $\pi $ and vice
versa, while preserving the relative phase of next-nearest-neighbors.
In this way, each solution in the defocusing case is linked to its counterpart in the
focusing model through this transformation. 
Following this relation, and adopting
the same matrix symbolic representation used for the 
focusing case in section \ref{attractive},
the branches of solutions are labeled as follows: 
$A1\equiv
\begin{pmatrix}
-1 & 1 \\
1 & -1%
\end{pmatrix}%
$, $A2\equiv
\begin{pmatrix}
-1 & 1 \\
-1 & 1%
\end{pmatrix}%
$, $A3\equiv
\begin{pmatrix}
1 & 1 \\
1 & 1%
\end{pmatrix}%
$, $A4\equiv
\begin{pmatrix}
-1 & 0 \\
0 & 1%
\end{pmatrix}%
$, $B1\equiv
\begin{pmatrix}
-1 & 1 \\
\varepsilon & -\varepsilon%
\end{pmatrix}%
$, $B2\equiv
\begin{pmatrix}
-1 & \varepsilon \\
\varepsilon & -1%
\end{pmatrix}%
$, $B3\equiv
\begin{pmatrix}
-1 & -1 \\
\varepsilon & \varepsilon%
\end{pmatrix}%
$, $C1\equiv
\begin{pmatrix}
-1 & \varepsilon \\
\varepsilon & -\varepsilon%
\end{pmatrix}%
$, $C2\equiv
\begin{pmatrix}
-1 & \varepsilon \\
1-\varepsilon & -1%
\end{pmatrix}%
$, $C3\equiv
\begin{pmatrix}
-1 & \varepsilon \\
-1-\varepsilon & -1%
\end{pmatrix}%
$, $C4\equiv
\begin{pmatrix}
-1+\varepsilon & 1 \\
1 & 1+\varepsilon%
\end{pmatrix}%
$, and $D1\equiv
\begin{pmatrix}
-1 & 1-\varepsilon \\
-1-\varepsilon & \varepsilon%
\end{pmatrix}$. 

Thus,  
% we realize that 
in this case, the symmetric ground state of the system is A3, which
is stable for arbitrary values of $N$. 
%and it inherits the stability of the A3 mode in the focusing model.
Branch A2 is immediately unstable, starting from the linear limit. 
B3 bifurcates from A2 and remains unstable before
getting stabilized through giving birth to D1 (and then becoming
destabilized again). Branch A1 is stable near the linear limit, but is
subsequently destabilized due to bifurcations that give rise to B1 and C1,
and an additional real eigenvalue pair arises at higher value of $N$ due to
the emergence of B2, from which another new branch, namely C2, arises in
turn. Branches C3 and C4 exist for a while (when $N$ is large enough), and
then collide at $\mu =0.389$ (for the values of parameters adopted herein).
The types of the bifurcations, the emergence of the corresponding solutions,
and the corresponding stability properties were found to be in direct
correspondence to the case of self-focusing nonlinearity, provided that 
one takes into account the staggering transformation 
%which relates 
relating the repulsive and attractive case models as indicated above.

\subsection{Dynamics}

\label{dynamics} 

\begin{figure}[tbph]
\centering
\includegraphics[width=.2\textwidth]{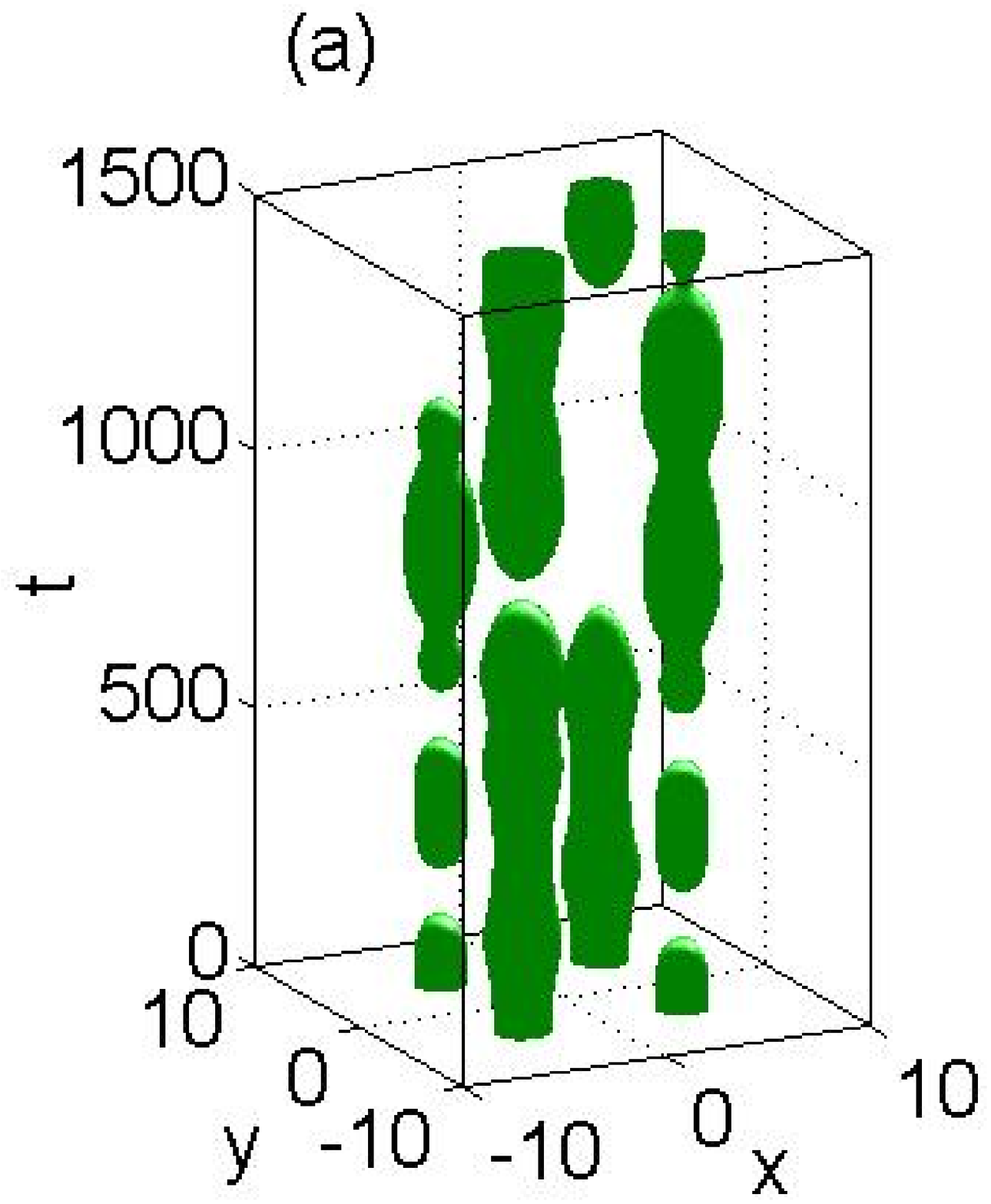} %
\includegraphics[width=.2\textwidth]{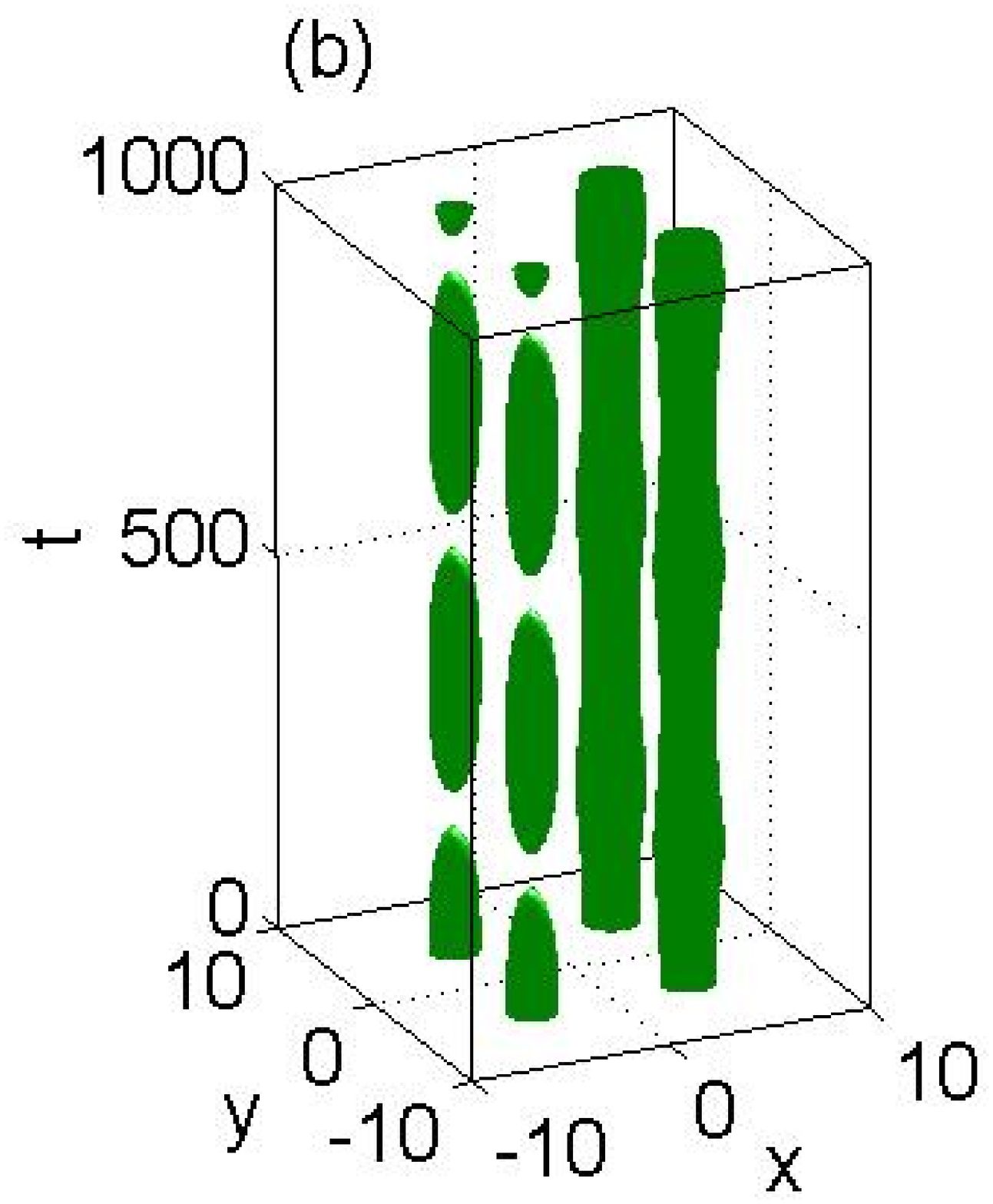} %
\includegraphics[width=.2\textwidth]{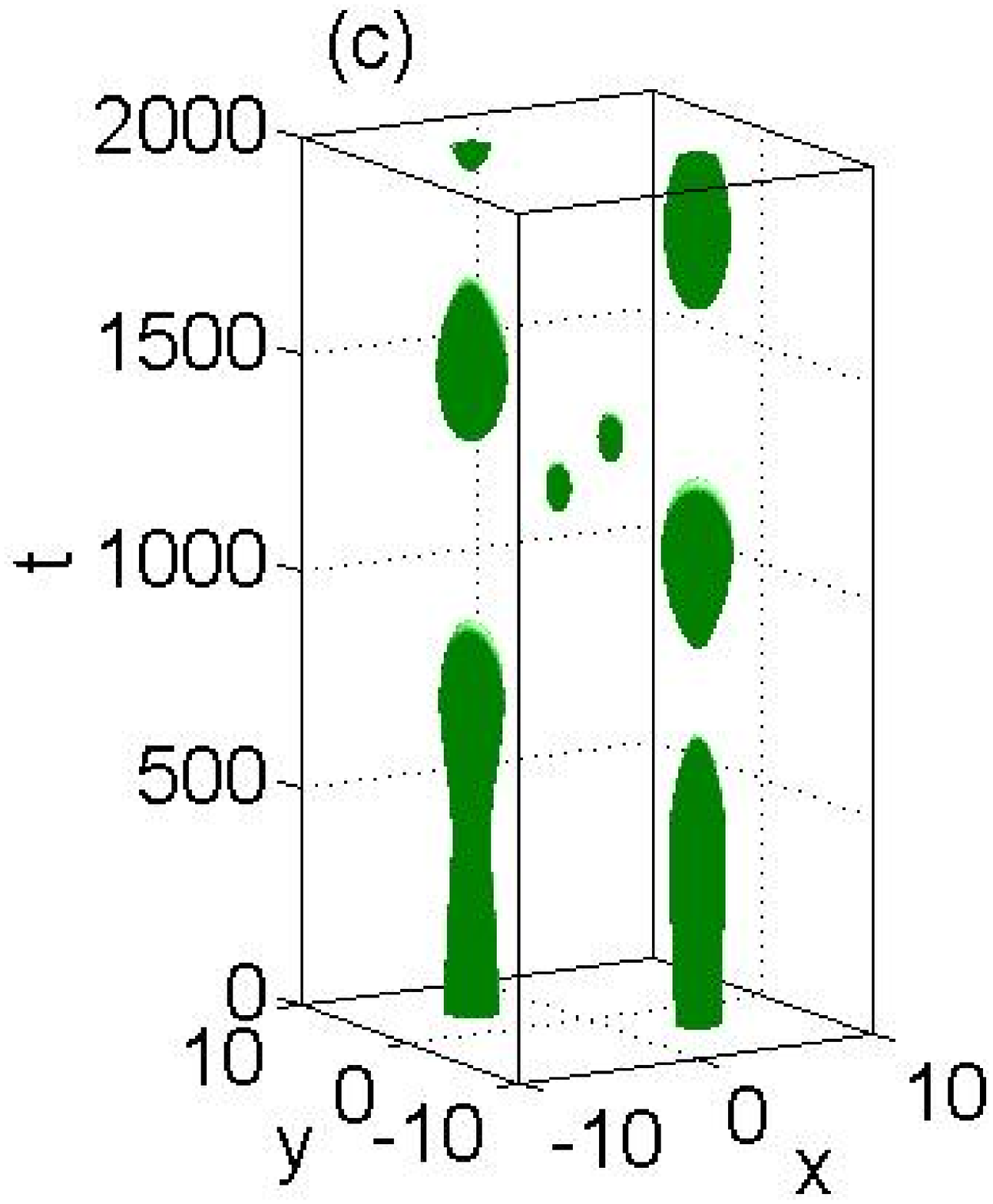}\newline
\includegraphics[width=.2\textwidth]{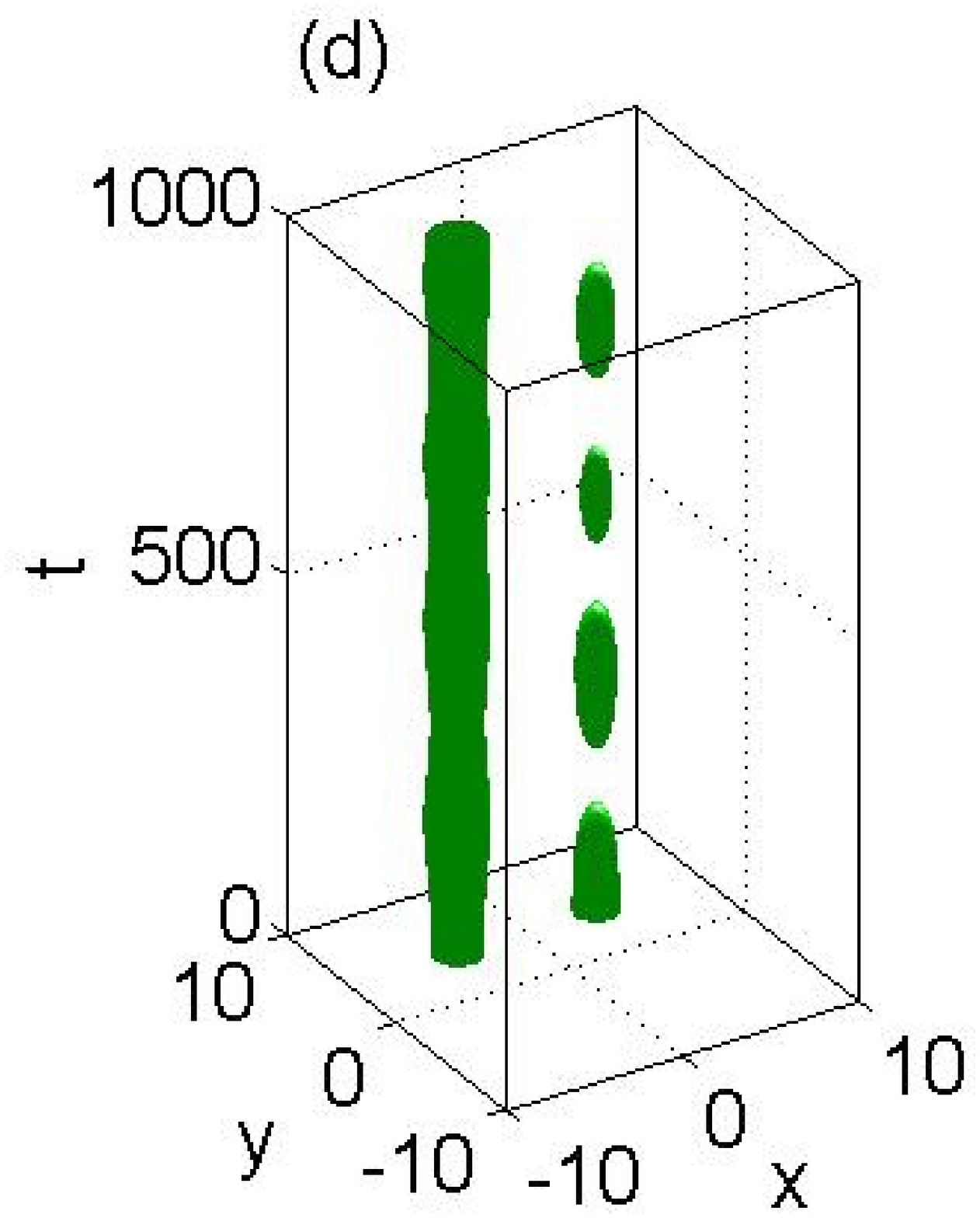} %
\includegraphics[width=.2\textwidth]{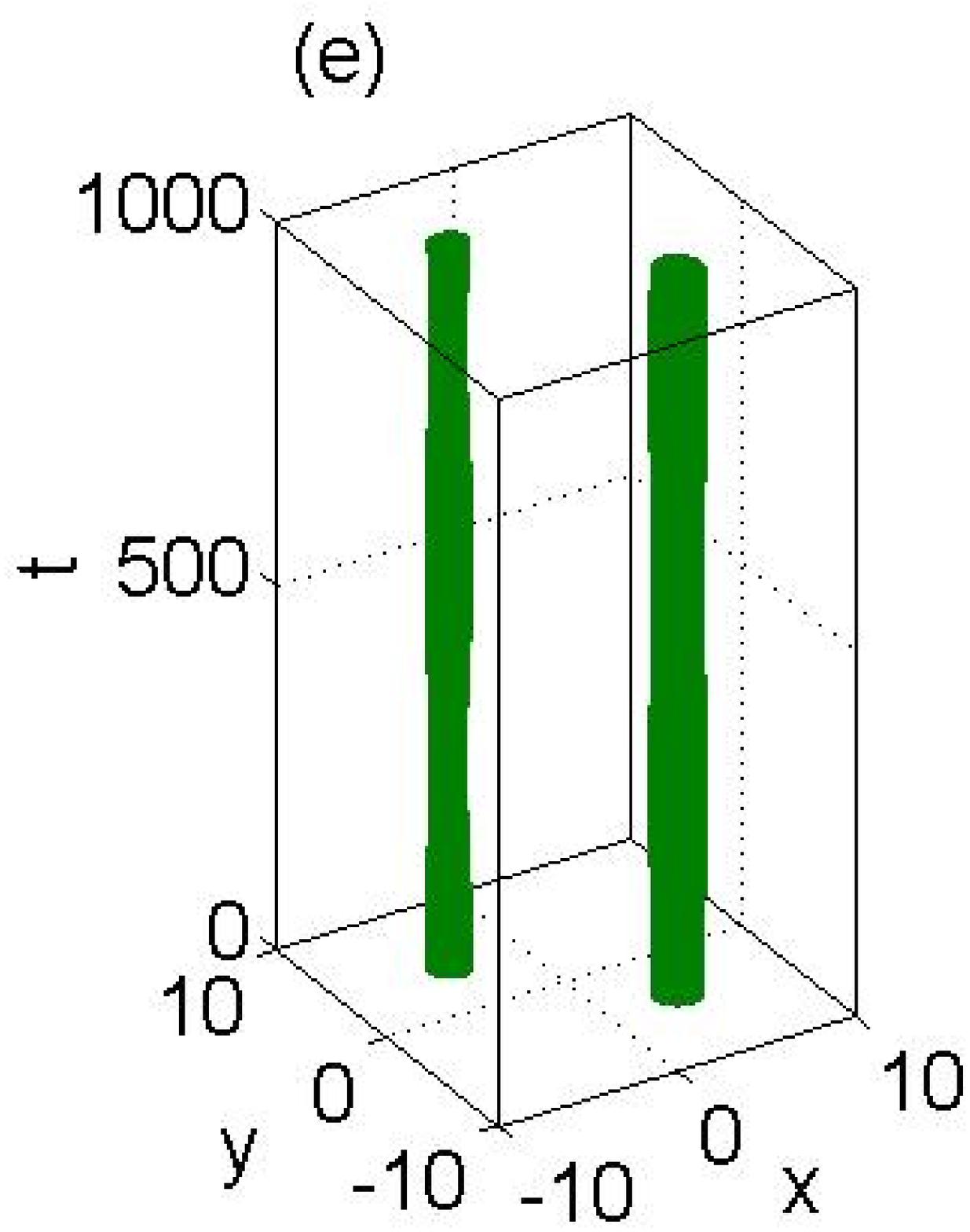} %
\includegraphics[width=.2\textwidth]{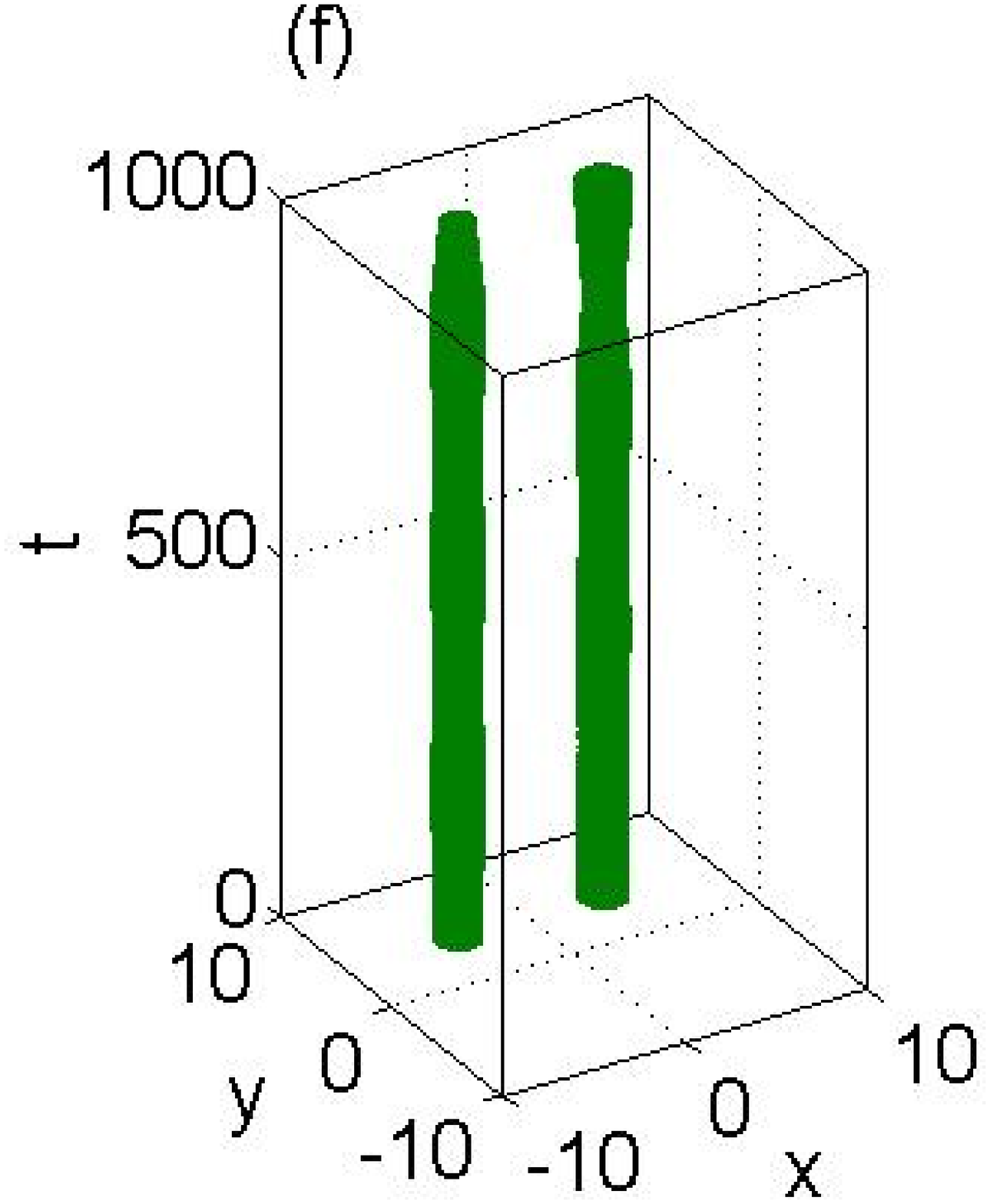}\newline
\includegraphics[width=.2\textwidth]{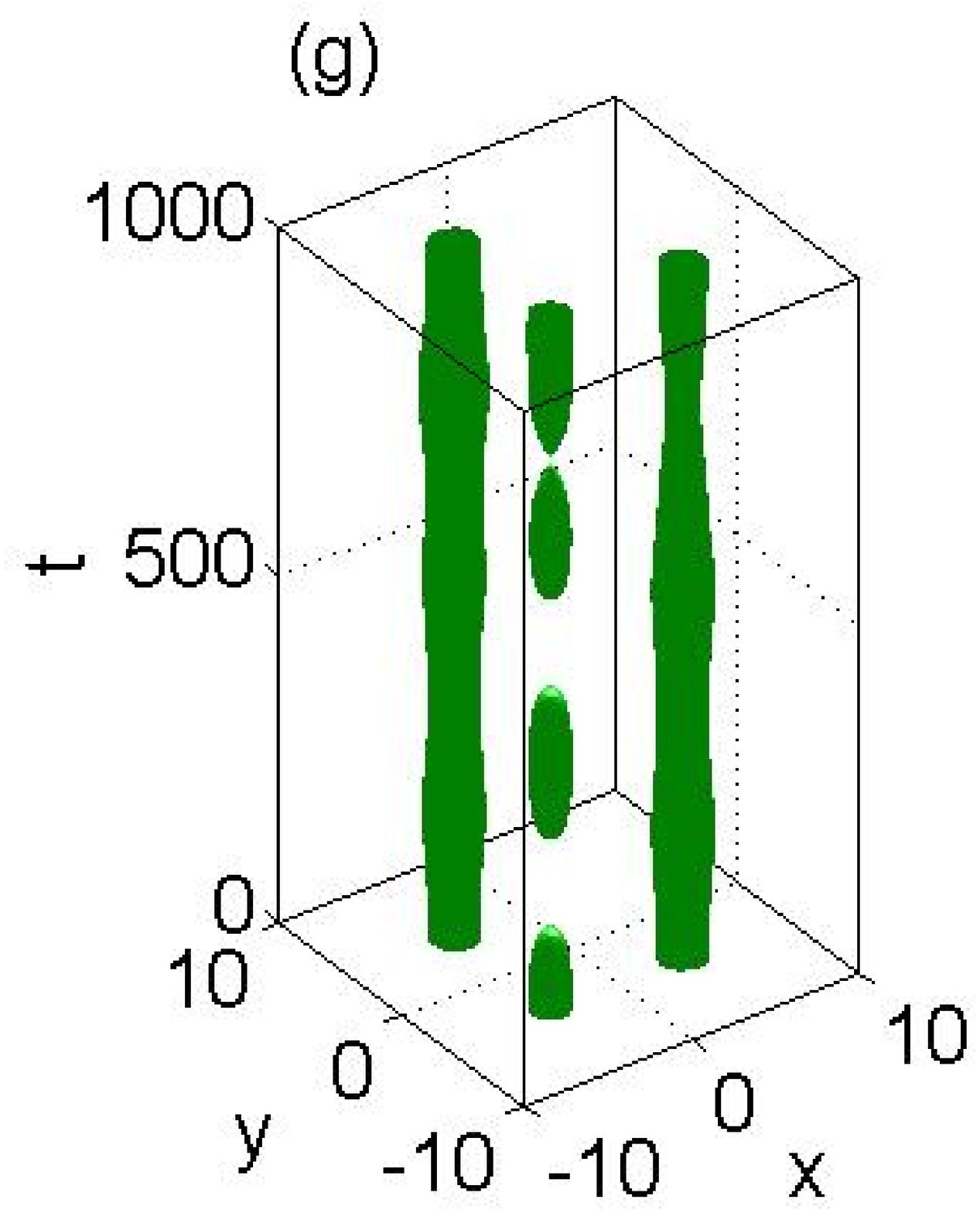} %
\includegraphics[width=.2\textwidth]{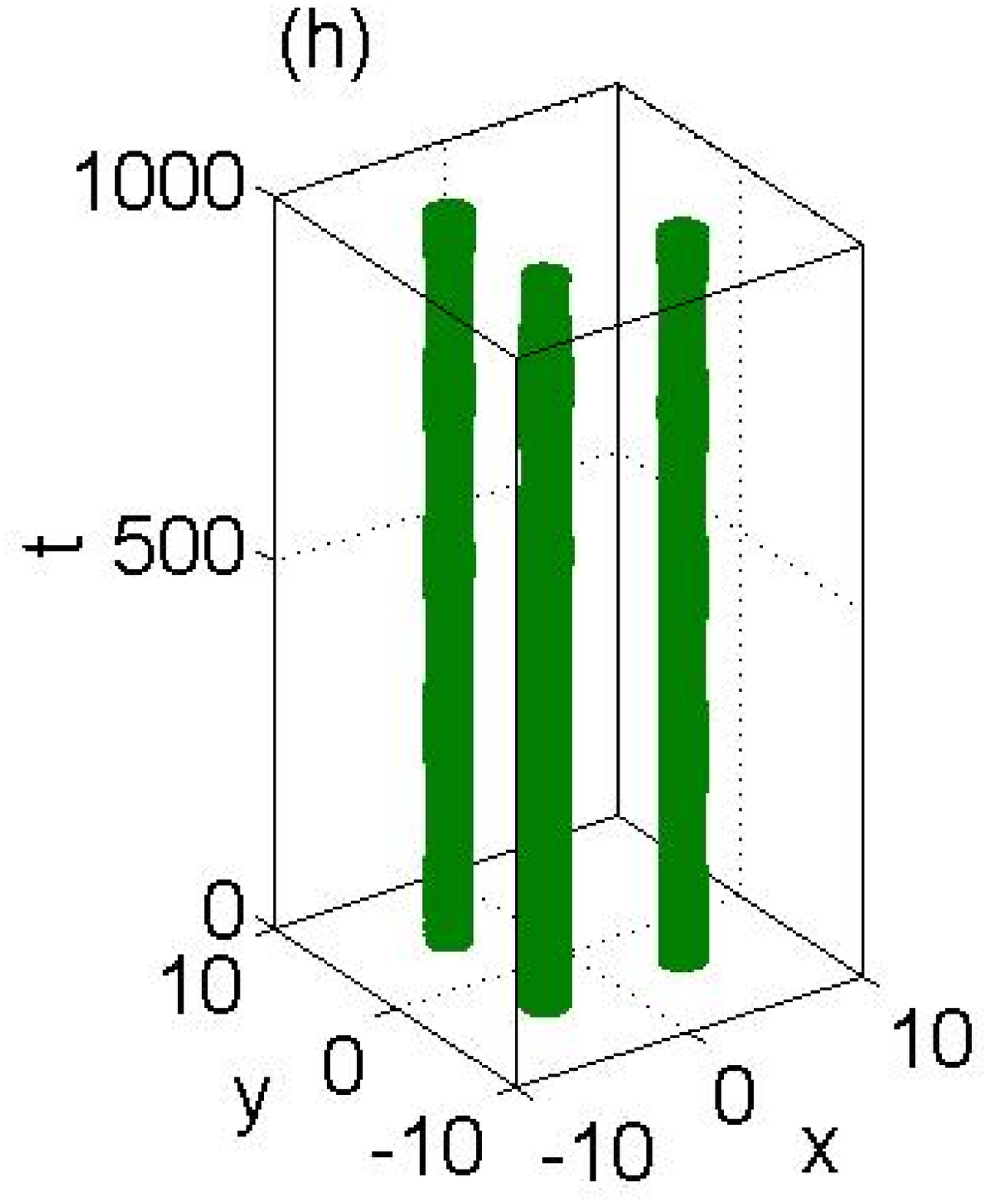} %
\includegraphics[width=.2\textwidth]{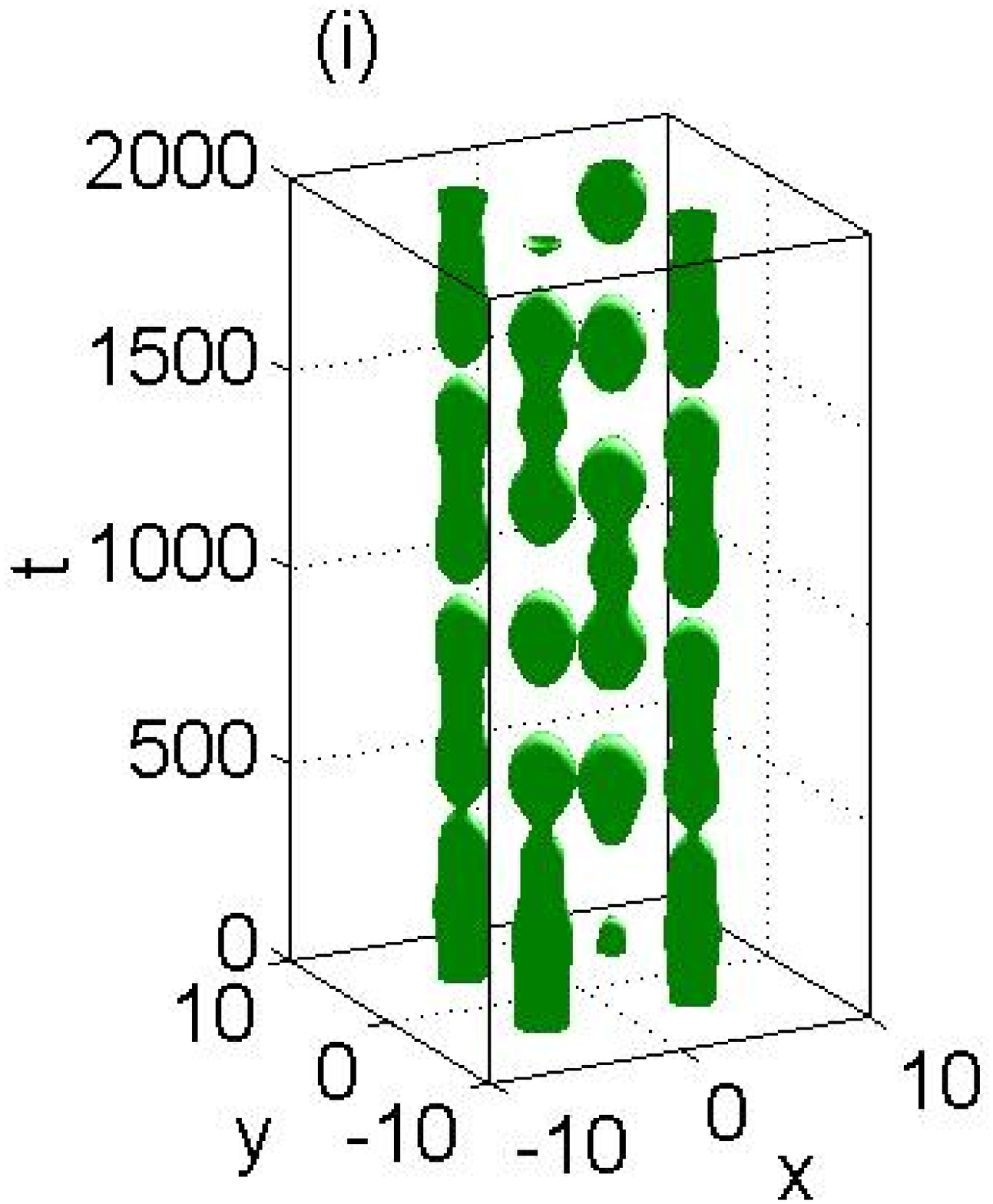} %
\includegraphics[width=.2\textwidth]{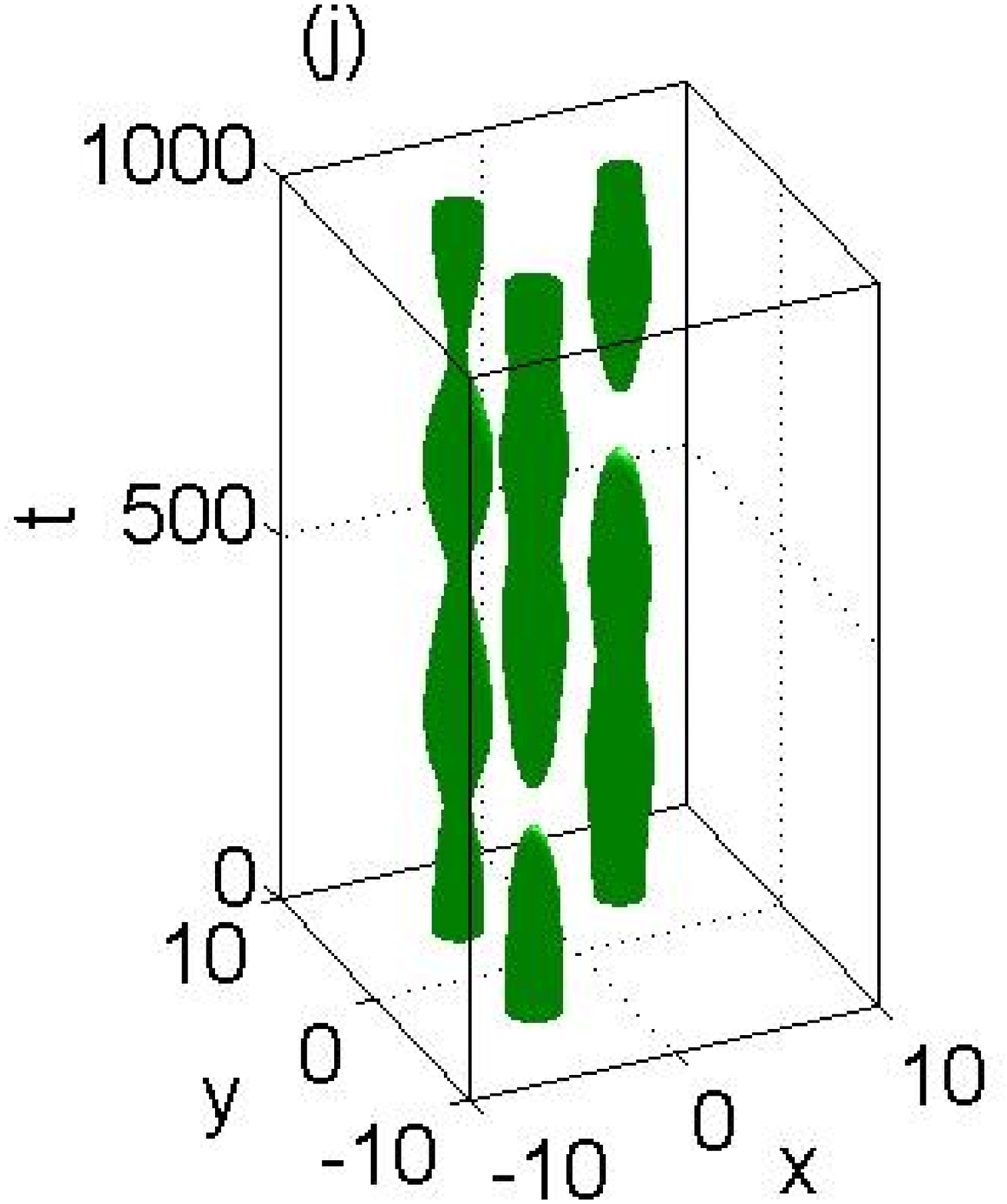}\newline
\caption{(Color online) The spatiotemporal evolution of unstable 
%solutions,
states, represented by the respective density isosurface, 
$|u(x,y,t)|^{2}=k$, where constant $k$ is taken as half the 
maximum value of the density distribution at $t=0$. 
The results are arranged as follows. Top panels: A1, A2, A4;
middle panels: B1, B2, B3; bottom panels: C2, C3, C4, D1.}
\label{evolution}
\end{figure}

%Lastly, we examine 
We now proceed to investigate the evolution of unstable 
%solutions 
states in the model with the self-focusing nonlinearity. 
To this end, for each unstable branch, a small 
perturbation is added to the most unstable eigendirection of the
linearization near the original stationary solution, at $\mu =0.335$.
Results of the simulations are presented in Fig. \ref{evolution}.

Panel (a) shows the behavior of solution A1, which, as a result of the
instability, starts oscillating between a state where all four wells are
populated and one in which only two diagonal wells are not empty. Panel (b)
depicts a periodic oscillatory behavior of A2, also between
%states (1 1; 1 1) and (0 1; 0 1),
four and two populated sites, but in this case the continuously populated
sites are adjacent to each other. Unstable mode A4 (panel (c)) features
only two non-empty wells, with 
the symmetry-breaking instability resulting in the
enhanced population of one of the two. As explained in section \ref%
{attractive}, modes B1, B2 and B3 have two very weakly populated wells, in
comparison with the other two. Since we employ isosurface $|u\left(
x,y,t\right) |^{2}=k$ (where $k$ is the half of the maximum density at $t=0$) 
to plot the dynamics in Fig. \ref{evolution}, the 
evolution in the weakly populated wells is not visible (which indicates 
that they play a minor role in the dynamics). Panel (d) shows that 
mode B1 sustains a symmetry breaking similar to that of A4, but between adjacent sites.
%oscillates between states (1 1; 0 0) and (1 0; 0 0).
On the other hand, modes B2 and B3 appear to be oscillating between the two
dominant wells roughly periodically, as shown in panels (e) and (f). Mode C2
[in panel (g)] oscillates between %(1 0; 1 1) and (1 0; 0 1).
three and two populated sites (the seemingly empty well is actually a weakly
populated one, similarly to the modes of type B, see above). Mode C3, whose
weak instability is caused by a quartet of eigenvalues, is also ``breathing''
within the respective set of three predominantly populated wells, as shown
in panel (h). Finally, mode C4 [shown in panel (i)] involves a complex
symmetry-breaking pattern, with different numbers of wells populated at
different times, while mode D1 (panel (j)) oscillates between three- and
two-well asymmetric configurations.

%Note that many of the general stability characteristics of the
%above-mentioned branches, which are valid at \textit{large }$N$, can be
%understood on the basis of a few simple principles developed in the context
%of discrete systems (from the so-called anti-continuum (AC)\ limit
%\cite{peli2d}) with the defocusing nonlinearity
%\cite{hadistuff}. In particular, branches of states which feature a single-site
%shape in the AC limit should be stable, those based on a set of two nearest
%neighbors with opposite signs produce real eigenvalue pairs, while
%out-of-phase sets lead to complex eigenvalue quartets. Finally, sets of
%in-phase next-nearest-neighbors lead to real pairs, while sets of the same
%type, but with the out-of-phase arrangement, may produce quartets.

%Naturally, these considerations do not apply to some of the asymmetric
%branches, such as C3 or D1, which cannot be examined in the AC setting.
%Nevertheless, this analysis provides a useful set of guidelines in
%understanding most of the stability features of the fundamental branches.

\section{Conclusion}

\label{conclusion}

In this work, we have studied stationary and dynamical properties of the
two-dimensional nonlinear Schr\"{o}dinger/Gross-Pitaevskii  
%model based on the GP/NLS\ (Gross-Pitaevskii/nonlinear Schr\"{o}dinger) 
equation, including a four-well external potential, with both 
signs of the nonlinearity, self-attractive (focusing) and self-repulsive
(defocusing). The model applies to a BEC confined in a highly anisotropic 
harmonic trap, with the transverse confining frequency being much smaller 
than the axial one, which results in a planar configuration. In this context, 
%, where 
the four-well potential can be generated by a combination of 
%an isotropic parabolic 
the transverse part of the harmonic trap and an optical lattice. The same 
model may describe the propagation of an optical beam in a bulk nonlinear 
medium with an embedded four-channel guiding structure. 

In our analysis, first we 
%have 
developed a four-mode approximation, which strongly simplifies 
the identification of 
%identifying 
stationary solutions. Using this approximation, we were able to find the 
four (two of which are identical) symmetric and antisymmetric linear modes, 
and all branches of asymmetric solutions emerging from them in the 
%self-focusing 
model with the self-focusing nonlinearity. The
linear-stability analysis 
%shows 
demonstrated how pitchfork and saddle-node bifurcations
change the stability of the branches. We have shown that 
in the limit of strong nonlinearity,
properties of localized modes in the model with either sign of the
nonlinearity can be, roughly, understood on the basis of earlier results 
%an understanding stemming from earlier works of the 
pertaining to the corresponding discrete NLS model. We have
also described the evolution of all unstable solutions, observing,
typically, the emergence of symmetry-breaking instabilities and the
emergence of respective oscillating solutions.

It would be interesting to investigate how these four-site configurations
may be embedded into a larger potential pattern, with $9$ or $16$ wells, and 
examine whether the symmetry-breaking bifurcations considered above are 
sustained (or how they are modified) within the larger pattern. In this 
context, a conjecture that would be worthwhile proving is that, in an 
infinite periodic lattice formed by potential wells, the nonlinearity can 
support 2D solitons and localized vortices with various symmetries, but not 
confined asymmetric states. This conjecture is suggested by results reported 
for infinite linear \cite{infinite} and nonlinear \cite{Barcelona} potential 
lattices.

The work of D.J.F. was partially supported by the Special Research Account 
of the University of Athens. P.G.K. gratefully acknowledges
support from NSF-DMS-0349023 and NSF-DMS-0806762, as well as from
the Alexander von Humboldt Foundation.

\end{document}